\documentclass{aa}  

\usepackage[switch]{lineno}
\usepackage{graphicx}
\usepackage{natbib}

\usepackage{supertabular}

\usepackage{txfonts}

\usepackage[]{hyperref}

\usepackage{lscape}

\usepackage[usenames]{color}

\newcommand{\bcep}{$\beta$~Cep }

\usepackage{amssymb}
\usepackage{mathrsfs}

\begin{document} 


   \title{Photometric detection of internal gravity waves in upper main-sequence stars}

   \subtitle{III. Comparison of amplitude spectrum fitting and Gaussian process regression using {\sc celerite2}}

   \titlerunning{GP regression of massive star light curves with {\sc celerite2}}
   
   \author{Dominic M. Bowman \inst{1,2} 
          \and
          Trevor Z. Dorn-Wallenstein \inst{3,4}
          }

    \institute{Institute of Astronomy, KU Leuven, Celestijnenlaan 200D, B-3001 Leuven, Belgium \\
              \email{dominic.bowman@kuleuven.be}
         \and
         Kavli Institute for Theoretical Physics, University of California, Santa Barbara, CA 93106, USA
         \and
         University of Washington Astronomy Department, Physics and Astronomy Building, 3910 15th Ave NE, Seattle, WA 98105, USA
         \and Observatories of the Carnegie Institution for Science, 813 Santa Barbara Street, Pasadena, CA 91101, USA
            }

   \date{Received 14 March 2022; accepted 14 November 2022}

  
 
  \abstract
   {Recent studies of massive stars using high-precision space photometry have revealed that they commonly exhibit stochastic low-frequency (SLF) variability. This has been interpreted as being caused by internal gravity waves (IGWs) excited at the interface of convective and radiative regions within stellar interiors, such as the convective core or sub-surface convection zones, or caused by dynamic turbulence excited in the sub-surface convection zones within the envelopes of main-sequence massive stars.}
   {We aim to compare the properties of SLF variability in massive main-sequence stars observed by the Transiting Exoplanet Survey Satellite (TESS) mission determined by different statistical methods, and confirm the correlation between the morphology of SLF variability and a star's location in the Hertzsprung--Russell (HR) diagram. We also aim to quantify the impact of data quality on the inferred SLF morphologies using both fitting methodologies.}
   {From a sample of 30 previously observed and characterised massive stars observed by TESS, we compare the resultant parameters of SLF variability, in particular the characteristic frequency, obtained from fitting the amplitude spectrum of the light curve with those inferred from fitting the covariance structure of the light curve using the {\sc celerite2} Gaussian process (GP) regression software and a damped SHO kernel.}
  {We find a difference in the characteristic frequency obtained from the amplitude spectrum fitting and from fitting the covariance structure of the light curve using a GP regression with {\sc celerite2} for only a minority of the considered sample. However, the trends among mass, age and the properties of SLF variability previously reported remain unaffected. We also find that the method of GP regression is more efficient in terms of computation time and, on average, more robust against the quality and noise properties of the input time series data in determining the properties of SLF variability.}
   {GP regression is a useful and novel methodology to efficiently characterise SLF variability in massive stars compared to previous techniques used in the literature. We conclude that the correlation between a star's SLF variability, in particular the characteristic frequency, and its location in the HR~diagram is robust for main-sequence massive stars. There also exists a distribution in the stochasticity of SLF variability in massive stars, which indicates that the coherency of SLF variability is also a function of mass and age in massive stars.}


   \maketitle


\section{Introduction}
\label{subsection: intro}

The treasure trove of time series photometric data assembled by space telescopes including CoRoT \citep{Auvergne2009}, Kepler/K2 \citep{Borucki2010, Koch2010, Howell2014} and TESS \citep{Ricker2015} in the last decade has revealed a remarkable range of diverse variability mechanisms in massive stars \citep{Bowman2020c}. A recent discovery is that essentially all early-type main-sequence stars show stochastic low-frequency variability (SLF) in time series photometry \citep{Bowman2019a, Bowman2019b, Bowman2020b}, in addition to other variability mechanisms including coherent heat-driven pulsation modes, rotational modulation, and the photometric signatures of binarity \citep{Degroote2009b, Blomme2011b, Pedersen2019a, Burssens2020a}. Similarly, SLF variability has also been reported in later evolutionary stages of massive stars (see e.g., \citealt{Bowman2019b, Dorn-Wallenstein2019a, Dorn-Wallenstein2020b, Naze2021b, Elliott2022a, Lenoir-Craig2022a}), which shows similar morphologies to main sequence stars. It currently remains unclear which of the several plausible mechanisms that can generate SLF variability dominate in particular parameter regimes. Hence the investigation of SLF variability has opened up a novel method to understanding the interiors of massive stars \citep{Bowman2020c}.

Coupled with spectroscopy, the analysis of time series photometry is able to extract detailed and precise constraints on the physics of stellar interiors thanks to asteroseismology \citep{ASTERO_BOOK, Aerts2021a}. In early-type stars, a heat-engine excitation mechanism operating in the iron opacity bump at 200\,000~K gives rise to pressure (p) and gravity (g) modes \citep{Dziembowski1993e, Dziembowski1993f, Miglio2007a, Szewczuk2017a}, which probe predominantly the envelope and near-core physics, respectively. The analysis of coherent pulsations has revealed the interior rotation profiles and a diversity in the amounts of convective boundary and envelope mixing profiles in late-B stars \citep{Degroote2009a, Degroote2010a, Daszy2013b, Moravveji2015b, Moravveji2016b, Szewczuk2018a, Szewczuk2022a, Pedersen2021a, Michielsen2021a, Bowman2021c}, and tight constraints on differential radial rotation profiles and (core) masses and ages for late-O and early-B stars \citep{Aerts2003d, Pamyat2004, Dziembowski2008, Burssens2022a**}.

Similarly to asteroseismology of coherent pulsation modes, the exploitation of SLF variability in time series photometry, which is seen across a wide range in mass, age and metallicity in massive stars, offers us the opportunity to improve our understanding of stellar structure and evolution \citep{Bowman2020c}. In their original methodology and characterisation study, \citet{Bowman2019a} used a sample of 35 early-type stars observed by the CoRoT mission to demonstrate how scaling laws based on solar granulation are inconsistent with the observed SLF variability\footnote{Sometimes referred to as a low-frequency power excess or `red noise' in the literature.} (Paper~I). Based on a much larger sample of more than 160 massive stars observed by Kepler/K2 and TESS, which included metal-rich galactic stars and metal-poor stars in the Large Magellanic Cloud (LMC), \citet{Bowman2019b} demonstrated that a common morphology in the SLF variability is observed for essentially all stars with $ M \gtrsim 3$~M$_{\odot}$. The SLF variability was determined to be dependent on the brightness of the star and by extension also its luminosity, and hence its location in the Hertzsprung--Russell (HR) diagram. More recently, \citet{Bowman2020b} used a sample of galactic O and B stars observed by the TESS mission and combined it with high resolution spectroscopy \citep{Simon-Diaz2011d, Burssens2020a} to show that the morphology of SLF variability is correlated with a star's mass, age, and amount of macroturbulence (Paper~II).

The detection of SLF variability in massive stars has inspired various theoretical works to explain this new type of observational signal. Motivated by the omnipresence of a convective core during the main-sequence evolution of a massive star led \citet{Aerts2015c} to postulate that SLF variability observed at the surface of three O-type stars could be explained by convectively driven internal gravity waves (IGWs) excited by core convection. This conclusion was supported based on the qualitatively similar frequency spectra observed for these stars with those predicted from 2D hydrodynamical simulations \citep{Rogers2013b, Rogers2015, Rogers2017c}. More recent 2D and 3D simulations of core convection with different numerical setups also show that IGWs excited by core convection produce detectable velocity and temperature perturbations \citep{Edelmann2019a, Horst2020a, Vanon2022a**, Herwig2022a**, Thompson_W_2022a**}. On the other hand, other studies have concluded that the cause of SLF variability in massive stars is IGWs excited by sub-surface convection zones \citep{Cantiello2009a, Cantiello2021b} since IGWs excited from core convection are likely damped before they reach the surface \citep{Lecoanet2019a, Lecoanet2021a}. Using simulations the dynamics of turbulent convection in massive star envelopes has also been shown to be an important cause of SLF variability \citep{Jiang_Y_2015, Jiang_Y_2018b, Schultz2022a}. However, the efficiency of convection in sub-surface zones, and hence the driving of IGWs from these regions, is strongly metallicity dependent \citep{Jermyn2022a}. Finally, another plausible mechanism for SLF variability, which is especially important for evolved (i.e post-main sequence) massive stars, is clumping in their optically thick winds (see e.g. \citealt{Krticka2018e, Krticka2021b}).

It remains an open question as to which excitation mechanism(s) dominates in specific parameter regimes of the HR~diagram, as the properties of (sub-surface) convection, winds, and the driving and propagation of IGWs depend on stellar properties such as mass, radius, metallicity and rotation. What is clear is that a mechanism that aims to unanimously explain SLF variability in main-sequence early-type stars must do so for stars that span a broad range of masses (i.e. $3 \lesssim M \lesssim 100$~M$_{\odot}$), ages from the zero-age main sequence (ZAMS) to the terminal-age main sequence (TAMS) and beyond, rotation rates between essentially zero and close to critical, and for both metal-poor (i.e. $Z \leq 0.002$) and metal-rich stars (i.e. $Z \geq 0.014$). A mechanism must also explain the broad frequency range of SLF variability and how its properties correlate with mass and age \citep{Bowman2019b, Bowman2020b}.

In this paper, we compare the difference in the measured properties of SLF variability when using different fitting methods: directly fitting the amplitude spectrum of the light curve as previously performed by \citet{Bowman2020b}, and covariance structure fitting of the light curve using the {\sc celerite2} Gaussian process (GP) regression software package \citep{Foreman-Mackey2017, Foreman-Mackey2018, Foreman-Mackey2021a}. Our goal is to demonstrate a more efficient method for extracting the characteristic frequency of the SLF variability for a large number of massive stars observed by the ongoing TESS mission. In section~\ref{section: method}, we outline the approach to using GP regression and briefly describe the stellar sample. In section~\ref{section: results}, we present the results of our comparison and we conclude in section~\ref{section: conclusions}.


\section{Method}
\label{section: method}

Given that SLF variability observed in massive stars is inherently a representation of a stochastic process, it is difficult to parametrise it with a physically motivated function in the time domain. This has led previous studies to characterise SLF variability instead in the Fourier domain \citep{Blomme2011b, Bowman2019a, Bowman2019b, Bowman2020b, Dorn-Wallenstein2019a, Dorn-Wallenstein2020b, Naze2021b, Elliott2022a}. However, there are various statistical tools and methodologies, such as GP regression \citep{Rasmussen2006, Aigrain2022a*}, low-order linear stochastic differential equations \citep{Koen2005e}, and continuous-time autoregressive moving average models \citep{Kelly2014}, that yield a model for stochastic signals in the time domain based on the statistical properties of time-series data.


	\subsection{Previous studies parameterising SLF variability}
	\label{subsection: previous studies}

	Previous efforts in parameterising SLF variability in massive stars have used a semi-Lorentzian function to fit the amplitude spectrum of the light curve with a from similar to:

	\begin{equation}
	\alpha \left( \nu \right) = \frac{ \alpha_{0} } { 1 + \left( \frac{\nu}{\nu_{\rm char}} \right)^{\gamma}} + C_{\rm w} ~ ,	
	\label{equation: red noise}
	\end{equation}
	
	\noindent where $\alpha_{0}$ represents the amplitude at a frequency of zero, $\gamma$ is the logarithmic amplitude gradient, $\nu_{\rm char}$ is the characteristic frequency, and $C_{\rm w}$ is a frequency-independent (i.e. white) noise term \citep{Bowman2019a}. We note that $C_{\rm w}$ is a parameter used to evaluate the white noise floor when fitting an amplitude spectrum, which is set by Poisson and instrumental errors and is not part of the spectrum of SLF variability. Hence, stars can have significantly different $C_{\rm w}$ values dependent on the quality of their light curves.
	
	Typically, to avoid potential bias in the resultant fit when comparing various stars using Eqn.~(\ref{equation: red noise}), the light curves of massive stars have undergone iterative pre-whitening to remove dominant periodicities a priori. This means that periodic variability caused by coherent pulsation modes or rotational modulation is removed from the light curve by means of subtracting sinusoids with optimised amplitudes, frequencies and phases to create a residual light curve \citep{Bowman2019a, Bowman2021c}. This is particularly important for B-type stars, such as \bcep and SPB stars, which are commonly multi-periodic pulsators in addition to having SLF variability \citep{Bowman2019b, Burssens2020a}. On the other hand, a two-step process of pre-whitening pulsation modes a priori and determining their frequencies, amplitudes, and phases may yield differences in the inferred morphology of the SLF variability compared to fitting the coherent modes and SLF variability simultaneously. Investigating the impact of this is beyond the scope of the current paper but will be explored in future work.
	
	 The fitting of the amplitude spectrum in previous studies has been achieved using a Markov chain Monte Carlo (MCMC) numerical framework (e.g. using the \texttt{python} {\sc emcee} package; \citealt{Foreman-Mackey2013, Foreman-Mackey2019a}). It was used to sample the posterior probability distribution to determine the optimum parameters for Eqn.~(\ref{equation: red noise}) using the amplitude spectrum of a star's light curve as observables. Yet the calculation of an amplitude spectrum and subsequent MCMC numerical framework is relatively computationally expensive and is also potentially quite sensitive to the explicit algorithm used to transform the light curve data from the time domain into the frequency domain. For example, the choices for the frequency over-sampling factor, frequency range, normalisation constant (e.g. the number of data points and/or the data set length when using the numerical implementation of a (discrete) Fourier transform (e.g. \citealt{Deeming1975, Kurtz1985b}) or (variant of) a Lomb-Scargle periodogram (e.g. \citealt{Scargle1982a, Press1989a}) should ideally not impact the results. Not all studies in the literature make comparable choices --- see discussion by \citet{Bowman2021c}. Moreover, differences in approaches to gap-filling a light curve to maximise its duty cycle, remove aliases and aim to minimise the noise floor in the amplitude spectrum, but different forms of post-processing data such as smoothing the amplitude spectrum prior to fitting may lead to different results. Although useful, these amplitude spectrum fitting methods implicitly include assumptions for the signal and noise components of a light curve and can affect subsequent (asteroseismic) inference.


	\subsection{Advantages of GP regression}
	\label{subsection: GP advantages}
	 
	 The application of GP regression has been highly successful in modelling radial velocity time series data, the light curves of transiting exoplanet host stars, and stellar variability including spot modulation in low-mass stars \citep{Brewer2009, Barclay2015a, Grunblatt2017a, Pereira2019, Nicholson2022}. In particular, a GP regression framework has been shown to be very effective at characterising stellar variability with periods shorter than the total time span of a time series, even in the case of degraded and noisy data sets (e.g. \citealt{Nicholson2022}). As pointed out by \citet{Pereira2019}, previous applications of GP regression in astrophysical contexts have typically focussed on determining accurate non-parametric models for time series data without particularly paying close attention to the physical processes that cause the variability. However, GP regression can be used under certain conditions to estimate physical parameters (e.g. such as characteristic timescales) that set the stochastic processes in time series data, such as modelling the signatures of granulation in the light curves of evolved low-mass stars \citep{Pereira2019}. In this work, we extend and test the application of covariance structure fitting of a light curve within a GP regression framework to extract physical parameters, in particular the characteristic frequency, of SLF variability in massive stars.
	 
	 Because of the potential impact of different choices when calculating amplitude spectra, it can be argued that fitting the covariance structure of a light curve with a GP regression framework, such as the {\sc celerite2} software package, is the more robust strategy since it is directly fitting the original measurement data set (i.e. the light curve). Relying on the light curve as the data set and avoiding transformations into the Fourier domain, a GP regression method avoids these potential discrepancies. See also the application and discussion of the advantages of GP regression regarding the window function of a time series by \citet{Nicholson2022}. On the other hand, covariance structure fitting methods also include assumptions, such as that the mean of the light curve is constant and stationary, hence have their own limitations.
	
	In this work, we investigated the efficacy of using GP regression for determining the properties of SLF variability in massive star light curves. In so doing, we restricted our sample to the 30 galactic O-type stars from the TESS sample previously studied by \citet{Bowman2020b} that do not have significant coherent pulsation modes and thus do not require iterative pre-whitening prior to fitting Eqn.~(\ref{equation: red noise}). Therefore, our comparison of fitting the amplitude spectrum and GP regression of the light curve was not influenced by any post-processing of the light curves, such as iterative pre-whitening.


	\subsection{Defining a GP kernel}
	\label{subsection: GP kernel}
	
	A GP regression model is a non-parametric model that describes correlated stochastic variability in a time series by fitting hyperparameters to define a covariance matrix. Each data point in a time series is described as a correlated random variable with a mean and a variance, for which any finite collection of variables has a multi-variate Gaussian distribution \citep{Rasmussen2006, Aigrain2022a*}. In GP regression, a function is not directly fitted to a data set, but rather hyperparameters are fit to define a covariance matrix. This yields a posterior distribution that is a representation of a distribution of functions with a given structure in their covariance matrices, which are then conditioned on the data to evaluate the best representation of the input time series. In this work, it is assumed that the variance of the input time series is constant and generally that the observed SLF variability is stationary.
 
	In the application of GP regression, a kernel to describe the correlation among data points is chosen, which is then applied to determine the best set of parameters to reproduce the observed time series (see e.g. \citealt{Pereira2019}). Posterior probability distributions for the relevant hyperparameters of the GP can be sampled using suitable priors, and the GP likelihood function (see equation 1 of \citealt{Foreman-Mackey2018}). Computing the likelihood function is typically the computational bottleneck, therefore the selection of an appropriate kernel, GP implementation, and sampling algorithm is crucial \citep{Foreman-Mackey2017}.
		
	In this work, we used the \texttt{python} {\sc celerite2} and {\sc exoplanet} software packages \citep{Foreman-Mackey2017, Foreman-Mackey2018, Foreman-Mackey2021a}\footnote{\url{https://github.com/exoplanet-dev/celerite2}} to fit a covariance structure for TESS light curves within a GP regression framework. Inspired by \citet{Pereira2019} in their application of GP regression to study the granulation signal of pulsating red giants and guided by the excellent documentation of the {\sc celerite2} software package, we utilised a GP kernel that corresponds to a damped simple harmonic oscillator (SHO), whose power spectral density is given by
	\begin{equation}
	S(\omega) = \sqrt{\frac{2}{\pi}} \frac{S_0 \omega^4_0}{\left(\omega^2 - \omega^2_0\right)^2 + \frac{\omega^2\omega^2_0}{Q^2}} ~ ,
	\label{equation: SHO}
	\end{equation}

	\noindent where $\omega_0$ is the natural frequency of the oscillator, $S_{0}$ is related to the amplitude of the variability, and $Q$ is the quality factor. For very large values of $Q$, Eqn.~(\ref{equation: SHO}) corresponds to an undamped SHO. As stated in equation~23 of \citet{Foreman-Mackey2017}, the kernel in the time domain is given by
	\begin{equation}
	\begin{array}{l}
	k_{\rm SHO}(\tau; S_0, Q, \omega_0) = S_0 \omega_0 Q e^{-\frac{\omega_0 \tau}{2Q}} \\
	\begin{cases}
		\cosh(\eta \omega_0 \tau) + \frac{1}{2 \eta Q} \sinh(\eta \omega_0 \tau), & 0 < Q < 1/2 \\
		2(1 + \omega_0 \tau), & Q=1/2 \\
		\cos(\eta \omega_0 \tau) + \frac{1}{2 \eta Q}\sin(\eta \omega_0 \tau), & 1/2 < Q \\
	\end{cases}
	\end{array}
	\label{equation: kernel}
	\end{equation}
	\noindent where $\eta = | 1 - (4Q^2)^{-1} |^{1/2}$, and $\tau$ is the (absolute) time separation between measurements needed to construct the covariance matrix (i.e., the $ij$ element of the covariance matrix uses $\tau = | t_i - t_j |$). In the limit of $Q = 1 / \sqrt{2}$ (see \citealt{Foreman-Mackey2017} for the derivation; omitted here for brevity), the power spectral density in Eqn.~\ref{equation: SHO} reduces to
	\begin{equation}
	S(\omega) = \sqrt{\frac{2}{\pi}} \frac{S_0}{\left(\frac{\omega}{\omega_0}\right)^4 + 1} ~.
	\label{equation: damped SHO}
	\end{equation}

	A GP regression of a time series can be parameterised using a characteristic variability timescale, $\rho_{\rm char} = 2\pi / \omega_0$, a characteristic amplitude\footnote{This is also sometimes referred to as the standard deviation of the variability process (e.g. \citealt{Pereira2019}).}, $\sigma_{\rm A} = \sqrt{S_0 \omega_0 Q}$, and a characteristic damping timescale, $\tau_{\rm damp} = 2Q / \omega_0$. The form of Eqn.~(\ref{equation: damped SHO}) is similar to the semi-Lorentzian function used to fit SLF variability in massive stars (cf. Eqn.~(\ref{equation: red noise})). Furthermore, a damped SHO kernel is a convenient starting point when a physically motivated kernel may not be known (see \citealt{Foreman-Mackey2017}). Given that SLF variability represents a stochastic oscillatory process, we deem this kernel appropriate for modelling TESS light curves of massive stars. Testing its efficacy to extract effective timescales (e.g. $\nu_{\rm char}$) from light curves is one of the goals of this work.


	\subsection{Input parameter guesses and priors}
	\label{subsection: priors}
	
	In our framework we used the kernel of a damped SHO and input guesses for its parameters as follows: $\omega_0$ as the characteristic angular frequency (i.e. $2\,\pi\,\nu_{\rm char}$); $Q = 1 / \sqrt{2}$; and $\sigma_{\rm A} = \alpha_{0} \sqrt{\omega_{0}\,Q} \times (\pi/2)^{0.25}$ based on roughly matching terms between Eqns.~(\ref{equation: red noise}) and (\ref{equation: damped SHO}). We note that our results are in fact not sensitive to the input guesses for these parameters as long as the domains of the priors are large enough to contain the true value. 
	
	Given the pristine quality of space photometry and the dominance of the astrophysical variability signal over measurement errors, in this work we assume that all possible sources of instrumental noise are white (see e.g. \citealt{Kallinger2014}). This may not be generally true for all stars observed by the TESS mission, but is a valid assumption for the bright massive stars dominated by SLF variability studied in this work. We emulated a white noise floor in the Fourier domain by including a noise jitter term in the GP regression following \citet{Pereira2019}. The jitter term is a constant added in quadrature to the diagonal of the covariance matrix and has a kernel function of:
	
	\begin{equation}
	k(t_i, t_j) = C_{\rm jitter}^{2} \, \delta_{i,j} ~ ,
	\end{equation}
	
	\noindent where $i$ and $j$ are matrix indices and $C_{\rm jitter}^{2}$ is the GP regression jitter term. Physically, the jitter term is used to represent uncorrelated noise in the data that cannot be captured by the GP regression, and hence represents a proxy of the measurement uncertainties in a light curve \citep{Foreman-Mackey2017, Foreman-Mackey2018, Foreman-Mackey2021a}. 
	
	In running a GP regression for a given star, there are five main parameters to be determined: (i) the characteristic timescale, $\rho_{\rm char}$ (from which the characteristic frequency can be determined: $\omega_0 = 2\pi\nu_{\rm char}$, where $\nu_{\rm char} = 1 / \rho_{\rm char}$); (ii) the characteristic damping timescale, $\tau_{\rm damp}$ (from which the characteristic damping frequency can be calculated: $\nu_{\rm damp} = 1/\tau_{\rm damp}$); (iii) the characteristic amplitude, $\sigma_{\rm A}$; (iv) a jitter term to emulate uncorrelated noise in the observations, $C_{\rm jitter}$; and (v) the mean of the light curve. We take the $\nu_{\rm char}$ parameters from \citet{Bowman2020b} multiplied by $2\pi$ to calculate $\omega_{0, {\rm guess}}$ as input guesses, and assign non-informative (i.e. uniform) priors in logarithmic space to all input parameter guesses except the mean of the light curve, which is assigned a normal prior based on the measured mean of the TESS light curve, which a priori has been set to be zero. We specified liberal lower and upper bounds for the four uniform priors as given in Table~\ref{table: priors}, but are results are not sensitive to these choices as long as the domains of the priors are sufficiently large. An input guess of the quality factor of the GP regression model is set as $Q_{\rm guess} = 1 / \sqrt{2}$, and is evaluated as $Q = \pi \tau_{\rm damp} / \rho_{\rm char}$.  
			
\begin{table}
\caption{Lower and upper limits of the four uniform logarithmic priors used in this work.} 
\begin{center}
\begin{tabular}{c c c}
\hline \hline
Parameter			&	Lower Bound	&	Upper Bound		\\
\hline
$\rho_{\rm char}$	&	$\ln\left(\frac{0.01}{\nu_{\rm char, guess}}\right)$							&	$\ln\left(\frac{100}{\nu_{\rm char, guess}}\right)$	\\
$\tau_{\rm damp}$	&	$\ln\left(0.01 \left(\frac{2Q_{\rm guess}}{\omega_{0, {\rm guess}}}\right)\right)$		&	$\ln\left(100\left(\frac{2Q_{\rm guess}}{\omega_{0, {\rm guess}}}\right)\right)$	\\
$\sigma_{\rm A}$	&	$\ln(\sigma_{\rm A}) = -10$											&	$\ln(\sigma_{\rm A}) = +10$	\\
$C_{\rm jitter}$		&	$\ln(C_{\rm jitter}) = -15$												&	$\ln(C_{\rm jitter}) = +15$		\\
\hline
\end{tabular} 
\end{center}
\label{table: priors}
\end{table}


	\subsection{Parameter and uncertainty estimation}
	\label{subsection: params}
	
	To estimate the best-fitting model parameters and their uncertainties we used the \texttt{python} {\sc pymc3} software package \citep{pymc3}\footnote{\url{https://github.com/pymc-devs/pymc}} and performed a Hamiltonian Monte-Carlo with a no U-turn sampler to determine parameter confidence intervals. In a standard MCMC approach, parameter chains make stochastic walks within the multi-dimensional parameter space, whereas a no U-turn sampler employs a Hamiltonian description of the probability distribution to sample the posterior probability distribution for the model parameters following Bayes' theorem. This makes a Hamiltonian Monte-Carlo no U-turn samplers typically much more efficient in finding global minima \citep{pymc3}.

	The best-fitting GP regression model is evaluated on the hyperparameters with maximum a posteriori probability. We use {\sc pymc3} and the maximum a posteriori parameters as an initial trial point to begin sampling the posterior. In our implementation of the Hamiltonian Monte-Carlo with a no U-turn sampler, we used four parameter chains, and at each iteration each chain is used to construct a model subject to a GP marginalised log likelihood function within the {\sc pymc3} model that also includes the log priors. A burn-in period of fewer than 1000 iterations was typically needed, which were discarded before parameter and uncertainty estimation were performed using another 1000 iterations. Convergence of the parameter chains is confirmed using the parameter variance criteria from \citet{Gelman1992}. We used the highest density interval (HDI) and 94\% of the posterior probability density to estimate the lower and upper confidence intervals for each marginalised parameter distribution. 

	Goodness-of-fit is assessed both by ensuring that good convergence statistics are obtained following parameter variance criteria \citep{Gelman1992} and visual inspection of the marginalised posterior distributions (i.e. corner plots; \citealt{Corner_2016}). We also visually inspect the posterior probability distribution, GP regression model and its corresponding power spectral density in the time and Fourier domain, respectively.
	
	We emphasise that our GP regression framework is a fit to the light curve and not a fit to an amplitude spectrum. But to aid in making our results visually comparable to the amplitude spectrum fitting results of \citet{Bowman2020b}, we calculated the power spectral density (i.e. amplitude squared per unit frequency as a function of frequency) of the TESS light curve and the resultant GP regression model, which was normalised using Parseval's theorem following \citet{Pereira2019}. Similarly, we also converted the fits from \citet{Bowman2020b} into units of power spectral density. We normalised all power spectral density figures to have the same amplitude at zero frequency (i.e. $\alpha_{0} = 1$), such that they converge at unity at zero frequency for ease of visual comparison. This is necessary to allow a consistent comparison of the power spectral density of the GP regression model including the jitter term, the power spectral density of the TESS data, and the converted fits of \citet{Bowman2020b} in the Fourier domain. Furthermore, this approach allowed us to perform a sanity check judgement of the quality of the best-fitting GP regression model to the data, since poor fits are more obvious in the Fourier domain\footnote{This is certainly true for an asteroseismologist who usually prefers to inspect amplitude spectra rather than light curves when analysing stellar variability.}. The conversion of the results into a power spectral density does not impact the inferred parameters, such as $\nu_{\rm char}$, as these are determined by the GP regression in the time domain.


\section{Results}
\label{section: results}

We applied our GP regression framework to the exact same TESS light curves of 30 massive stars previously analysed by \citet{Bowman2020b}. For each star, we provide a summary figure in Appendix~\ref{section: appendix: figures}, which contains the TESS observations and the best-fitting GP regression model in the top panel, and the power spectral densities of the TESS observations, the converted best-fitting model from \citet{Bowman2020b}, and the GP regression model in the bottom panel. A full summary of our parameter results from the GP regression and the upper and lower confidence intervals for each parameter from the Hamiltonian Monte-Carlo with no U-turn sampler is given in Table~\ref{table: params}. We note that for all summary figures in Appendix~\ref{section: appendix: figures} that the minimal informative frequency of the data set (i.e. resolution set by the inverse of the TESS light curve time span of $\geq 28$~d) is significantly smaller than the chosen 0.1~d$^{-1}$ lower limit of the x-axes.

\begin{figure}
\centering
\includegraphics[width=0.99\columnwidth]{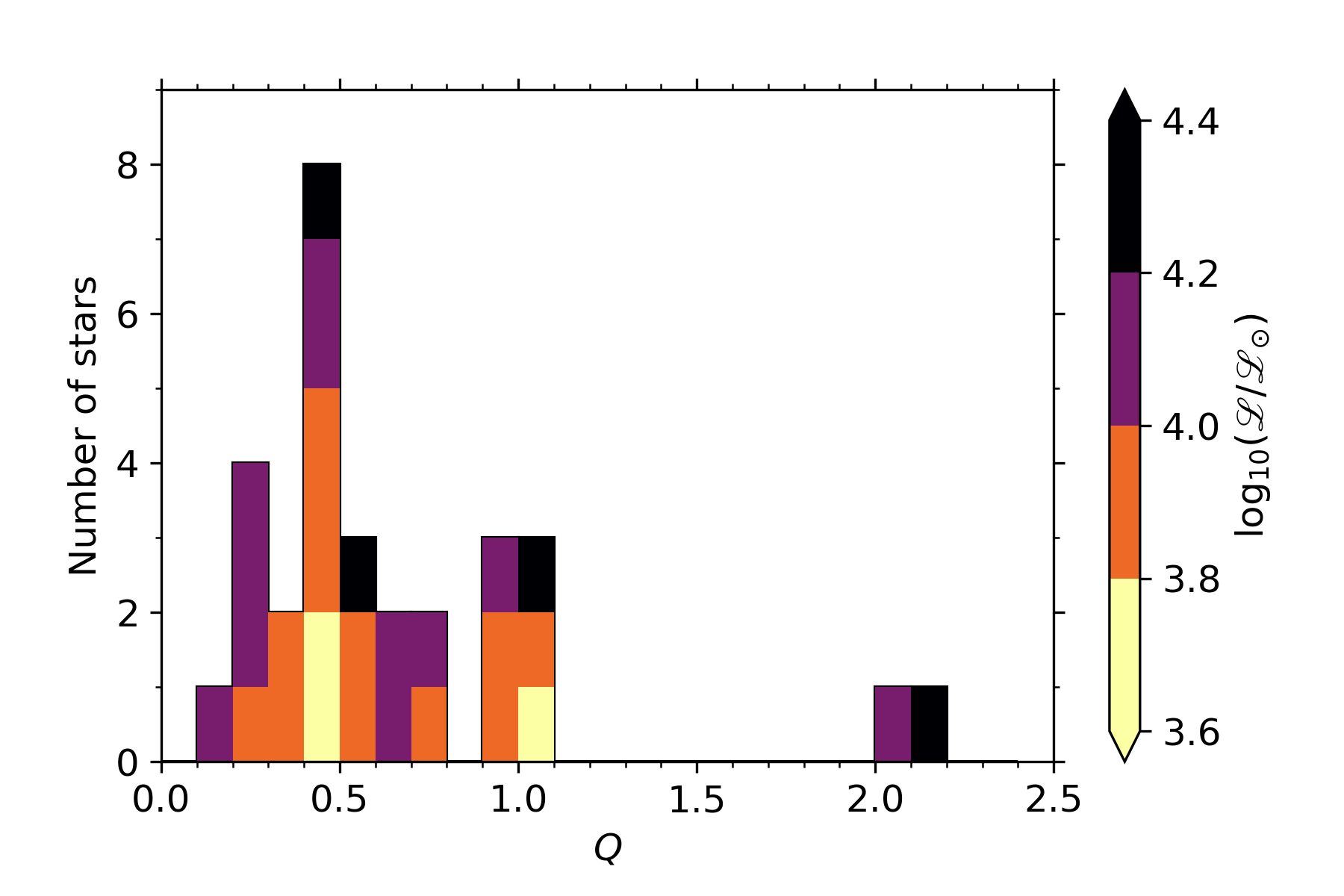}
\caption{Histogram of inferred $Q$ values from the GP regression colour coded by their spectroscopic effective luminosity. The determined values are clustered around $Q = 1 / \sqrt{2}$.}
\label{figure: Q histogram}
\end{figure}

An immediate major advantage of using a damped SHO model within a covariance structure fitting framework with the GP regression {\sc celerite2} software is that it provides additional useful inferred parameters of the SLF variability in massive stars, such as $\nu_{\rm damp}$ and $Q$, which are not provided by fitting Eqn.~(\ref{equation: red noise}) to an amplitude spectrum. Therefore, in addition to the characteristic timescale of the SLF variability, we are able to make inferences on its coherency and quasi-periodicity. We show the histogram of the inferred $Q$ values for the sample of 30 massive stars in Fig.~\ref{figure: Q histogram}, which has been colour-coded by the spectroscopic effective luminosity of each star defined as $\mathscr{L} := T_{\rm eff}^{4} / g$ \citep{Langer2014a}. We find a range of $Q$ values spanning from 0.18 to 2.15 that cluster around a value approximately equal to $1/\sqrt{2}$ and have a median value of 0.48. The inferred Q values represent a wide range in the coherency of SLF variability of massive stars. The two stars with the largest $Q$ values are HD~154368 (TIC~41792209) and HD~152424 (TIC~247267245). Both of these stars have large amplitude and quasi-coherent SLF variability in their light curves (see Fig.~\ref{figure: A3}) and are among the most massive and luminous stars in our sample.

	\subsection{Comparison of \citet{Bowman2020b} and GP regression}
	\label{subsection: comparison}
	
	\begin{figure*}
	\centering
	\includegraphics[width=0.99\columnwidth]{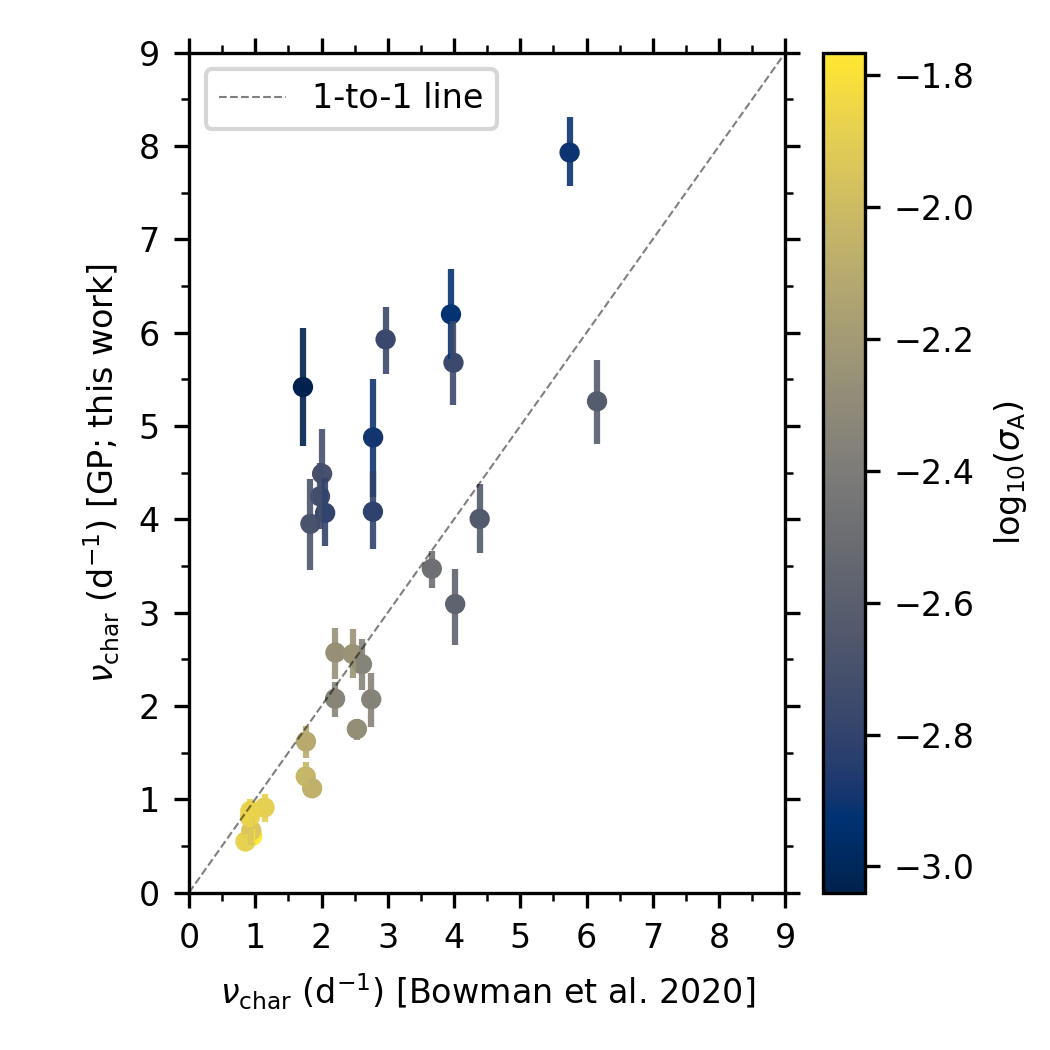}
	\includegraphics[width=0.99\columnwidth]{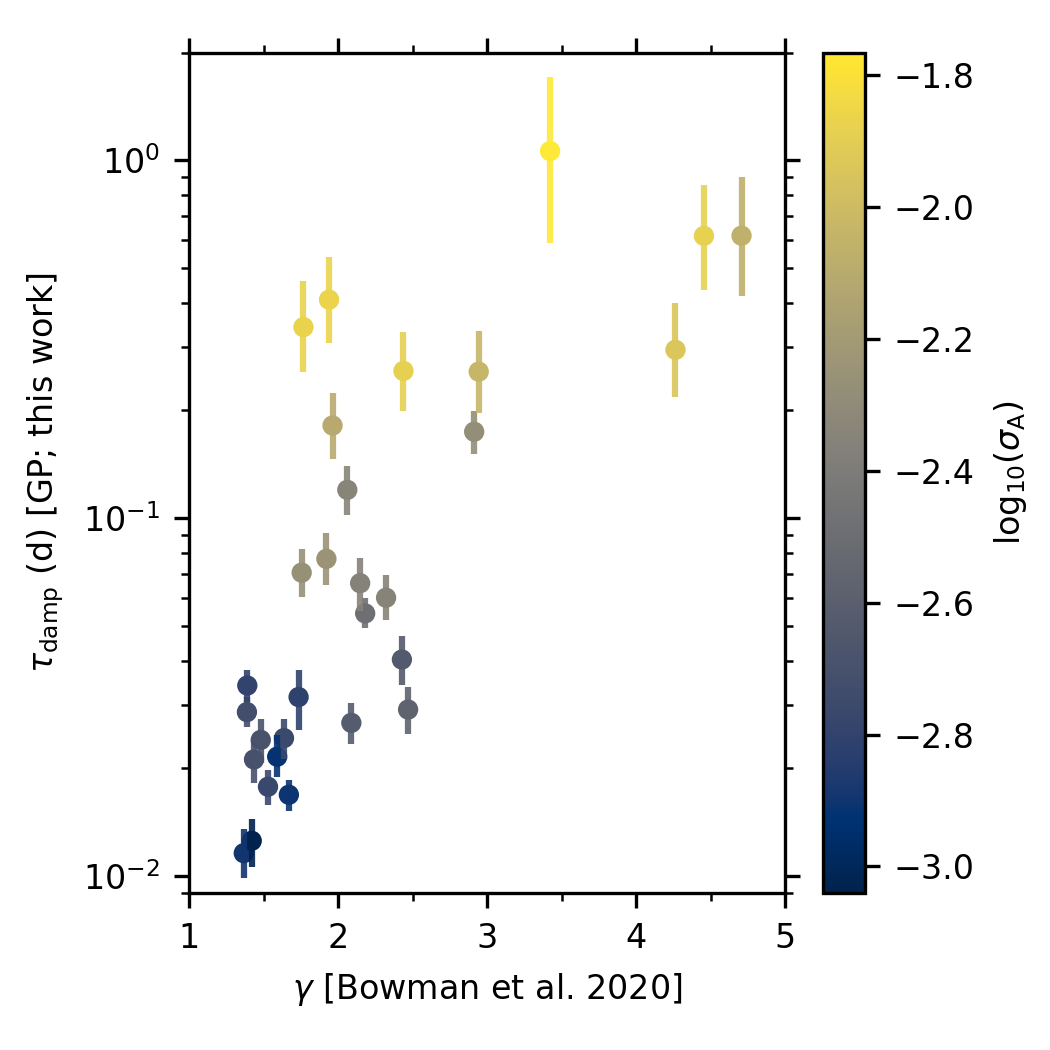}
	\includegraphics[width=0.99\columnwidth]{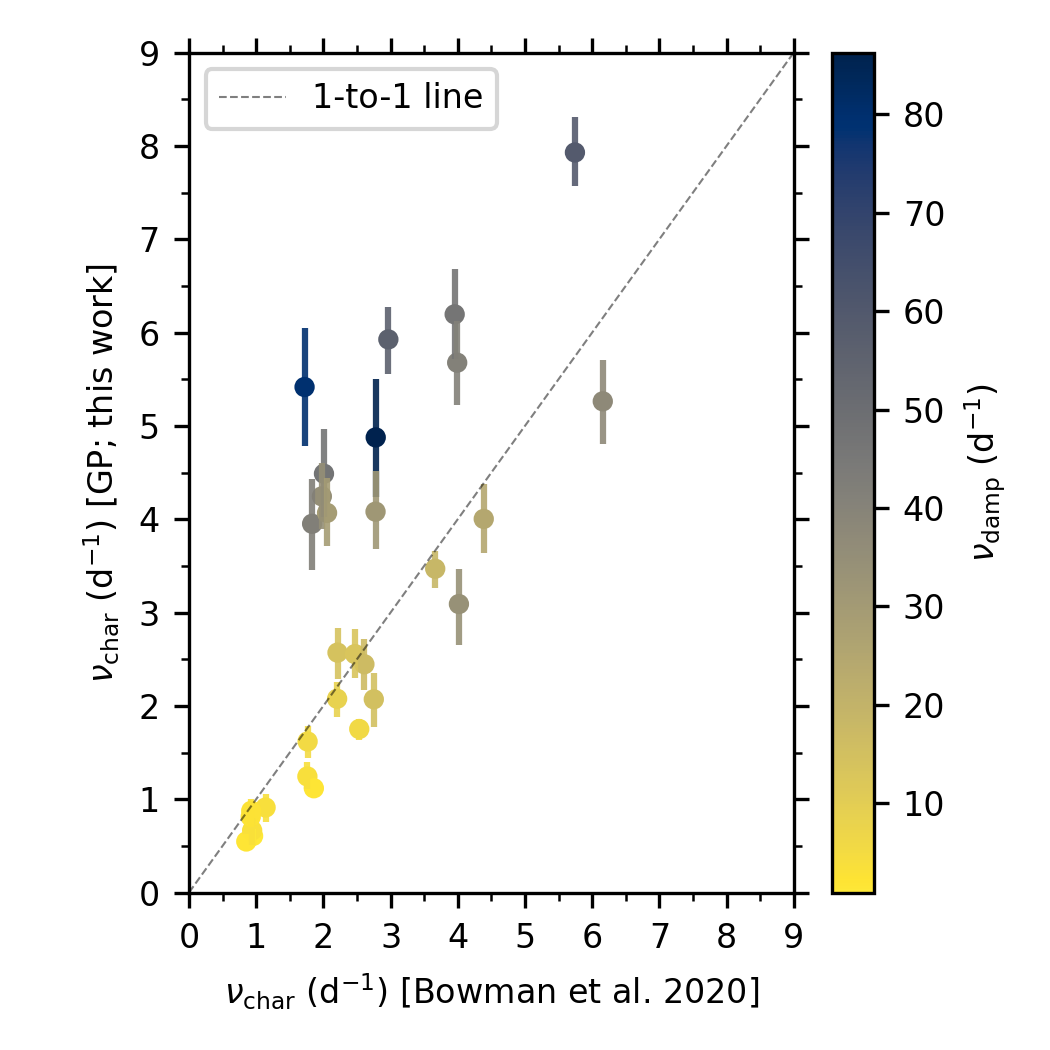}
	\includegraphics[width=0.99\columnwidth]{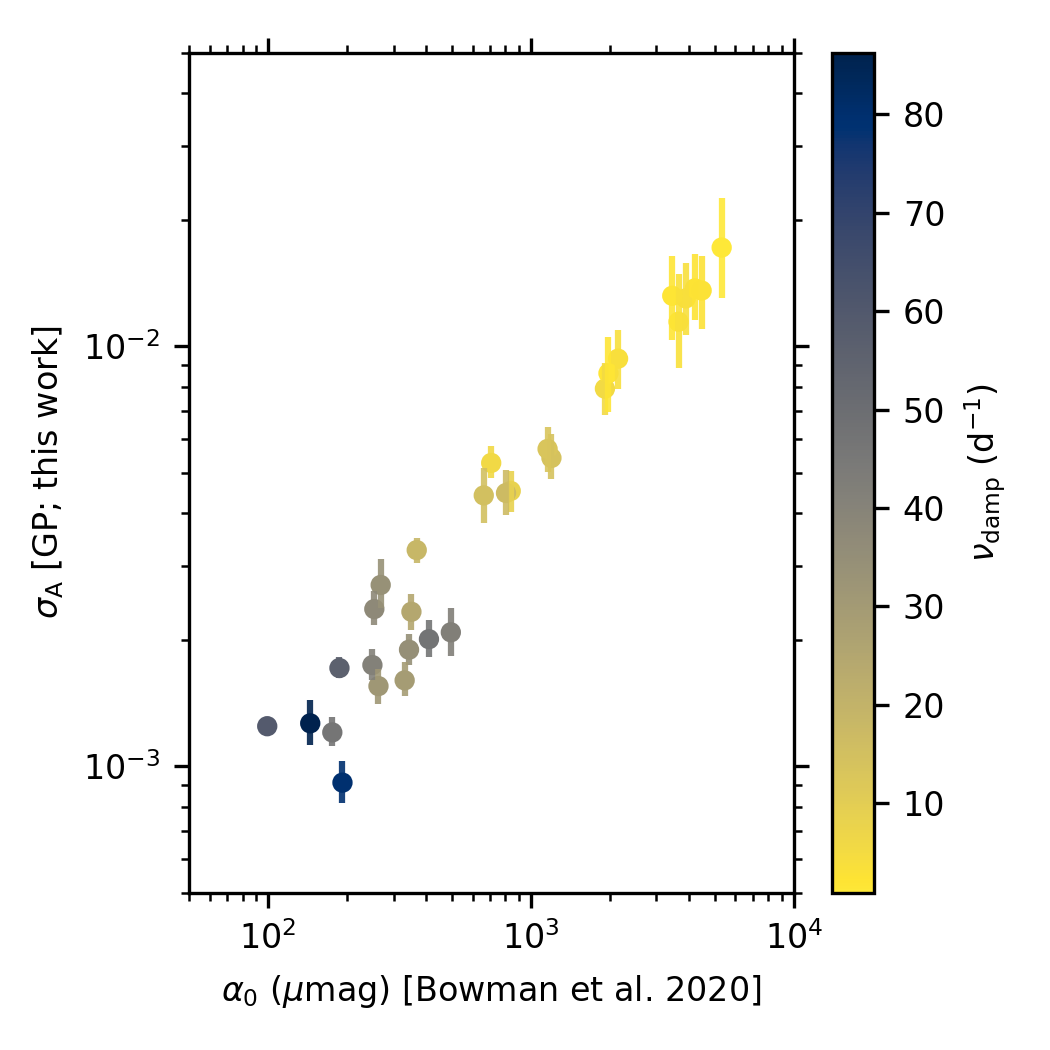}
	\caption{Comparison of GP regression parameters determined in this work and equivalent parameters from \citet{Bowman2020b}. {\it Top-left panel:} $\nu_{\rm char}$ values colour coded by $\sigma_{\rm A}$. {\it Top-right panel:} $\tau_{\rm damp}$ values from the GP regression and the steepness of the amplitude spectrum, $\gamma$, colour coded by $\sigma_{\rm A}$. {\it Bottom-left panel:} $\nu_{\rm char}$ values colour coded by $\nu_{\rm damp}$. {\it Bottom-right panel:} $\sigma_{\rm A}$ values from the GP regression and $\alpha_0$ colour coded by $\nu_{\rm damp}$.}
	\label{figure: pairwise}
	\end{figure*}
	
	In the top- and bottom-left panels of Fig.~\ref{figure: pairwise}, we show the pairwise comparison of the $\nu_{\rm char}$ values determined in this work using GP regression and those determined by \citet{Bowman2020b}, which are colour coded by the $\sigma_{\rm A}$ and $\nu_{\rm damp}$ GP regression parameters, respectively. On average, a fairly good agreement is seen between the two sets of $\nu_{\rm char}$ values. Yet a small fraction of the sample of stars have $\nu_{\rm char}$ values that differ by more than a factor of two. In some cases this leads to significant differences when considering the relatively small uncertainties on this parameter, but this varies from star to star. The discrepancies in some stars are not necessarily systematic with $\nu_{\rm char}$. However, we note that stars with a significant difference in $\nu_{\rm char}$ values between the two methods have lower $\sigma_{\rm A}$ and higher $\nu_{\rm damp}$ values (i.e. the SLF variability is relatively low amplitude and more stochastic). Also, in cases when a significant difference is found, the GP regression value for $\nu_{\rm char}$ is typically larger than that of \citet{Bowman2020b}. This is possibly related to the difference between a GP regression incorporating a different model and being a fit to the light curve rather than a fit to the amplitude spectrum. In other words, the sensitivity of the GP regression to the window function of the time series data is in the time domain rather than the Fourier domain, such that stars with lower amplitude SLF variability have larger discrepancies between the two methods. However, a much larger sample than the 30 stars considered in this work is needed to further investigate this if this effect is astrophysical or methodological in origin.
	
	In the top-right panel of Fig.~\ref{figure: pairwise}, we show the relationship between the steepness of the SLF variability in the amplitude spectrum, $\gamma$, as determined by \citet{Bowman2020b} and the $\tau_{\rm damp}$ timescale colour-coded by $\sigma_{\rm A}$ as determined using GP regression in this work. Whilst there is considerable scatter between these two parameters, a positive correlation between $\gamma$ and $\tau_{\rm damp}$ can be seen. This correlation is caused by stars with SLF variability characterised with longer damping timescales, $\tau_{\rm damp}$, being less stochastic and more quasi-periodic (i.e. higher values of $Q$). In turn this leads to larger values of $\gamma$, and the majority of power in the SLF variability being concentrated near $\nu_{\rm char}$ as opposed to being spread over a wide range in frequency.
	
	In the bottom-right panel of Fig.~\ref{figure: pairwise}, we show the relationship between the amplitude of the SLF variability at zero frequency, $\alpha_0$, as determined by \citet{Bowman2020b} and the characteristic amplitude, $\sigma_{\rm A}$, in this work. A clear correlation between these parameters is seen, which demonstrates that they can serve as proxies of each other. We note that the relationship is almost linear in log-log space, with a linear regression of $\alpha_0$ and $\sigma_{\rm A}$ yielding a Pearson correlation coefficient of $r = 0.97$ ($p < 10^{-10}$) and a gradient of $0.73 \pm 0.03$. This is expected given the equivalence of these parameters in both methodologies, but the gradient of order unity and the extremely high degree of correlation is a useful sanity check.
	
	Our comparison of the GP regression parameters and those of \citet{Bowman2020b} leads us to define two sub-groups of stars: (i) those with high $\alpha_0$ (i.e. high $\sigma_{\rm A}$), low $\nu_{\rm char}$ (i.e. high $\tau_{\rm char}$), and low $\nu_{\rm damp}$ values; and conversely (ii) those with low $\sigma_{\rm A}$, high $\nu_{\rm char}$, and high $\nu_{\rm damp}$ values. We term these the `yellow' and `blue' sub-groups, respectively, as indicated by the colour coding in Fig.~\ref{figure: pairwise}. In other words, stars in the `yellow' sub-group typically have high-amplitude, quasi-periodic low-frequency variability, and stars in the `blue' group have low-amplitude, stochastic variability spanning a broad range in frequency, on average. These are not distinct populations and our sample in fact covers the transition between these two extremes. However, the definition of these two sub-groups serves to highlight the differences in the properties of SLF variability among a sample of massive stars.


	\subsection{Correlations with mass and age in HR~diagram}
	\label{subsection: HRD}
	
	\begin{figure*}
	\centering
	\includegraphics[width=0.99\columnwidth]{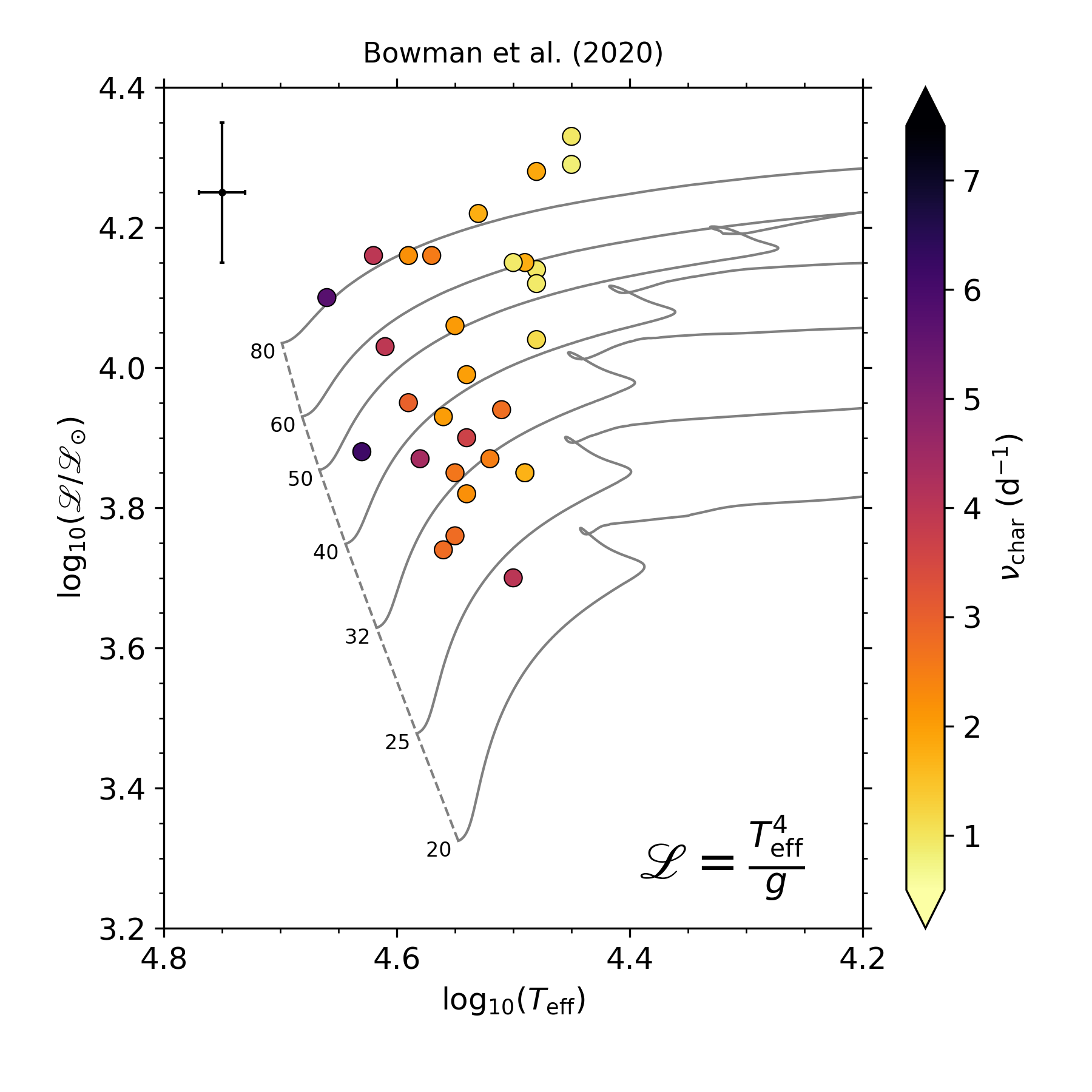}
	\includegraphics[width=0.99\columnwidth]{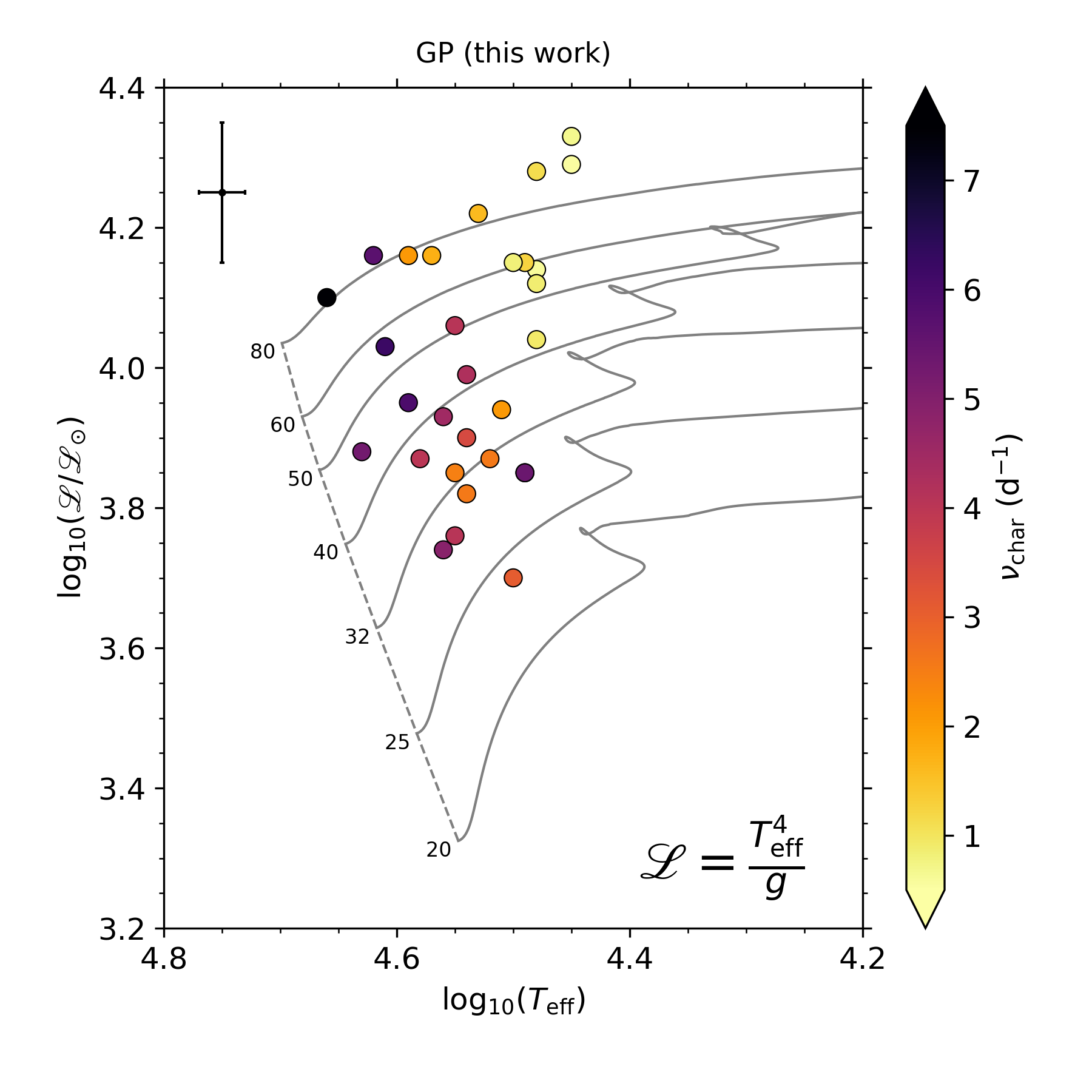}
	\caption{Spectroscopic HR~diagram of 30 O~stars determined by \citet{Bowman2020b} to be dominated by SLF variability in their TESS light curves ({\it left panel}) that we re-model using GP regression in this work ({\it right panel}), which are colour-coded by their corresponding $\nu_{\rm char}$ values. Evolutionary tracks from \citet{Burssens2020a} are shown as solid grey lines and labelled in units of M$_{\odot}$, and the dashed grey line represents the ZAMS. A typical spectroscopic error bar for the sample is shown in the top-left corner.}
	\label{figure: HRD}
	\end{figure*}
	
	In Fig.~\ref{figure: HRD} we show the spectroscopic HR~diagram for the 30 galactic O~stars studied by \citet{Bowman2020b}, which we have re-modelled using GP regression in this work. As discussed in section~\ref{subsection: comparison}, there are some star-to-star differences in the determined $\nu_{\rm char}$ parameters between the results of \citet{Bowman2020b} and from our GP regression. However, an important result from this work is that the dependence of $\nu_{\rm char}$ on the mass and age of a star remains unaffected since both independent fitting methods yield the same correlation. Therefore, as demonstrated by comparing the left and right panels of Fig.~\ref{figure: HRD}, our new modelling approach further strengthens the conclusion of \citet{Bowman2020b} that the light curve of a massive star encodes valuable constraints on its mass and age. More specifically, younger stars near the ZAMS have larger $\nu_{\rm char}$ values of approximately 7~d$^{-1}$, on average, whereas older stars near the TAMS have smaller $\nu_{\rm char}$ values of approximately 1~d$^{-1}$. 
		
\begin{figure}
	\centering
	\includegraphics[width=0.99\columnwidth]{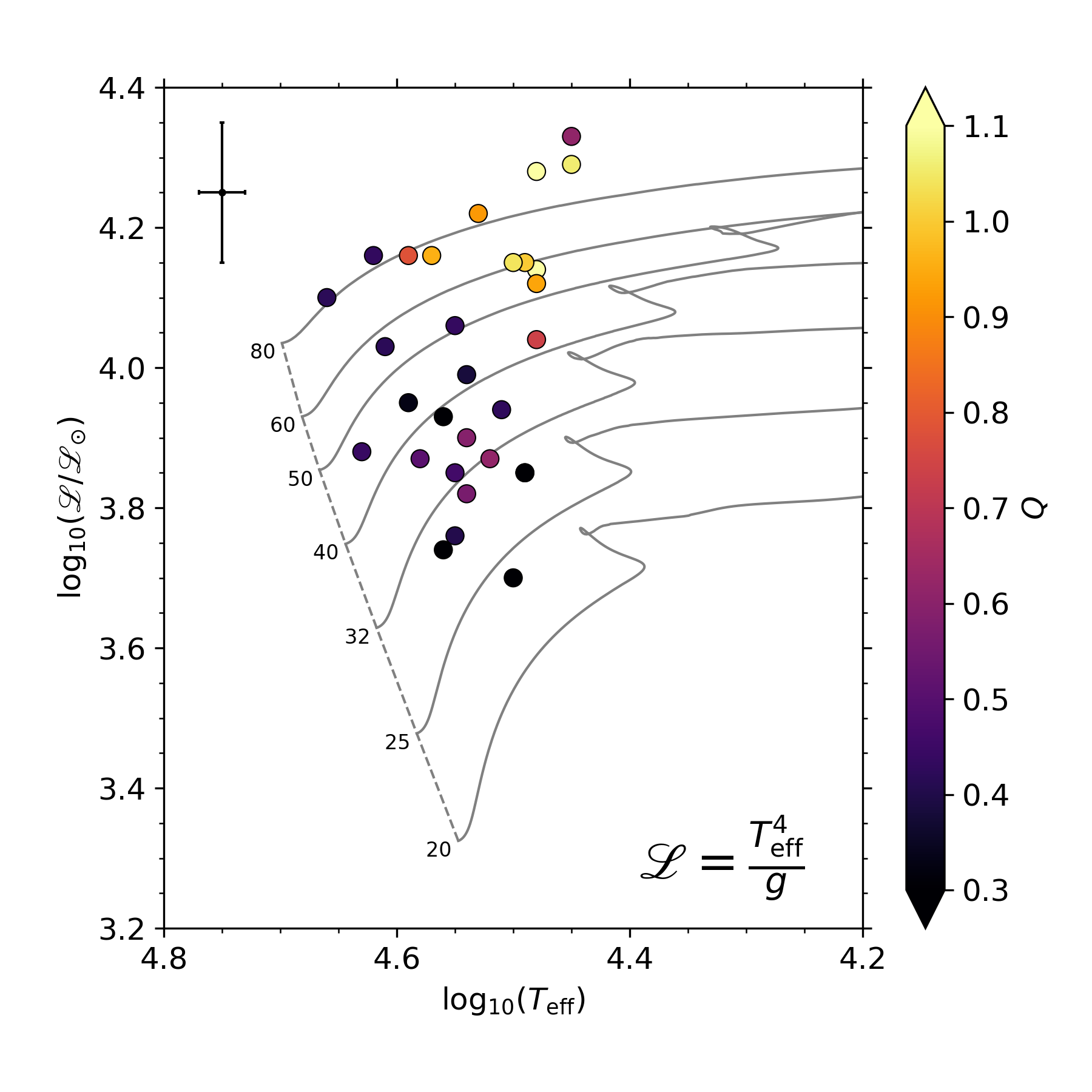}
	\caption{Spectroscopic HR~diagram of the same O~stars as in Fig.~\ref{figure: HRD} but colour-coded by the quality factor, $Q$, of the variability as determined in our GP regression.}
	\label{figure: HRD Q}
	\end{figure}

	Furthermore, we note that the `yellow' sub-group of stars (cf. section~\ref{subsection: comparison}) are typically higher mass and more evolved and the `blue' sub-group of stars are closer to the ZAMS on average. This is a new result of our work. Not only does the value of $\nu_{\rm char}$ of a star's SLF variability probe its mass and age, but the degree of coherency and stochasticity of a massive star's light curve is also a function of age. More specifically, more massive and more evolved stars typically have larger $Q$ values, as shown in Fig.~\ref{figure: HRD Q}. 
	
	The larger the $Q$ value in an oscillator, the narrower and sharper the peak in the power spectral density becomes because the system is more selective of what range of frequencies are driven to resonance. For example, systems with low quality factors (i.e. $Q < 1/\sqrt2$) are referred to as `overdamped' because damping dominates over driving. After a perturbation, such a system is returned asymptotically to its equilibrium steady state by means of an exponential decay. Whereas systems with high quality factors (i.e. $Q > 1/\sqrt2$) are referred to as `underdamped', which means they continue to oscillate at their resonant frequency whilst experiencing a decaying signal amplitude. This is because an underdamped system has driving that is able to overcome damping. For a perfect oscillation with zero damping the quality factor is infinite. Hence, coherent pulsators driven by the heat-engine mechanism, would have extremely high $Q$ values. 
	
	Given the measured $Q$ values of the stars shown in Fig.~\ref{figure: HRD Q}, our results demonstrate that SLF variability in massive stars transitions from being well described as an overdamped SHO to an underdamped SHO between the ZAMS and the TAMS in the HR~diagram. Therefore, the coherency of SLF variability as inferred using GP regression is a new avenue for constraining the responsible physical mechanism in different parameter regimes of the HR~diagram.
	
	
	\subsection{Is SLF variability quasi-periodic or stochastic?}
	\label{subsection: fixed Q}
				
	Two prime examples of stars in the `yellow' sub-group are HD~152424 (TIC~247267245) and HD~154368 (TIC~41792209), which have $Q = 2.0_{-0.8}^{+1.3}$ and $Q = 2.2_{-0.7}^{+1.0}$, respectively (see Table~\ref{table: params}). We note that there are relatively large uncertainties for the $Q$ values, the largest of all the stars, but place the stars in the `underdamped' regime. These stars have almost `perfect' GP regression fits to their light curves (see Appendix~\ref{section: appendix: figures}), and have light curves that are clearly dominated by astrophysical signal and negligible instrumental noise. Stars in the `yellow' sub-group are those that deserve special attention for the purpose of identifying and extracting quasi-coherent pulsation modes from their apparent SLF variability for the purpose of asteroseismology (see e.g. \citealt{Bowman2020c}). However, such an attempt would require long-term light curves spanning dozens of TESS sectors to resolve individual modes owing to their apparent intrinsically damped nature. This is beyond the scope of our current paper but is the subject of future work.
	
	We also perform a second modelling setup and perform GP regression for all stars in which we fix the quality factor to be $Q = 1/\sqrt2$. As a consequence, the damping timescale, $\tau_{\rm damp}$, is no longer included as a parameter. This allows for a better understanding of the measured $Q$ values when it is an inferred parameter and of the power spectral density predictions derived from the GP regression. We provide the parameters from the fixed-$Q$ GP regression in Table~\ref{table: params fixed Q}. Fixing $Q$ as a parameter forces the shape of the power spectral density prediction to be similar to that of Eqn.~\ref{equation: red noise}. This has the potential to explain the differences in $\nu_{\rm char}$ values between those from \citet{Bowman2020b} and from the non-fixed $Q$-value GP regression. However, the $\nu_{\rm char}$ values from the GP regressions with fixed and non-fixed $Q$ values are within 1$\sigma$ of each other, which demonstrates that we are somewhat insensitive to $Q$ when using GP regression or there is significant degeneracy between the $Q$ and $\nu_{\rm char}$ parameters.
	
	We also calculated and compared the Bayesian Information Criteria (BIC) values, which penalises models based on their complexity, using the GP log likelihood value returned by {\sc celerite2} software for the GP regression models when $Q$ is a free and a fixed parameter. In most cases, 20 of the 30 TESS light curves, the BIC values indicate that the solutions with $Q$ as a free parameter are strongly preferred. This is understandable since the added complexity is needed to more accurately describe the morphology of SLF variability in the time domain. A damped SHO model with a fixed Q value is less flexible, with the light curves of massive stars containing added complexity that is not well captured in this overly simplistic model. Moreover, the 10 stars in which the model with a fixed $Q$ value is preferred are generally the stars with noisier light curves (i.e. `blue' subgroup members) and have generally poorer fits, as can be seen when comparing the power spectral density of the GP solution and the TESS data in the Fourier domain. Therefore, in order to accurately capture the shape of SLF variability we conclude that the use of GP regression with $Q$ (and thus also $\tau_{\rm damp}$) as a free parameter is the more robust approach, on average.


	\subsection{Testing the impact of poorer quality observations}
	\label{subsection: quality}
	
	\begin{figure*}
	\centering
	\includegraphics[width=0.45\textwidth]{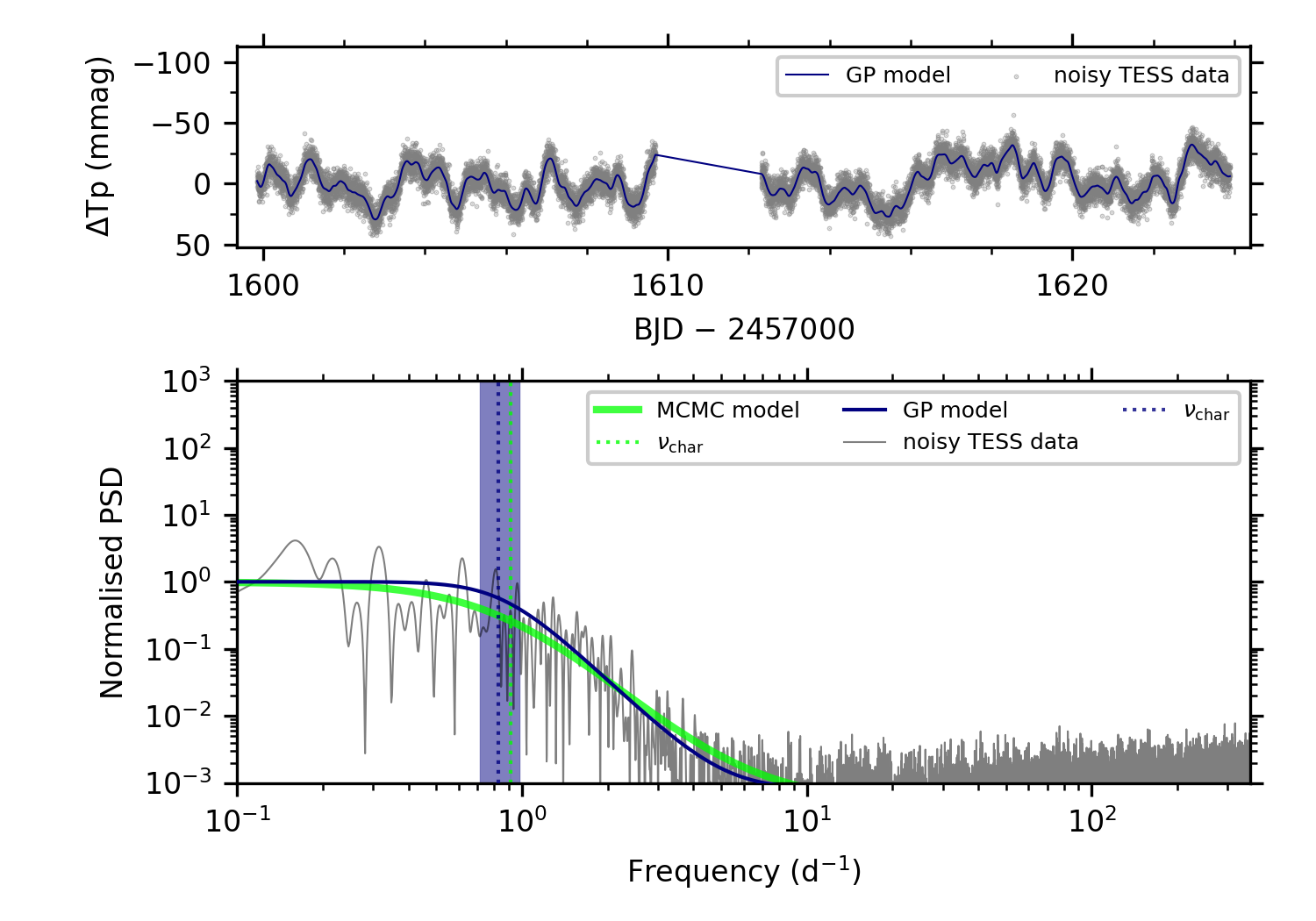}
	\includegraphics[width=0.45\textwidth]{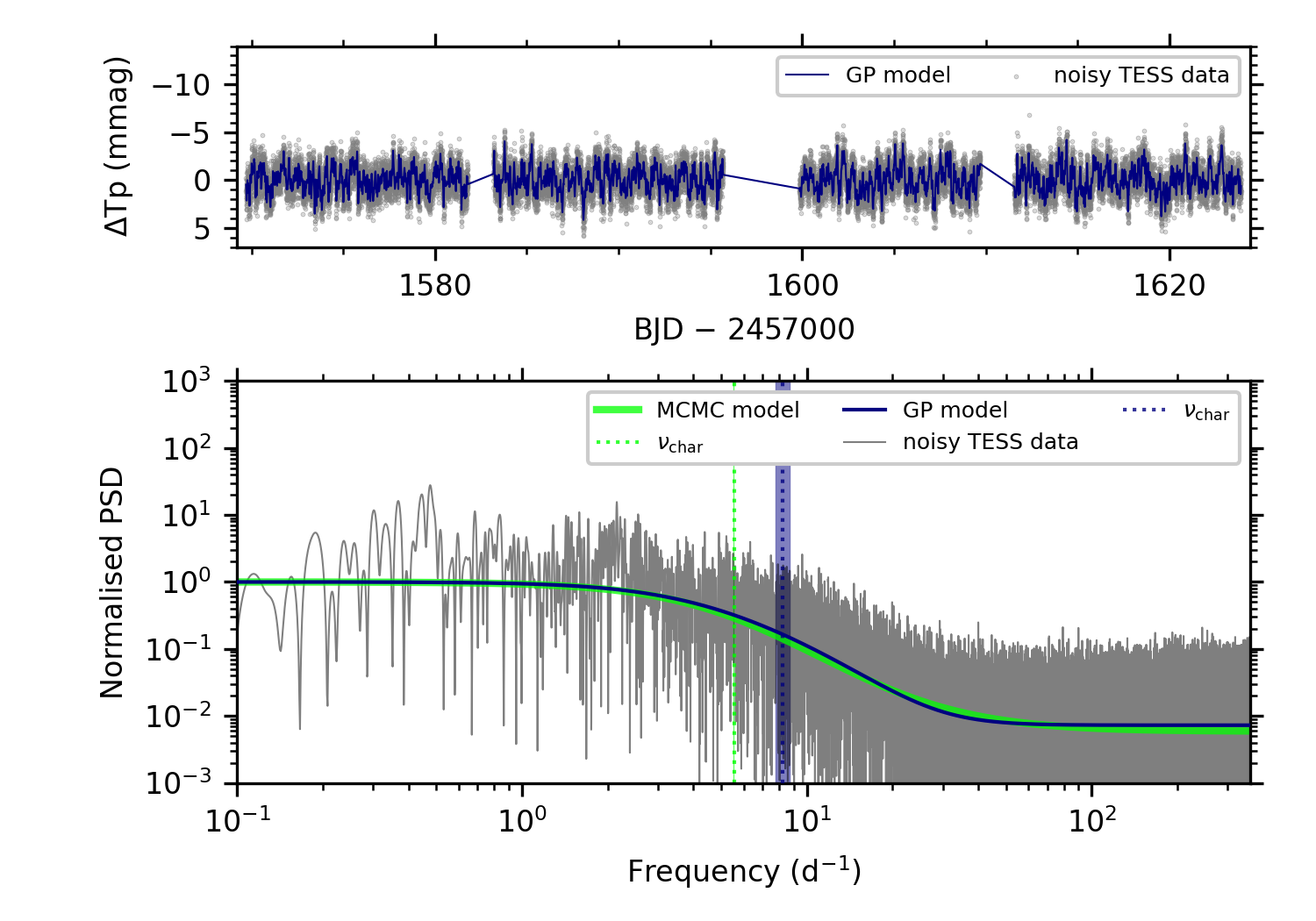} \\
	\includegraphics[width=0.45\textwidth]{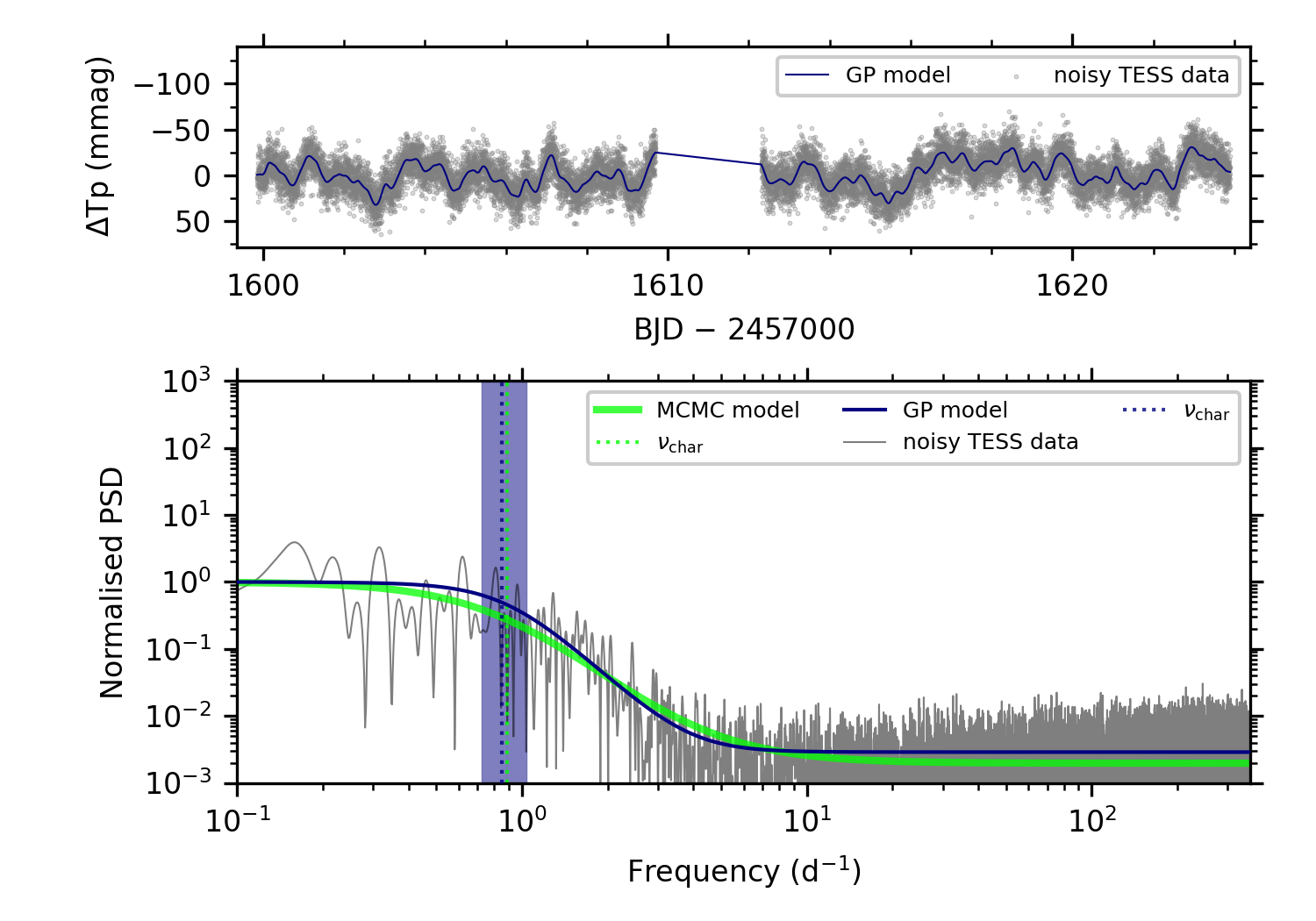}
	\includegraphics[width=0.45\textwidth]{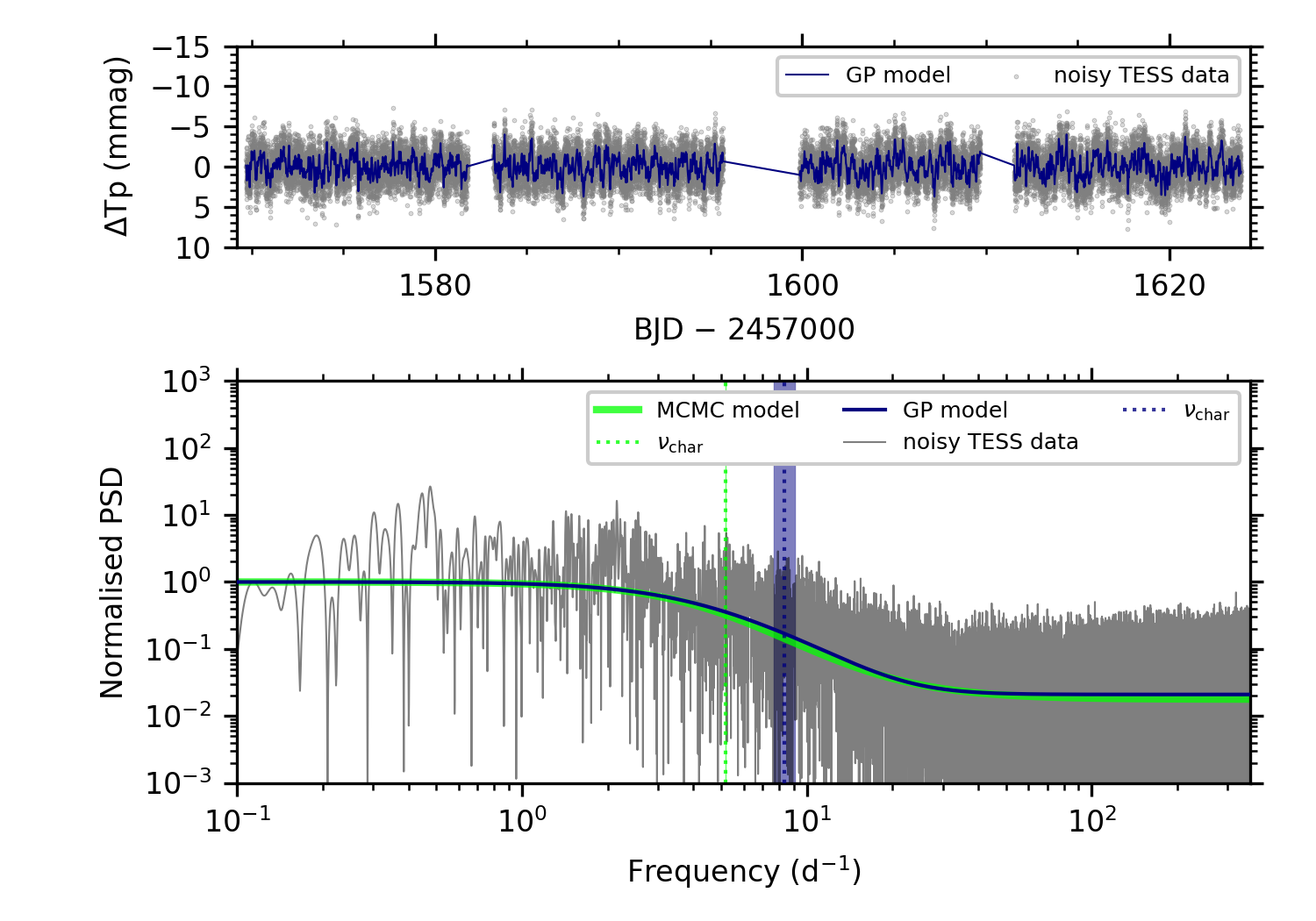} \\
	\includegraphics[width=0.45\textwidth]{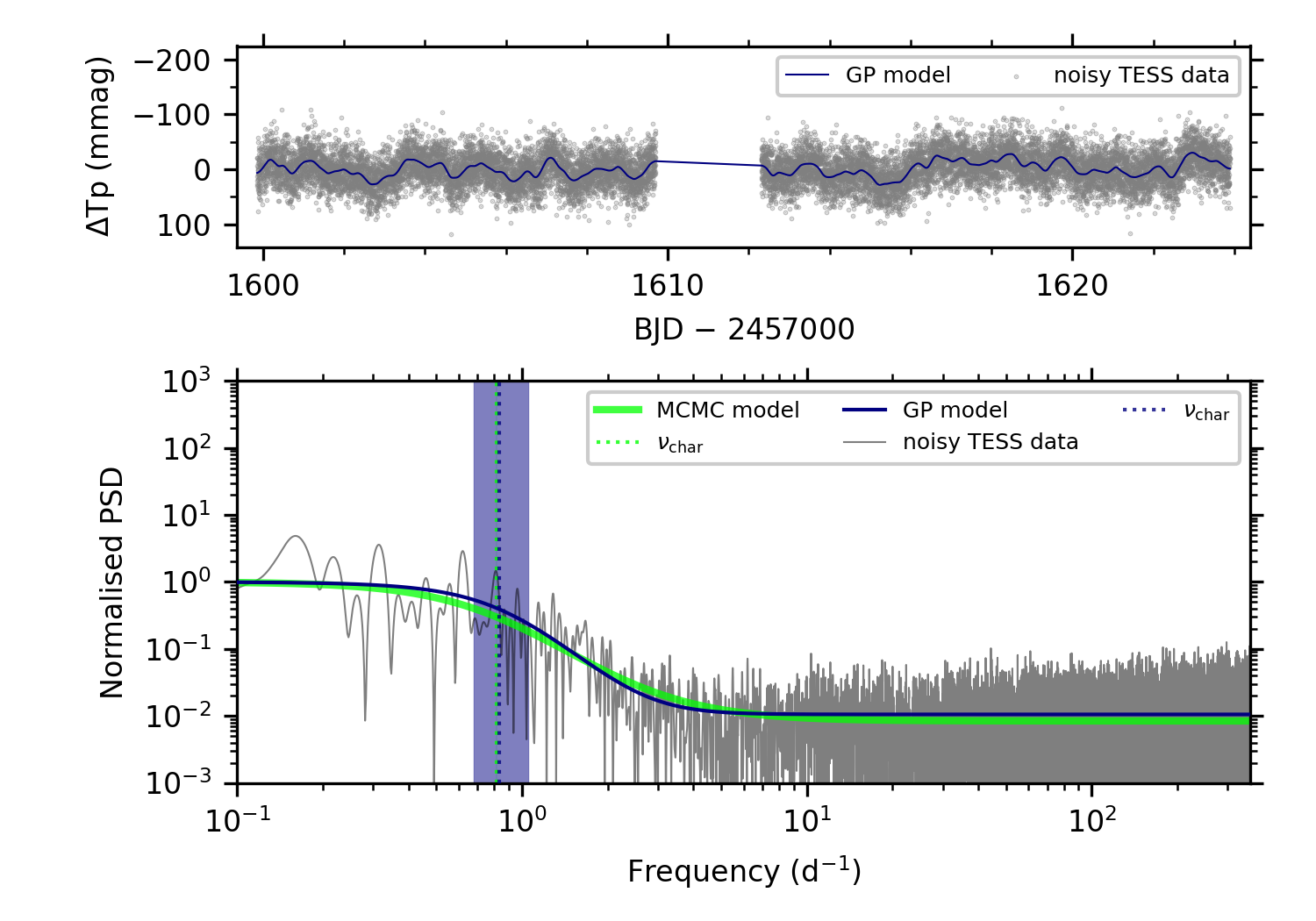}
	\includegraphics[width=0.45\textwidth]{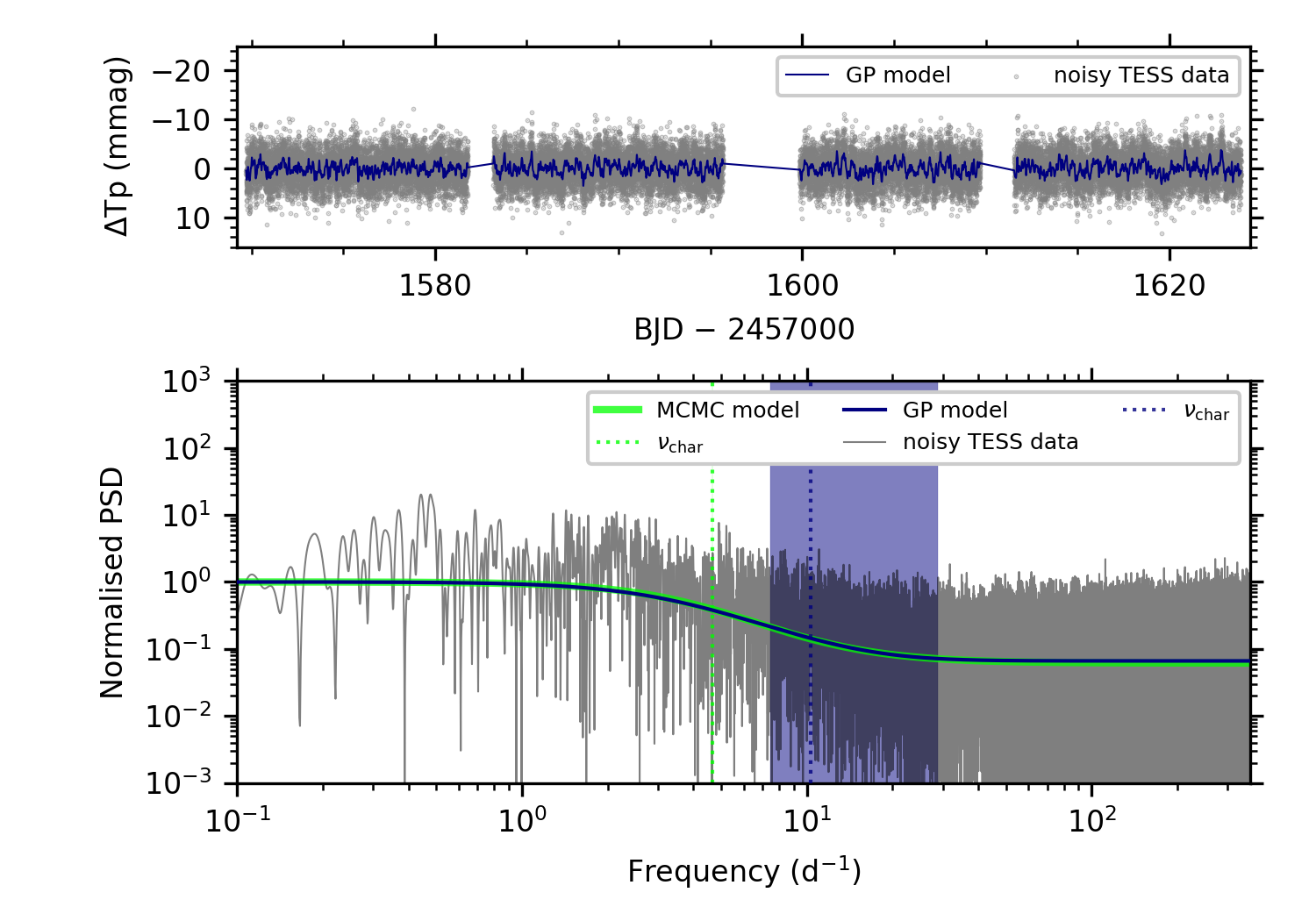} \\
	\includegraphics[width=0.45\textwidth]{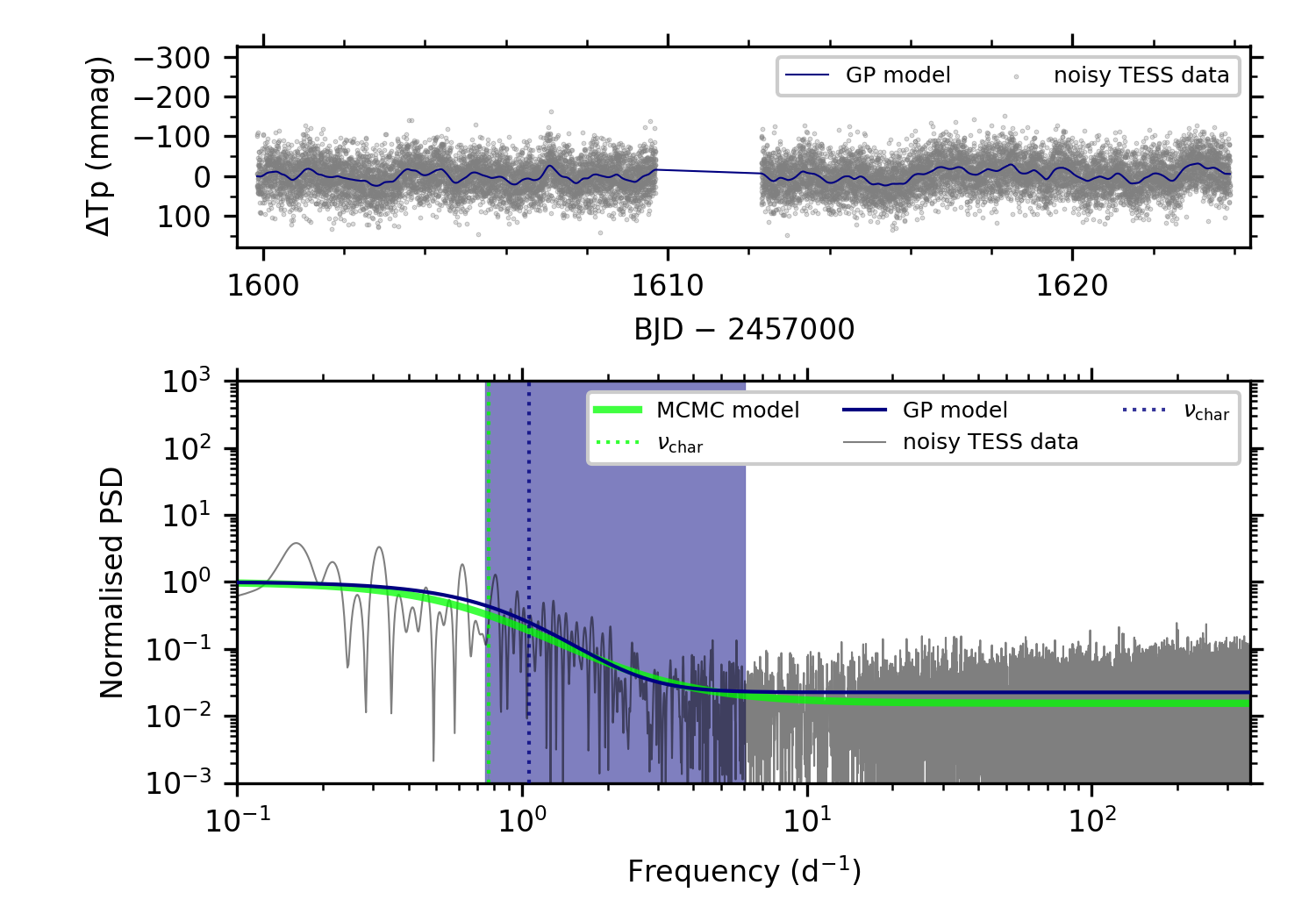}
	\includegraphics[width=0.45\textwidth]{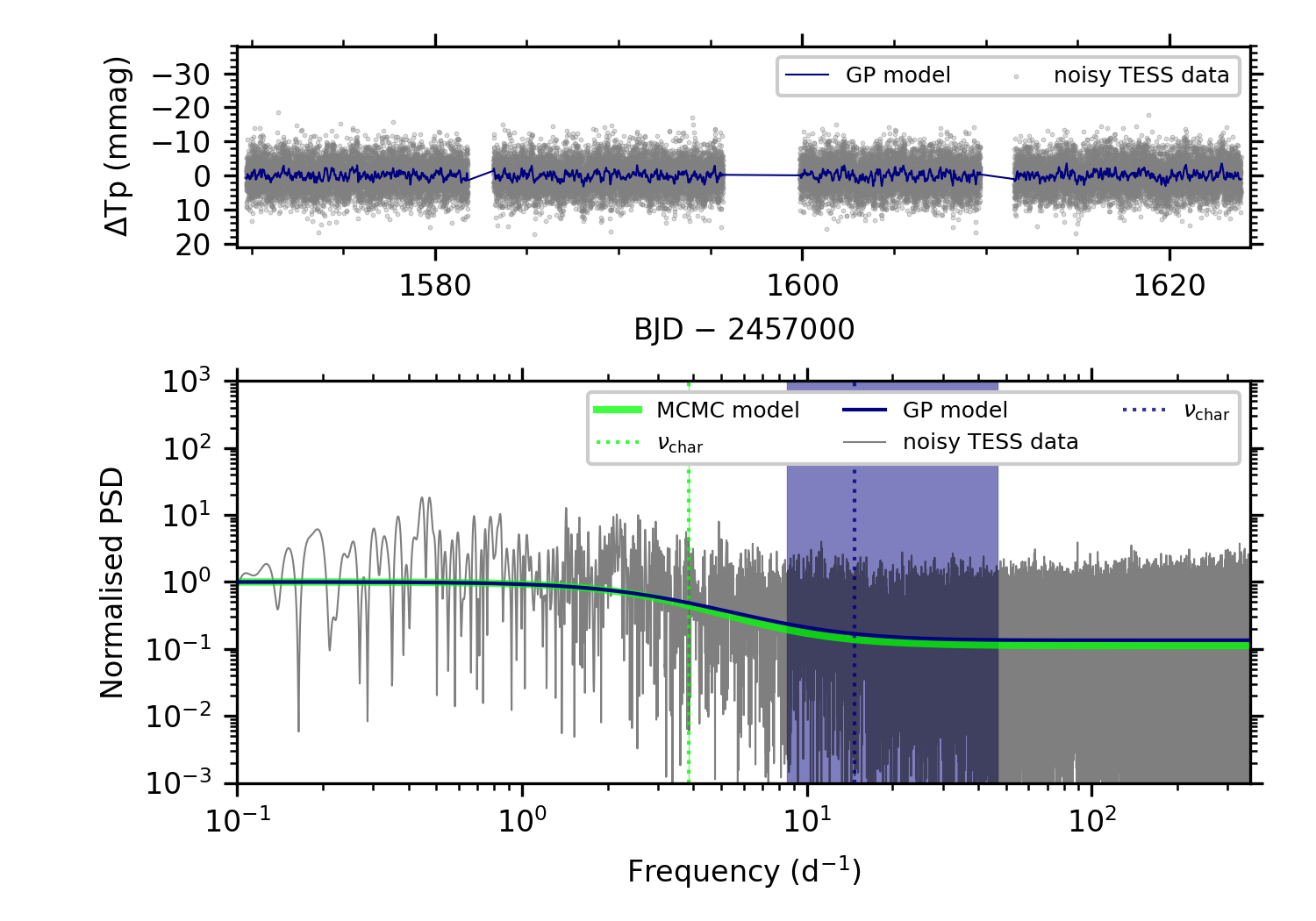}
	\caption{Results of GP regression and amplitude spectrum fitting for the noise-enhanced light curves of HD~112244 (TIC~406050497) and HD~96715 (TIC~306491594), which are typical `yellow' and `blue' subgroup members, are shown in the left and right columns, respectively. From top to bottom, increasingly larger amounts of white noise has been added. The blue and green shaded regions denote the 94\% confidence interval for $\nu_{\rm char}$ for each method. The $\nu_{\rm char}$ values from the GP regression are more consistent, on average, than those from fitting the amplitude spectrum.}
	\label{figure: noisy GP}
	\end{figure*}	
	
	We have demonstrated that the best-fitting parameters characterising SLF variability in massive stars are sometimes dependent on the choice of methodology, but this affects each star differently. This is expected because we are comparing the results of two different models each with different fitting methods. On the other hand, there is a general consistency in $\nu_{\rm char}$ values obtained from the two methods, but it generally depends on the amplitude of the SLF variability and the noise properties of the TESS light curve. For the `yellow' sub-group of stars, the astrophysical signal (i.e. SLF variability) strongly dominates over the noise. Whereas, some stars in the `blue' sub-group have SLF variability that is comparable to the white noise, and only detected because of the high precision and long duration of TESS light curves. It is the `blue' sub-group of stars that have, on average, larger inconsistencies in their inferred $\nu_{\rm char}$ values between the two methods (cf. Fig.~\ref{figure: pairwise}). In order to study SLF variability across a wide range of masses and ages, thus including the full range of SLF variability amplitudes (i.e. tens of $\mu$mag to tens of mmag; \citealt{Bowman2019a, Bowman2019b}), high-photometric precision and short-cadence space light curves are necessary to yield the necessarily low instrumental noise.

	For the sample at large, there is fairly good agreement between the results in this work and the alternative approach taken previously by \citet{Bowman2020b}. This leads us to wonder which of the two approaches is more sensitive to the quality of the observations and perhaps less robust considering the notable difference in $\nu_{\rm char}$ values for some stars. For example, not all stars in our sample are the same brightness, such that the underlying random and instrumental noise contributions in their light curves are different. Perhaps the covariance structure fitting of a light curve is more sensitive to these noise properties compared to the amplitude spectrum fitting in the Fourier domain.
	
	To test the sensitivity of the GP regression to the quality of the input time series, we degraded the quality of the TESS light curves of the 30 massive stars by adding additional frequency independent (i.e. white) noise. This increased the value of $C_{\rm w}$ (cf. Eqn.~\ref{equation: red noise}) in the amplitude spectrum and consequently also the contribution of the jitter term in the GP regression. For each star in our sample, we created four additional light curves which had 0.5, 1, 2 and 3 times the standard deviation of the original light curve added to them as white noise to emulate increasingly noisier data sets. The latter of which can be considered an extremely and unrealistic noise-dominated light curve, but we include it for completeness. We then re-modelled these noisy light curves using our GP regression framework and the method of \citet{Bowman2020b} and compare the resultant $\nu_{\rm char}$ values.
	
	In Fig.~\ref{figure: noisy GP}, we show examples of the results of the `yellow' (left) and `blue' (right) sub-group members,  HD~112244 (TIC~406050497) and HD~96715 (TIC~306491594), respectively. The GP regression typically returns the same $\nu_{\rm char}$ value for all noisy light curves of a star within the respective uncertainties, which are indicated by shaded regions that denote the 94\% confidence interval for $\nu_{\rm char}$ from each fitting method. Naturally, the confidence intervals become larger for noisier light curves, but we note that they are always much smaller for the MCMC amplitude spectrum method compared to those from the GP regression framework. Importantly, there is considerably larger scatter in $\nu_{\rm char}$ values determined from the amplitude spectrum fitting methodology of \citet{Bowman2020b} to the same artificially noise-enhanced data sets. The values of $\nu_{\rm char}$ of the latter are significantly different from each other because of the typically much smaller uncertainties returned by the method of amplitude spectrum fitting. In almost all cases, and as demonstrated by the example stars in Fig.~\ref{figure: noisy GP}, the $\nu_{\rm char}$ value returned by the amplitude spectrum fitting methodology systematically decreases to lower values when the white noise contribution is larger. Whereas, the $\nu_{\rm char}$ value from the GP regression stays the same within its (much larger) uncertainties, hence not significantly different among the noise-enhanced data sets given its relatively large uncertainties. 
	
	Given the consistency of the $\nu_{\rm char}$ values of the GP regression, regardless of the underlying signal-to-noise ratio of SLF variability over the white noise, we conclude that it is the more robust methodology. The reason for this is because of the fundamental difference between the two fitting methods. The GP regression is a fit of the covariance structure in the light curve and incorporates a different model, in which the implemented jitter term more effectively handles the included white noise compared to directly fitting the amplitude spectrum. Therefore, large differences in the $C_{\rm w}$ term in Eqn.~(\ref{equation: red noise}) can lead to significantly different $\nu_{\rm char}$ values, especially given the small uncertainties typically returned from this method.
	

	\subsection{Application to synthetic data}
	\label{subsection: simulations}
	
	Considering that our tests in section~\ref{subsection: quality} were applied to observed data sets of real stars with unknown underlying frequency spectra, it is possible that the GP regression fitting and amplitude spectrum include different biases. In this section, we test this hypothesis using synthetic light curve data sets generated from different combinations of hyperparameters for the damped SHO kernel generated using {\sc celerite2}. We applied both fitting methods and compared the derived $\nu_{\rm char}$ values to investigate which approach is more robust in returning the true input value for $\nu_{\rm char}$.
	
\begin{figure*}
	\centering
	\includegraphics[width=0.45\textwidth]{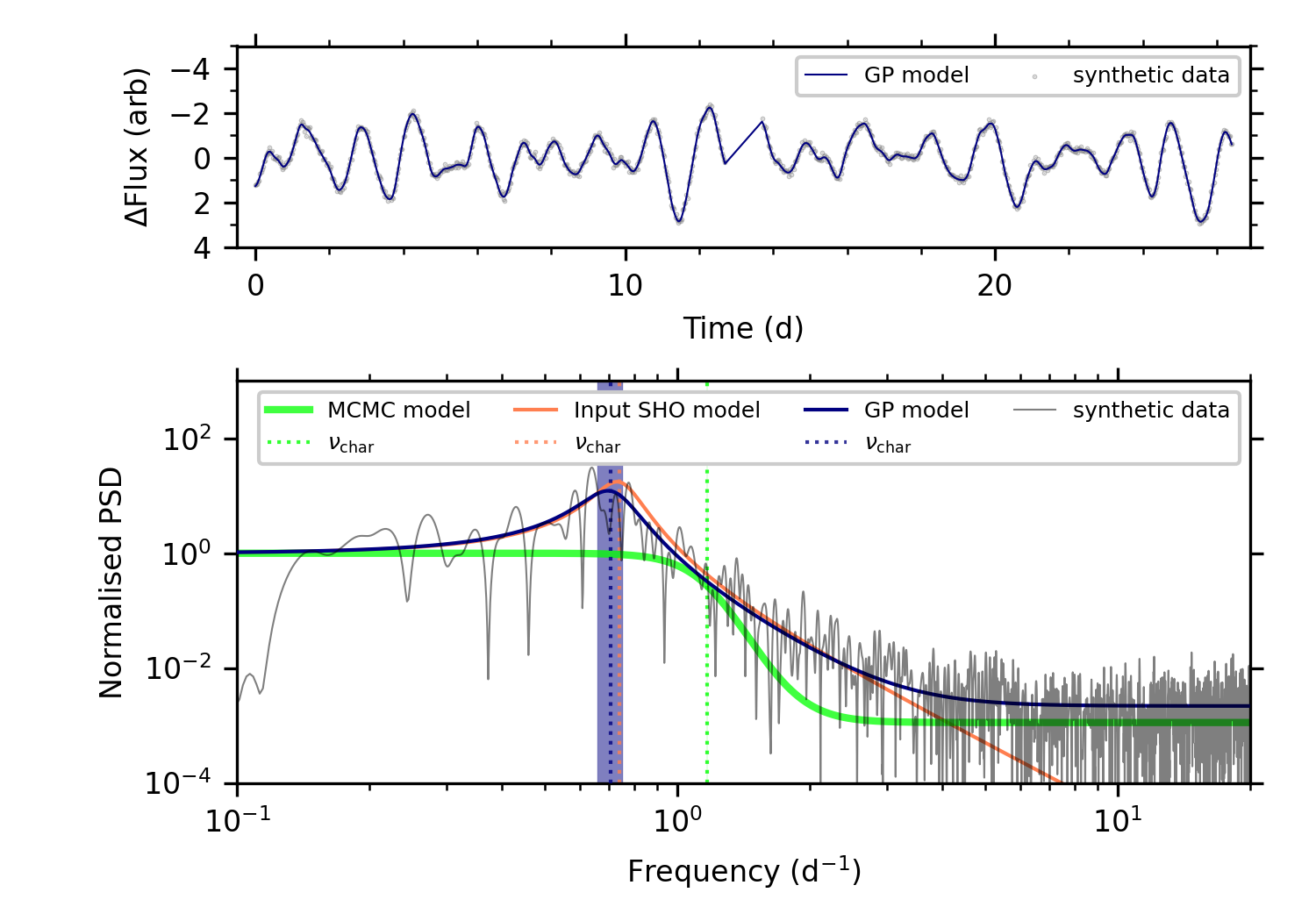}  
	\includegraphics[width=0.45\textwidth]{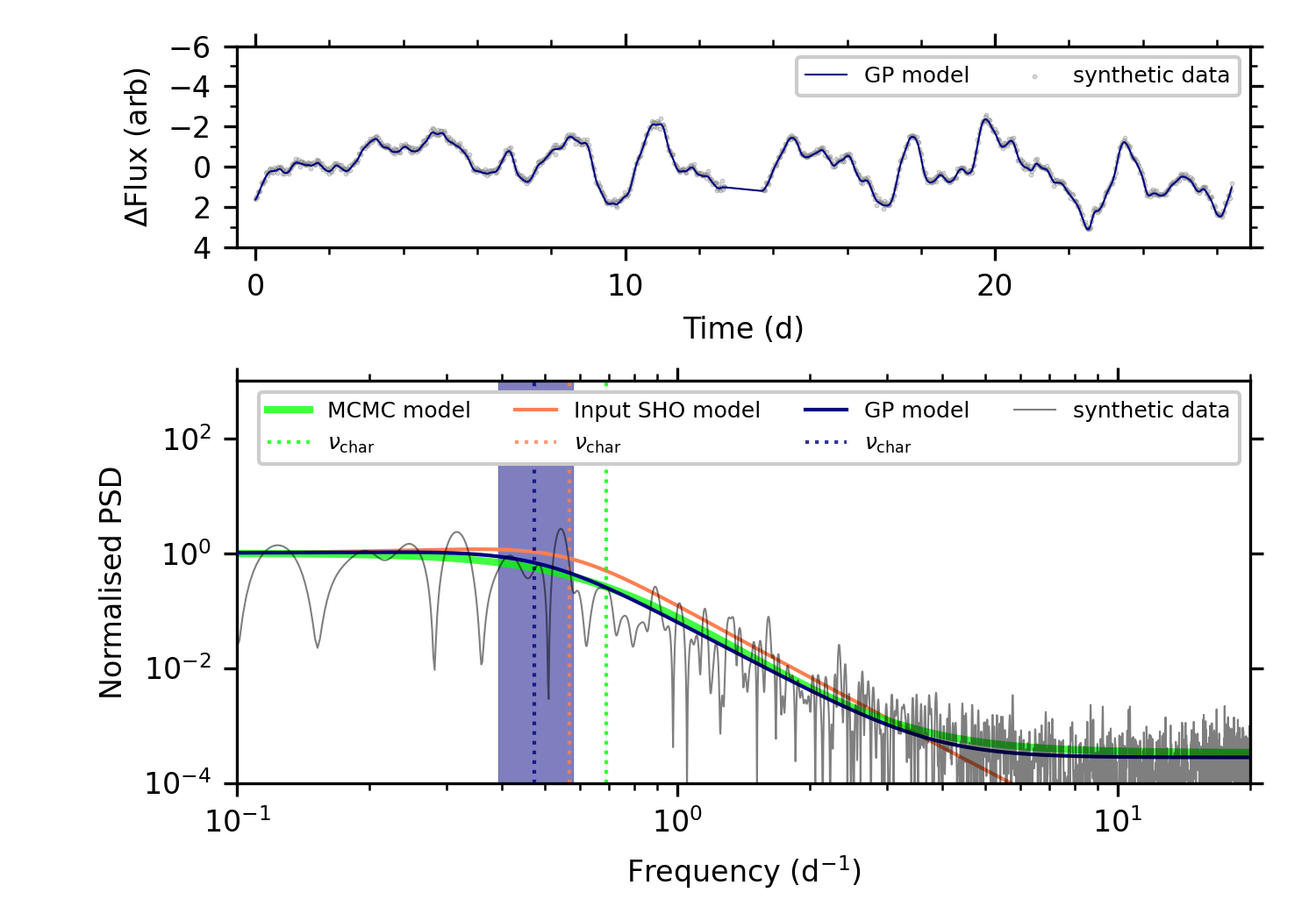} \\  
	\includegraphics[width=0.45\textwidth]{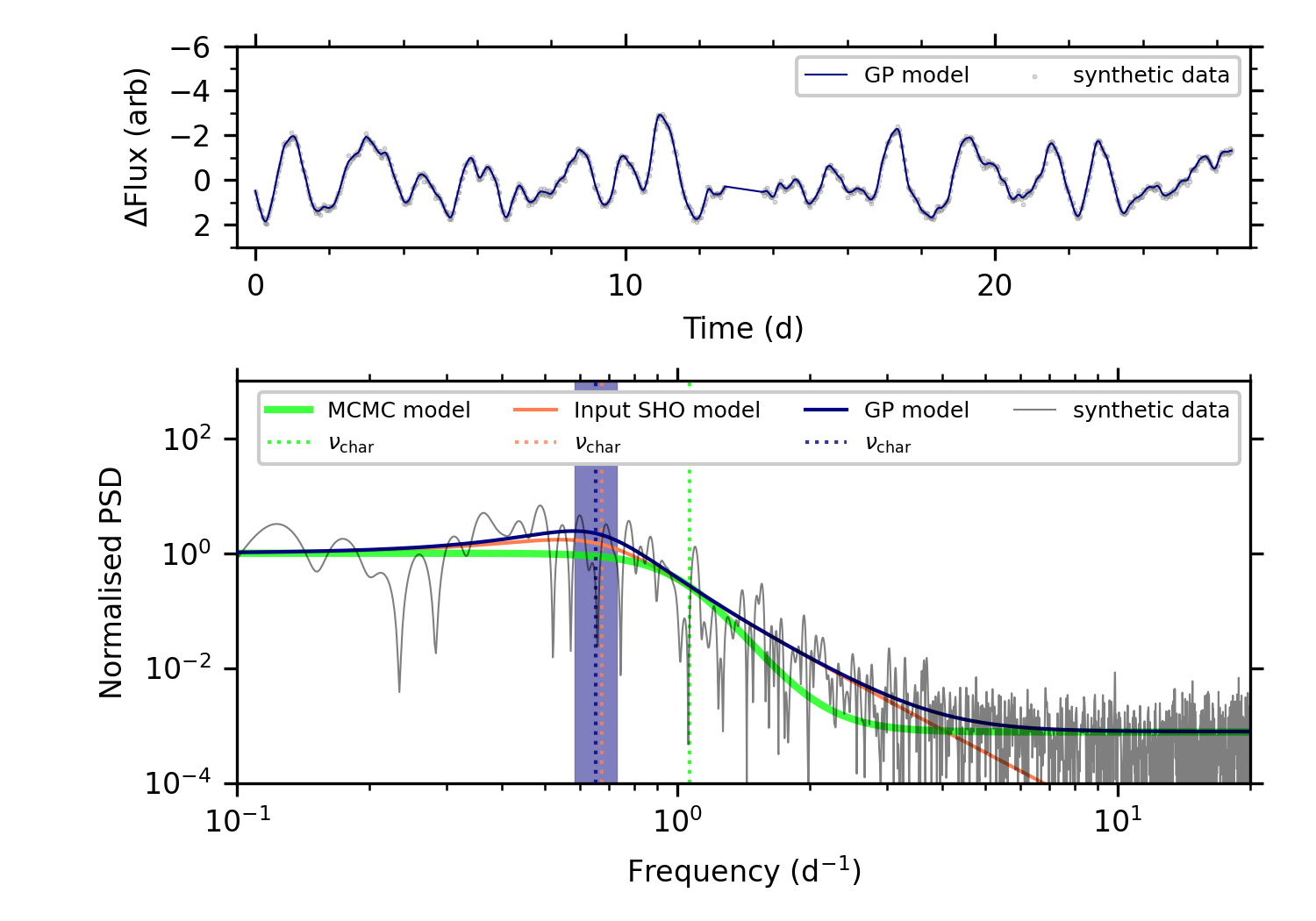}  
	\includegraphics[width=0.45\textwidth]{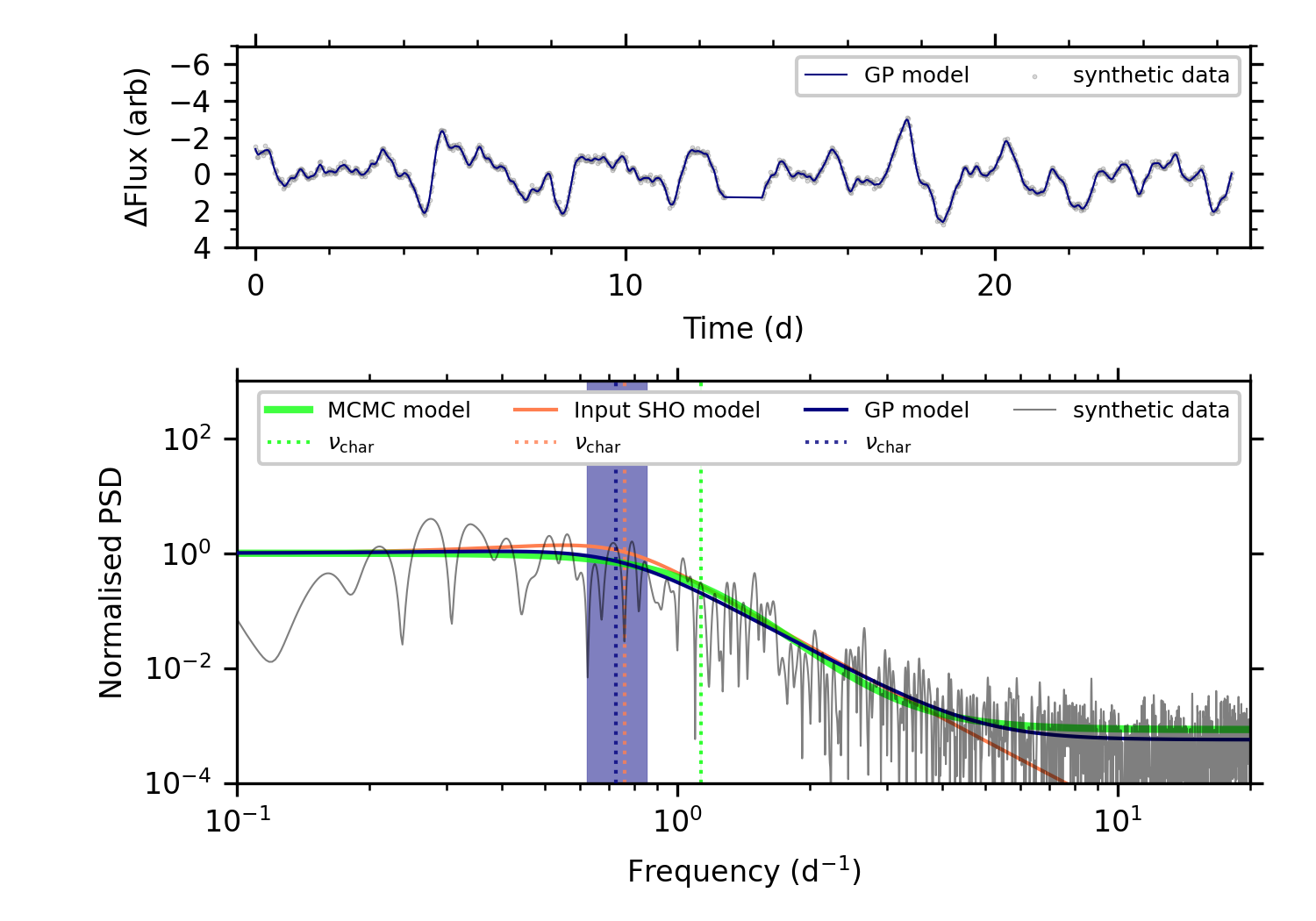} \\  
	\includegraphics[width=0.45\textwidth]{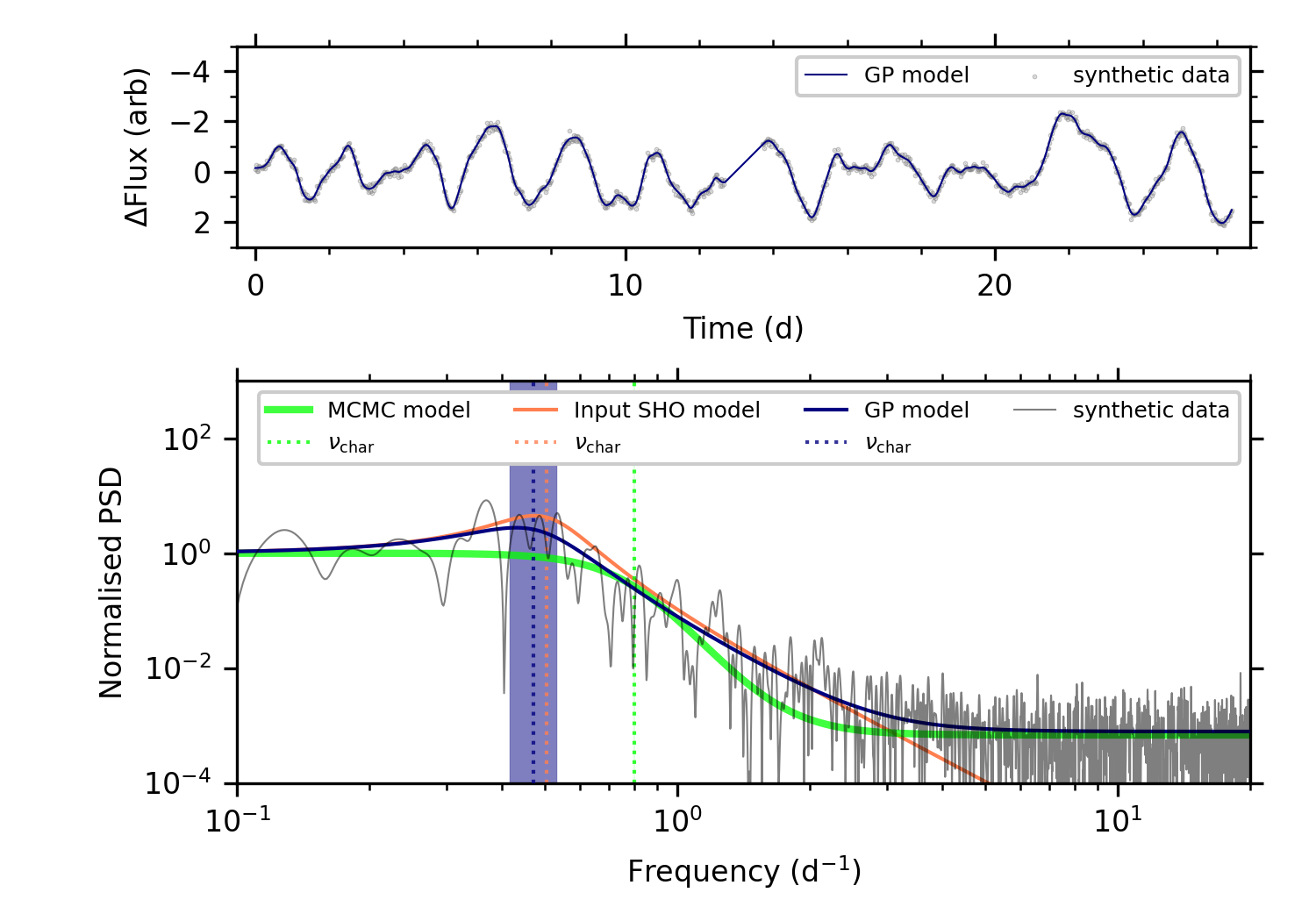}  
	\includegraphics[width=0.45\textwidth]{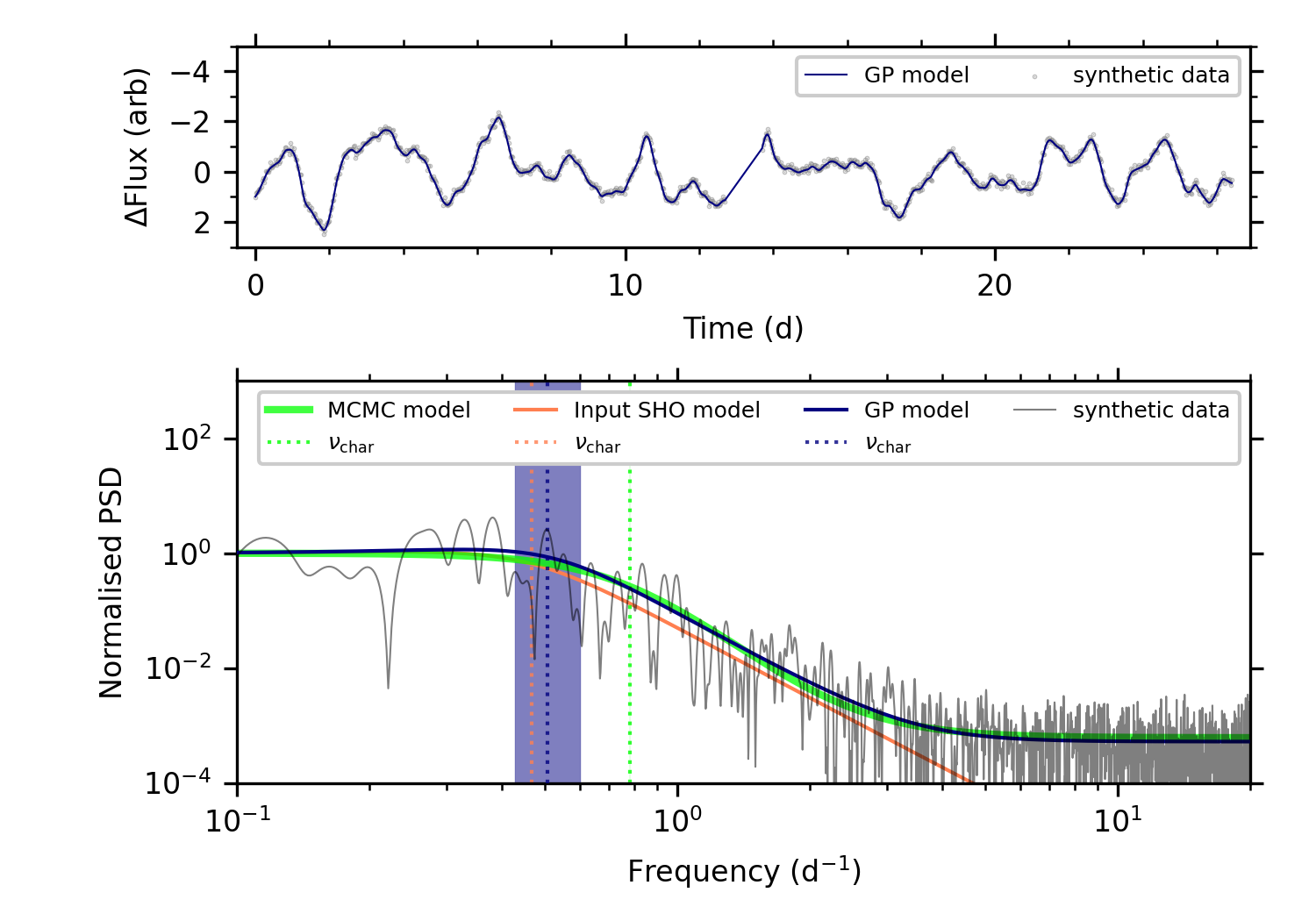} \\  
	\caption{Examples of results of GP regression and amplitude spectrum fitting for synthetic light curves. The left and right columns show three examples that resemble `yellow' and `blue' subgroup members in terms of their quality factor. The blue and green shaded regions denote the 94\% confidence interval for $\nu_{\rm char}$ for each method. In all cases, the true input value of $\nu_{\rm char}$ is recovered well by the GP model with {\sc celerite2}, whereas the MCMC method of \citet{Bowman2020b} sometimes overestimates the value of $\nu_{\rm char}$ because of the more rigid profile imposed in the Fourier domain, as demonstrated by a comparison with the noiseless input SHO model.}
	\label{figure: synthetic data}
	\end{figure*}

	We generated 100 artificial light curves that span 26~d in length, consist of 1000 data points, and contain a 1-d gap in the middle to emulate one sector of TESS data. Each light curve was generated containing different noise properties and hyperparameters for a damped SHO kernel. We applied our covariance structure fitting method within a GP regression framework and the amplitude spectrum fitting method of \citet{Bowman2020b} to these 100 synthetic light curves. For all cases, the derived $\nu_{\rm char}$ values are comparable between the two fitting methods and are approximately within a factor of two of each other, as was the case in section~\ref{subsection: quality} when applied to real and noise-enhanced TESS light curves. However, the $\nu_{\rm char}$ values determined from the methodology of \citet{Bowman2020b} are sometimes up to approximately a factor of 1.4-1.7 larger than the true input values. As before, this difference is caused by the two approaches employing two different models and fitting methods. Moreover, the GP regression returns $\nu_{\rm char}$ values that are completely consistent with the true input values. Therefore, we conclude that the GP regression methodology is generally more robust, especially in cases of noisier input light curves. We show several examples of the fitted synthetic light curves and corresponding power density spectra in Fig.~\ref{figure: synthetic data}, which include the power spectral density of the noiseless damped SHO input model for comparison to the two fitting methods.
	
	We note that even though the GP regression is a fit to the light curve, and the method of \citet{Bowman2020b} is a fit to the amplitude spectrum, the normalised power density spectra from the best-fitting GP regression models are typically better (visual) fits in the cases when $Q>1$ (i.e. left panels of Fig.~\ref{figure: synthetic data}). This is because the shape of Eqn.~(\ref{equation: red noise}) is quite rigid and unable to capture the complete morphology of the SLF variability in the Fourier domain. For low-$Q$ data sets, the form of Eqn.~(\ref{equation: red noise}) and the resultant power density spectrum of the best-fitting GP regression model are similar in shape, as shown in the right column panels of Fig.~\ref{figure: synthetic data}. However, the amplitude spectrum fitting methodology typically overestimates the true input value of $\nu_{\rm char}$ by a factor of $\sim$\,1.5. We refer the interested reader to \citet{Foreman-Mackey2017} for similar hare-and-hound exercises with {\sc celerie2} using simulated time series data.
		
\section{Discussion and Conclusions}
\label{section: conclusions}

The vast majority of massive stars have SLF variability in time series photometric observations, if not all when sufficiently high enough quality data are available, which probes the physical properties of the star, such as mass and age. In this work, we have tested the use of GP regression, specifically by fitting a covariance structure with the {\sc celerite2} software package and a damped SHO kernel \citep{Foreman-Mackey2017, Foreman-Mackey2018, Foreman-Mackey2021a}, to the time series photometric observations for a sample of 30 massive stars observed by the TESS mission and previously studied by \citet{Bowman2020b}. We use the same input light curves from the TESS mission and compare our GP regression results, paying particular attention to the characteristic frequency of the SLF variability, $\nu_{\rm char}$, to the previous amplitude spectrum fitting method by \citet{Bowman2020b}. We find that the $\nu_{\rm char}$ determinations by the two different methods are the same, within their uncertainties, for the majority of massive stars observed by TESS. This is encouraging and demonstrates that the previously found correlation of $\nu_{\rm char}$ with mass and age is robust given that the GP regression method is a fit to the light curve using a damped SHO model, and the method of \citet{Bowman2020b} is a fit to the amplitude spectrum using a semi-Lorentzian (i.e. Harvey-like; \citealt{Harvey1985}) model that is not necessarily physically motivated. A minority of stars have different and systematically higher values of $\nu_{\rm char}$ by up to a factor two when using the GP regression compared to amplitude spectrum fitting, as demonstrated by the examples shown in Fig.~\ref{figure: synthetic data}. Specifically, the damped SHO model of the GP regression is more flexible in fitting the SLF variability in the time domain compared to the rigid shape imposed by the semi-Lorentzian profile in the Fourier domain given in Eqn.~(\ref{equation: red noise}).

We have demonstrated that GP regression is a viable method for characterising SLF variability in time series photometry of massive stars. Our work lends further support to the conclusion of \citet{Bowman2020b} that SLF variability probes the mass and age of a massive star, demonstrated by the correlation between $\nu_{\rm char}$ and a star's location in the HR~diagram, as shown in Fig.~\ref{figure: HRD}. Therefore, we conclude that the absence of coherent, heat-driven pulsation modes is not necessarily a prerequisite for asteroseismology of massive stars, given the probing power of $\nu_{\rm char}$ on bulk properties like mass and age. Furthermore, we have shown that a fraction of massive stars with SLF variability exhibit the characteristics of underdamped quasi-periodic variability, suggesting the presence of damped pulsations (i.e. stochastic IGWs). Although a range of quality factors exist in the SLF variability of massive stars, a transition from overdamped to underdamped variability is found to coincide with a transition from the ZAMS to the TAMS, as shown in Fig.~\ref{figure: HRD Q}. We have also demonstrated that a GP regression is more robust against noisier input time series data, especially in the cases where the SLF variability is comparable in amplitude to the white noise.

Several recent theoretical studies have focussed on understanding the possible physical causes that give rise to SLF variability in time series photometry. One possible mechanism is that stochastically excited IGWs from the convective core reach the surface with detectable amplitudes and produce a broad frequency spectrum of damped modes \citep{Rogers2013b, Rogers2015, Rogers2017c, Edelmann2019a, Horst2020a, Vanon2022a**, Herwig2022a**, Thompson_W_2022a**}. However, such an interpretation has been challenged by other studies employing numerical simulations showing that core-excited IGWs are damped before they reach the surface \citet{Lecoanet2019a, Lecoanet2021a}. This latter result hence supports instead that SLF variability originates from sub-surface convection zones \citep{Cantiello2021b, Schultz2022a}. The presence and location of sub-surface convection zones and inferring their existence in stellar evolution models, especially for stars of low metallicity, is clearly important for understanding IGW excitation mechanisms \citep{Jermyn2022a}. From stellar evolution theory, it is expected that the size of sub-surface convection zones grow from being negligible to relatively large fractions of a star's radius as it evolves from the ZAMS to the TAMS. Hence, the contribution to SLF variability from sub-surface convection zones is plausibly also expected to increase as stars evolve. On the other hand, the increasing radius as stars evolve produces different contributing factors to the driving, damping and propagation of IGWs. Detailed (rotating) hydrodynamical simulations of massive stars at different masses and ages are needed to explore this transition and the relative importance of IGWs from both the core and sub-surface convection zones.

In this work we have tested and implemented the use of GP regression to infer the characteristic frequency, $\nu_{\rm char}$ of massive stars from TESS light curves using the {\sc celerite2} software package. We have demonstrated that a transition from a broad range of low-amplitude SLF variability to high-amplitude quasi-periodic variability exists between the ZAMS and TAMS. We postulate that the observed transition discovered in our work corresponds to an astrophysical scenario in which the dominant cause of SLF variability in massive stars transitions from core convection to sub-surface convection as stars evolve from the ZAMS to the TAMS.

Other complementary methods to study SLF variability in astronomical time series data include low-order linear stochastic differential equations (e.g. \citealt{Koen2005e}), and continuous-time autoregressive moving average models (see \citealt{Kelly2014}). There are thus a variety of statistical tools discussed in the literature that can aid in unlocking the large quantity of TESS light curves of massive stars. In the future, we will expand our analysis to include such methods for inferring properties of the SLF variability in massive stars. We will also apply our methodology to additional stars observed by TESS to extract the morphology of their SLF variability. This will allow us to further populate Fig.~\ref{figure: HRD} and study the origin of SLF variability and its correlation to stellar properties such as mass and age, but also macroturbulence as shown by \citet{Bowman2020b}. The TESS mission is ongoing and currently providing high precision light curves of thousands of massive stars, which also span much longer time spans than those considered in this initial study. In our next study, we will expand our method to apply it to a much larger sample and use it test if masses and radii can be reliably inferred from the light curves of massive stars.


\begin{acknowledgements}

The authors acknowledge that this work was partially conducted on the traditional land of the first people of Seattle, the Duwamish People, past and present, and honours with gratitude the land itself and the Duwamish Tribe.

We thank the anonymous referees for the constructive reports that improved the clarity of this work.

DMB gratefully acknowledges funding from the Research Foundation Flanders (FWO) by means of a senior postdoctoral fellowship (grant agreement No.~1286521N) and an FWO long stay travel grant (agreement No.~V411621N). DMB is also grateful to the staff and scientists at the Kavli Institute for Theoretical Physics, University of California, Santa Barbara, and to the organisers of the `Probes of Transport in Stars' program, which hosted him during when part of this work was conducted. This research was supported in part by the National Science Foundation (NSF) under Grant Number NSF PHY-1748958. TZD-W gratefully acknowledges support from Grant Number NSF AST-1714285.

The authors thank the TESS science team for the excellent data, which were obtained from the Mikulski Archive for Space Telescopes (MAST) at the Space Telescope Science Institute (STScI), which is operated by the Association of Universities for Research in Astronomy, Inc., under NASA contract NAS5-26555. Support to MAST for these data is provided by the NASA Office of Space Science via grant NAG5-7584 and by other grants and contracts. Funding for the TESS mission is provided by the NASA Explorer Program. This research has made use of the SIMBAD database, operated at CDS, Strasbourg, France; the SAO/NASA Astrophysics Data System; and the VizieR catalog access tool, CDS, Strasbourg, France.

This research has made use of the following \texttt{python} software packages: {\tt MATPLOTLIB} \citep{Matplotlib_2007}, {\tt Seaborn} \citep{Seaborn_2021}, {\tt Numpy} \citep{Numpy_2006, Numpy_2011, Numpy_2020}, {\tt astropy} \citep{Astropy_2013, Astropy_2018}, {\tt SciPy} \citep{SciPy_2007, SciPy_2020}, {\tt Pandas} \citep{Pandas_2010}, {\tt celerite2} \citep{Foreman-Mackey2017, Foreman-Mackey2018, Zenodo_celerite_v0.3.1}, {\tt exoplanet} \citep{Zenodo_exoplanet_v0.2.4, Foreman-Mackey2021a}, {\tt arviz} \citep{arviz_2019}, {\tt corner} \citep{Corner_2016}, and {\tt pymc3} \citep{pymc3}.

\end{acknowledgements}


\bibliographystyle{aa}
\bibliography{/Users/dominic/Documents/RESEARCH/Bibliography/master_bib.bib}


\newpage 

\begin{appendix}
\onecolumn


\section{Light curve and power density spectra summary figures}
\label{section: appendix: figures}

\begin{figure*}  
\centering
\includegraphics[width=0.33\textwidth]{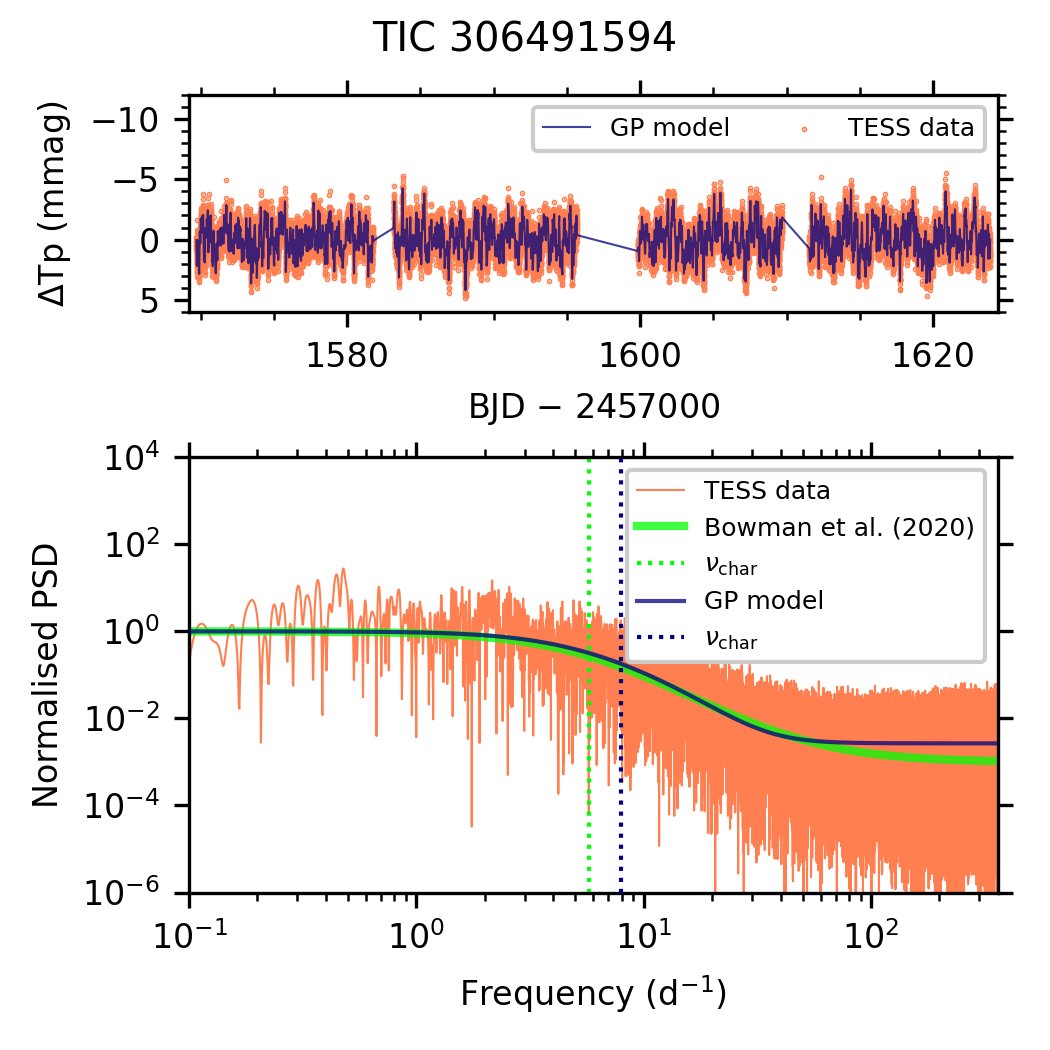}
\includegraphics[width=0.33\textwidth]{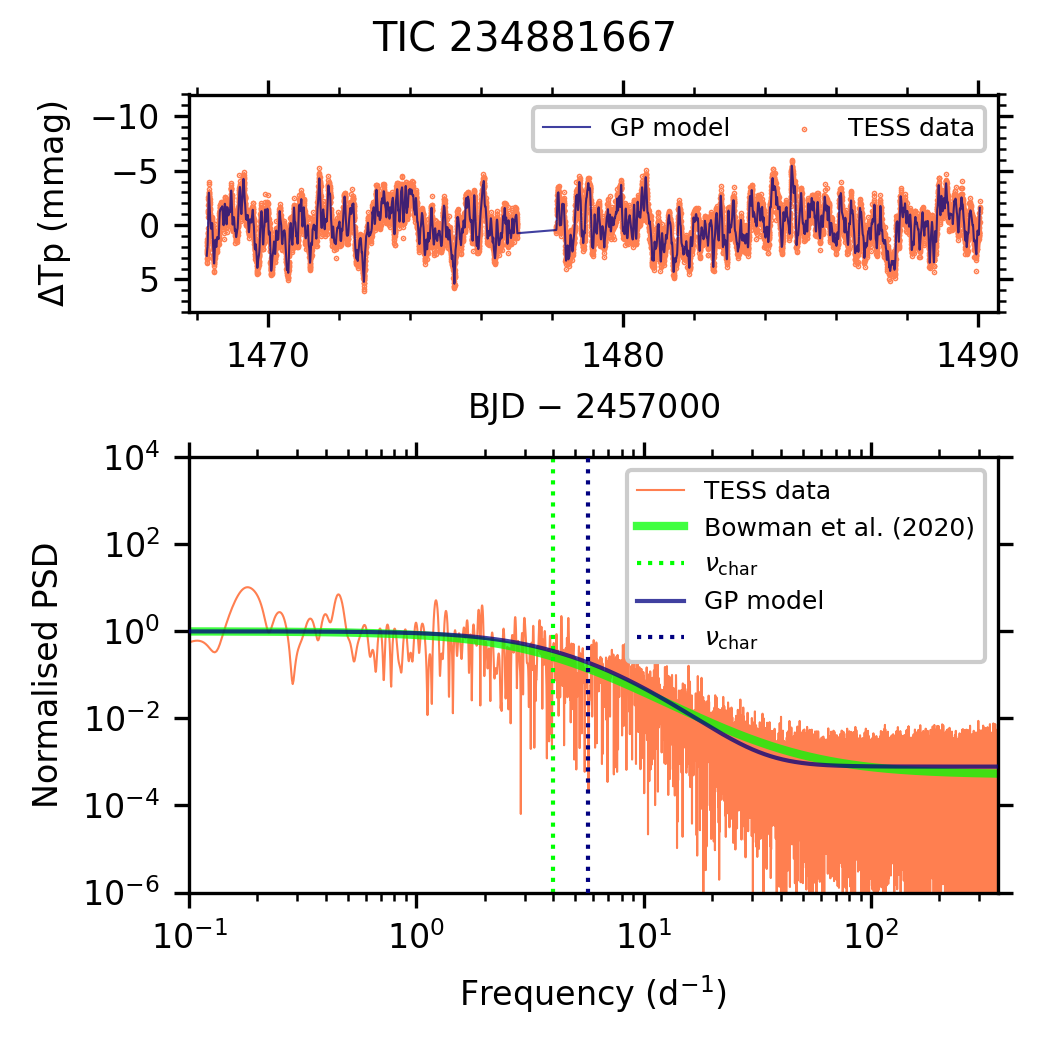}
\includegraphics[width=0.33\textwidth]{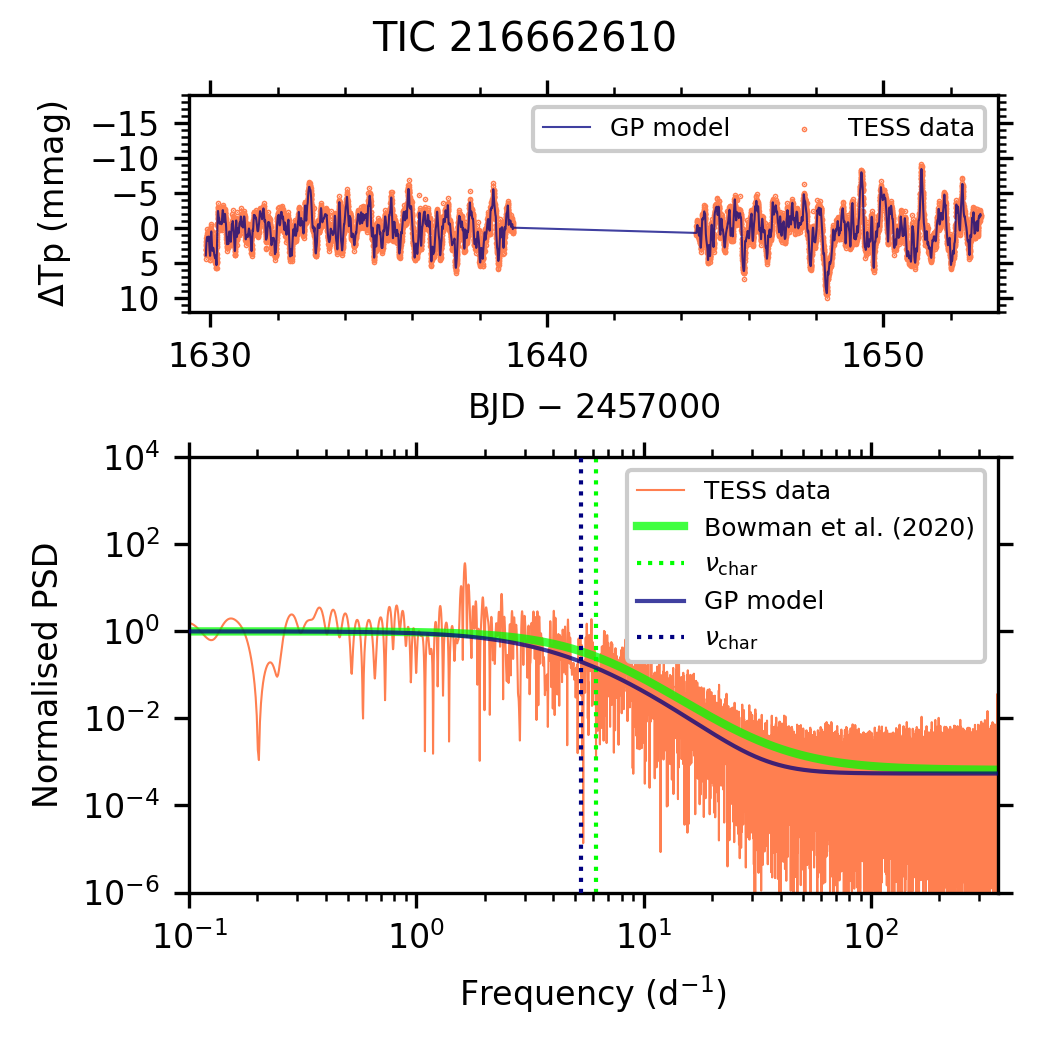}
\includegraphics[width=0.33\textwidth]{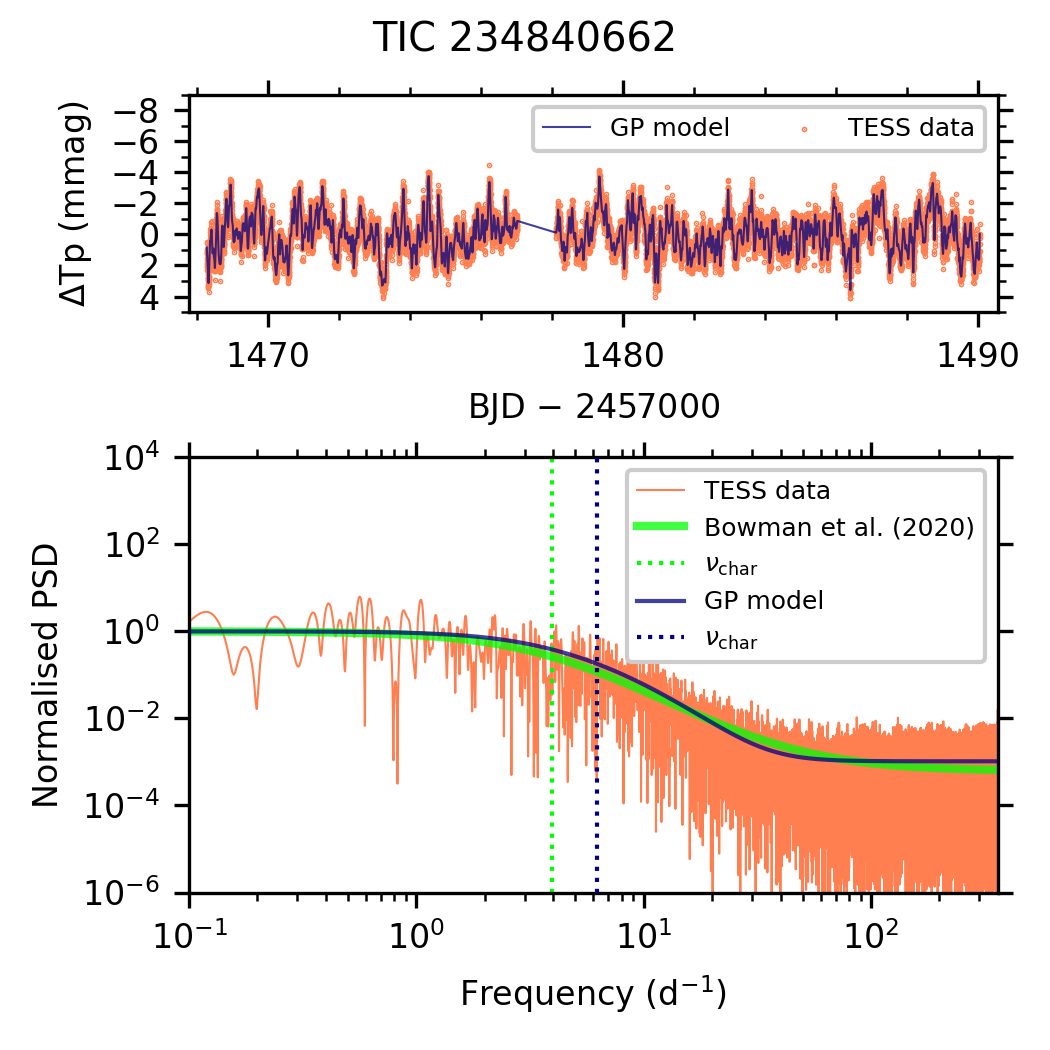}
\includegraphics[width=0.33\textwidth]{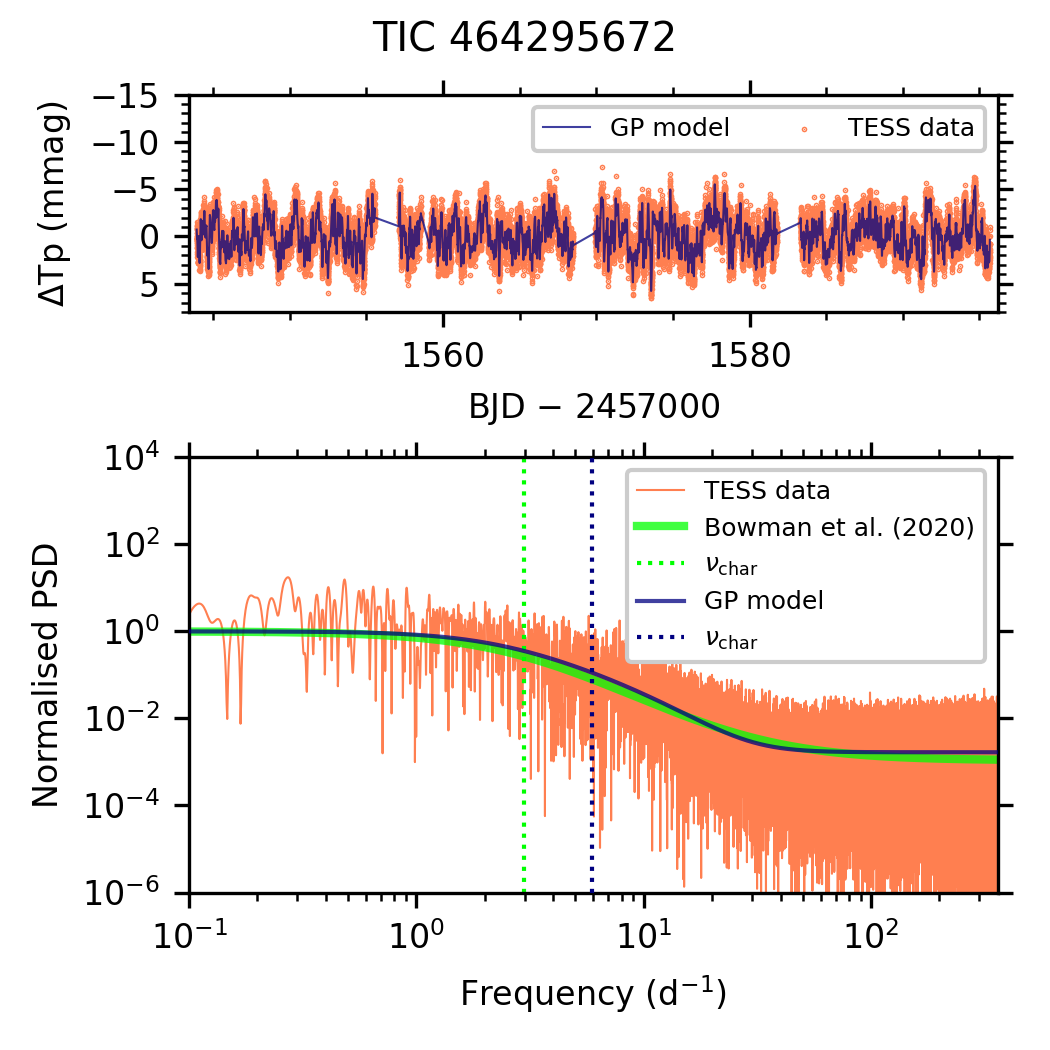}
\includegraphics[width=0.33\textwidth]{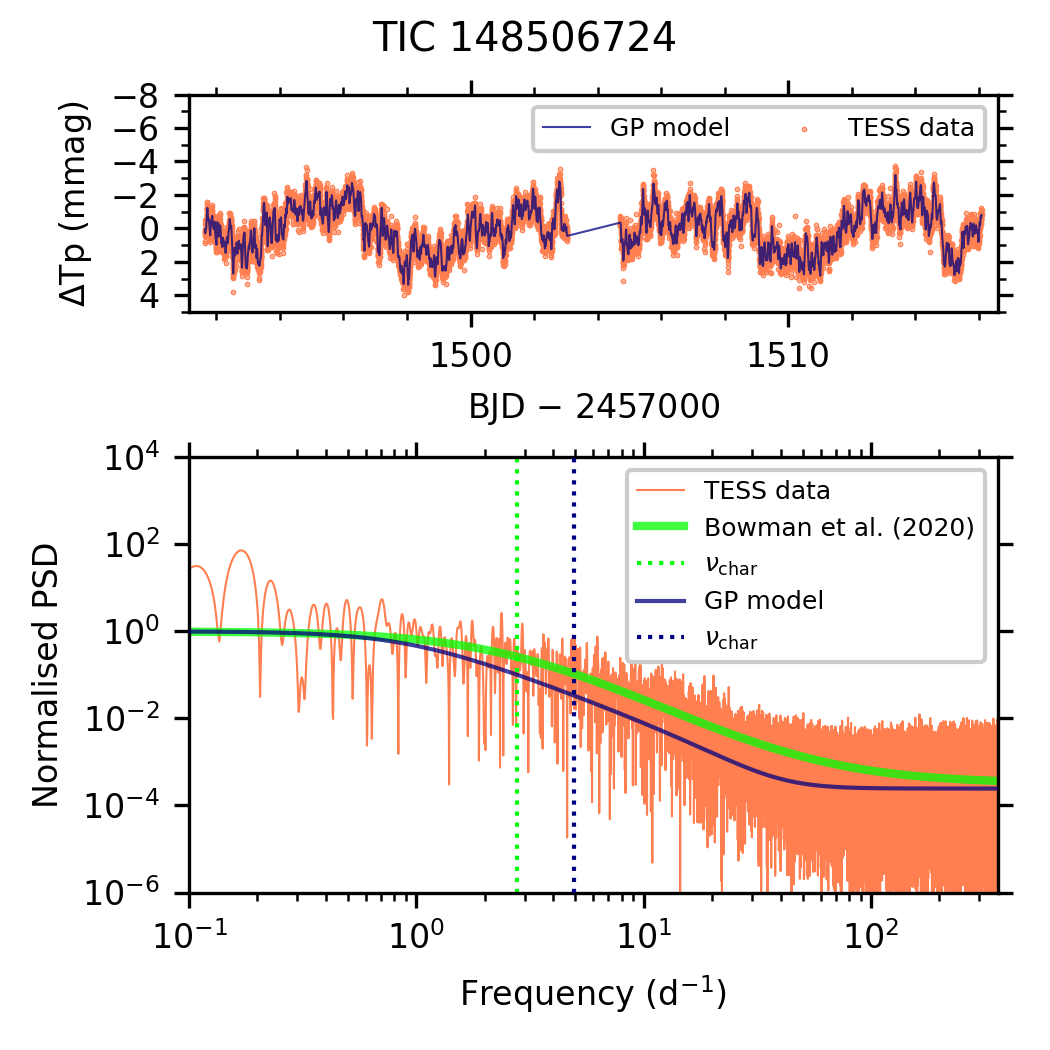}
\includegraphics[width=0.33\textwidth]{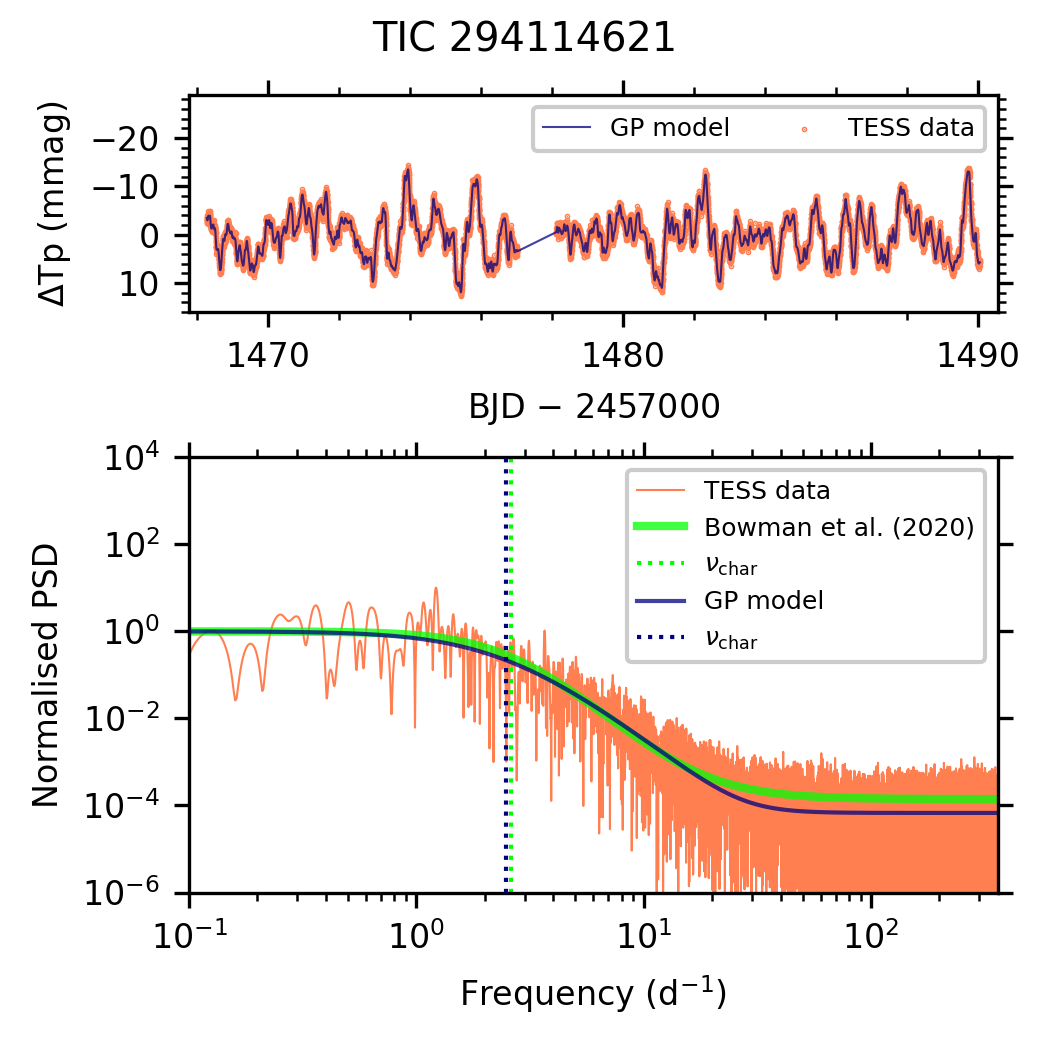}
\includegraphics[width=0.33\textwidth]{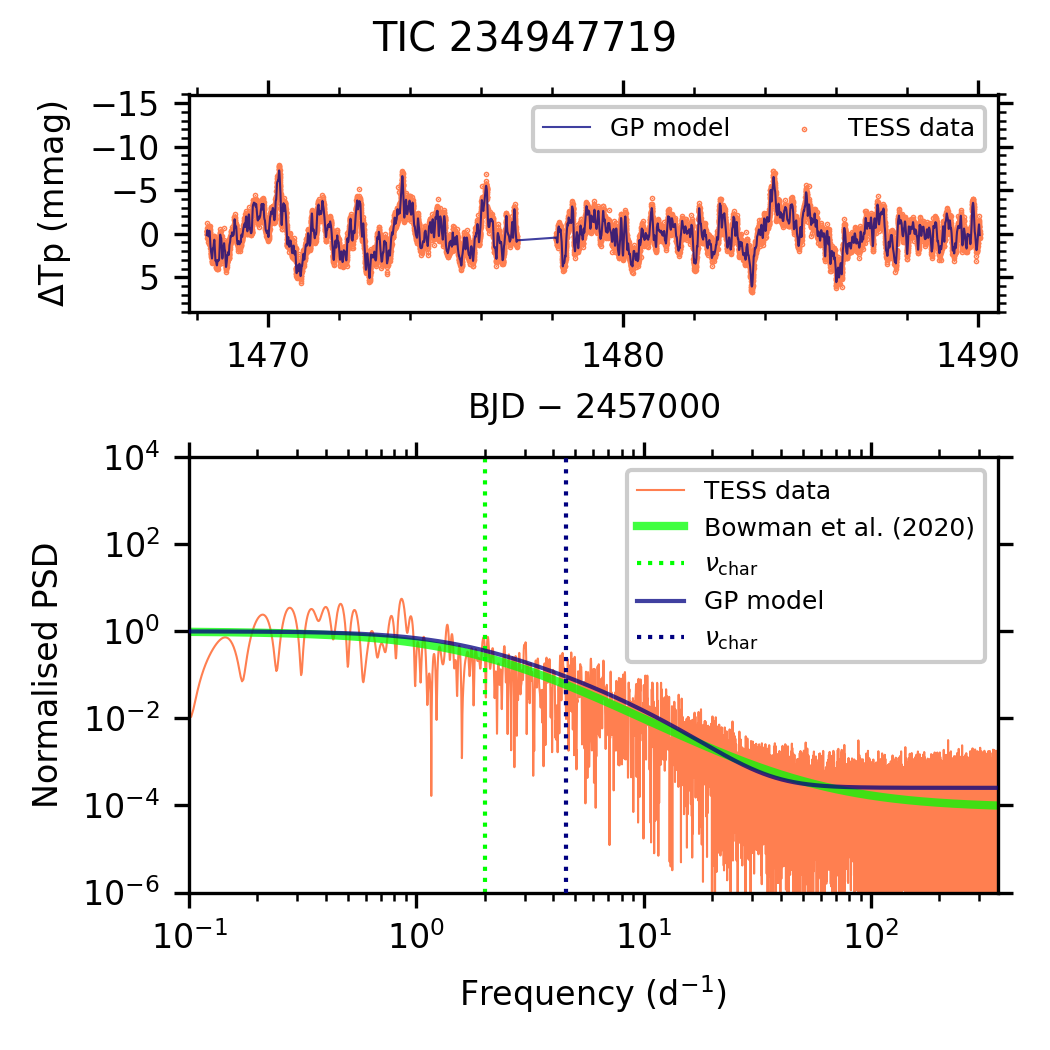}
\includegraphics[width=0.33\textwidth]{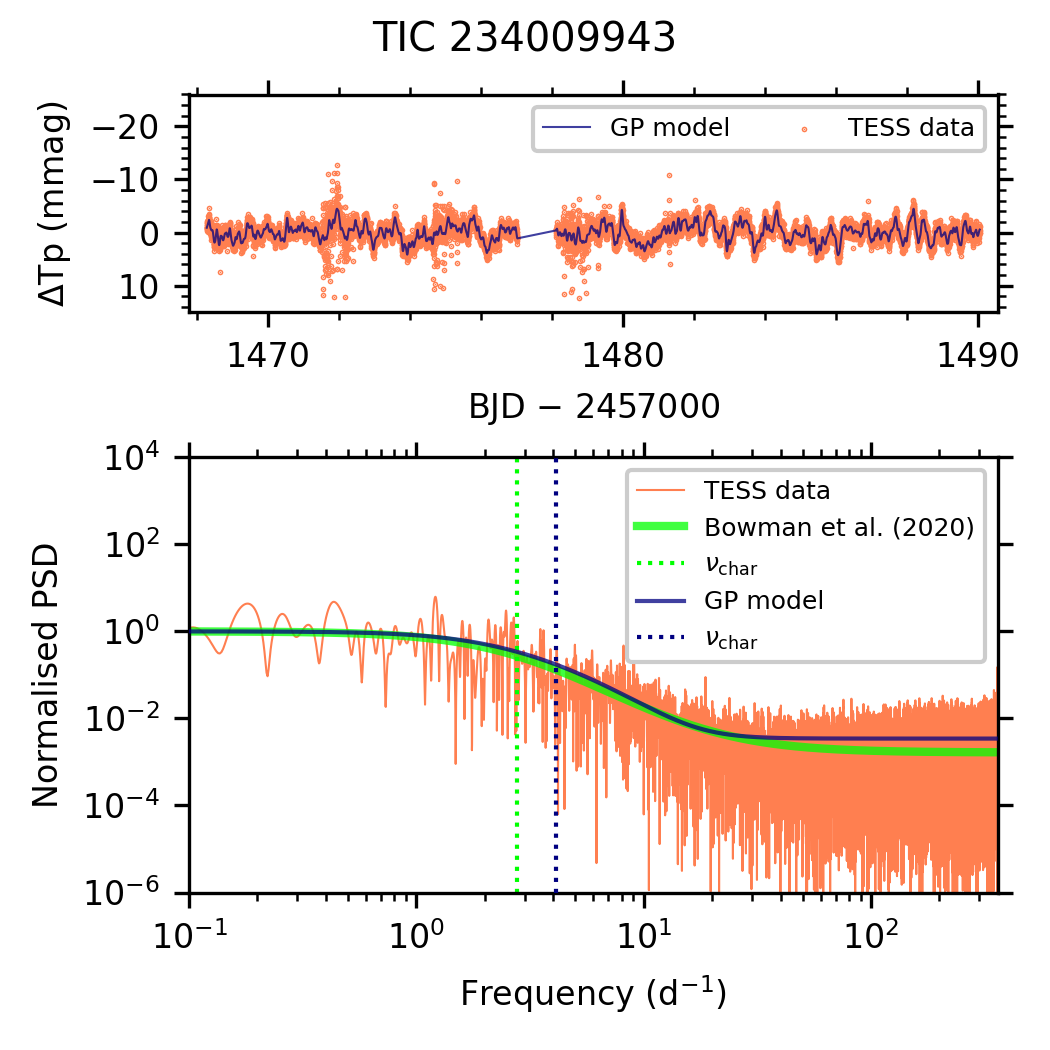}
\caption{Summary figures for massive stars analysed in this work.}
\label{figure: A1}
\end{figure*}

\begin{figure*}  
\centering
\includegraphics[width=0.33\textwidth]{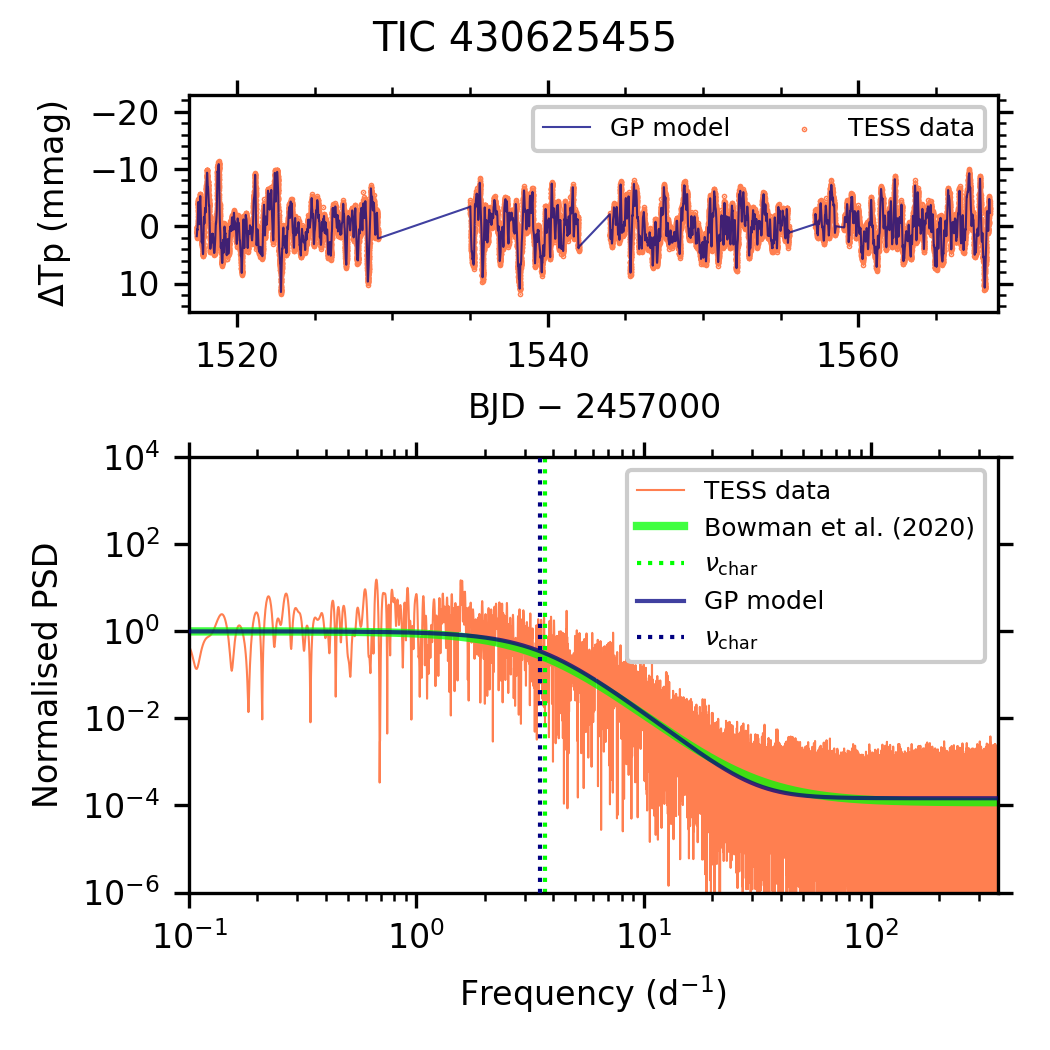}
\includegraphics[width=0.33\textwidth]{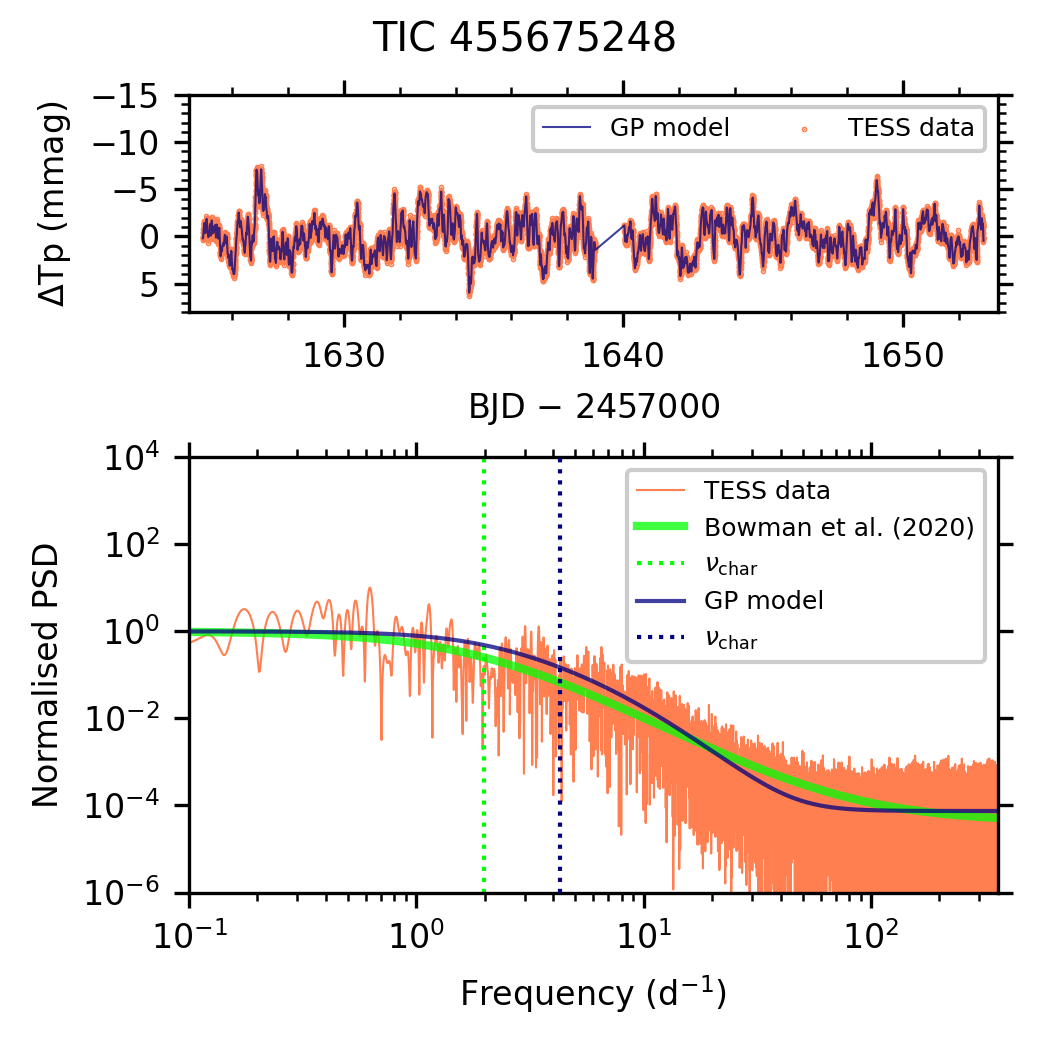}
\includegraphics[width=0.33\textwidth]{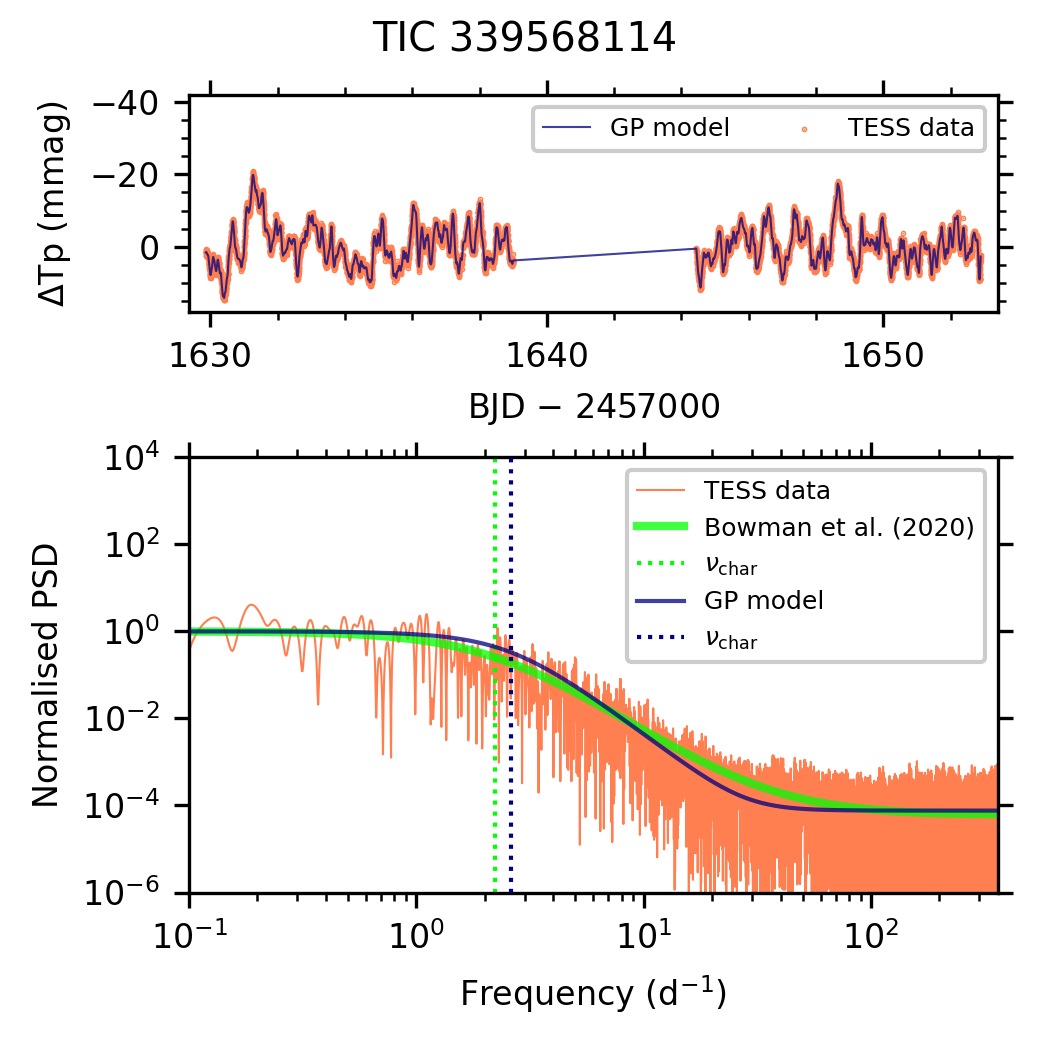}
\includegraphics[width=0.33\textwidth]{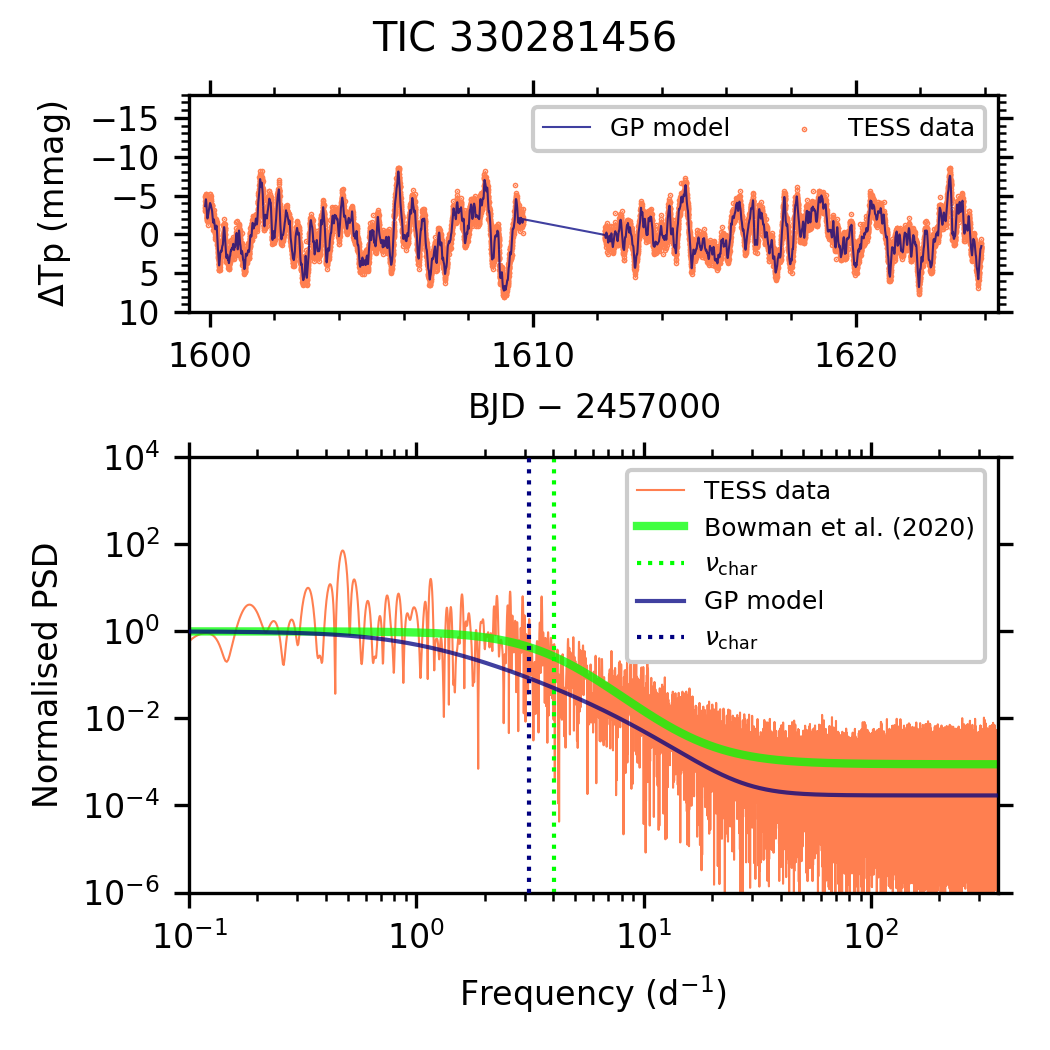}
\includegraphics[width=0.33\textwidth]{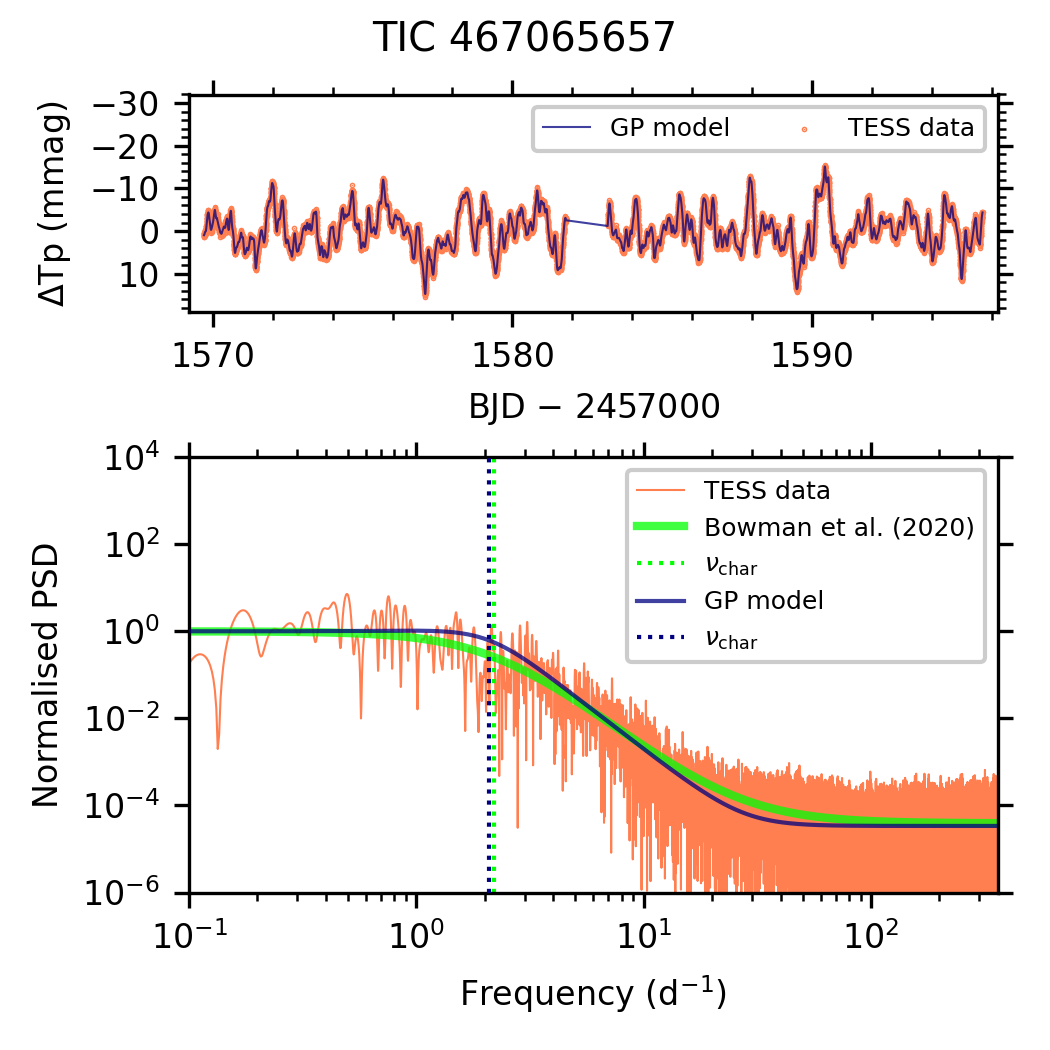}
\includegraphics[width=0.33\textwidth]{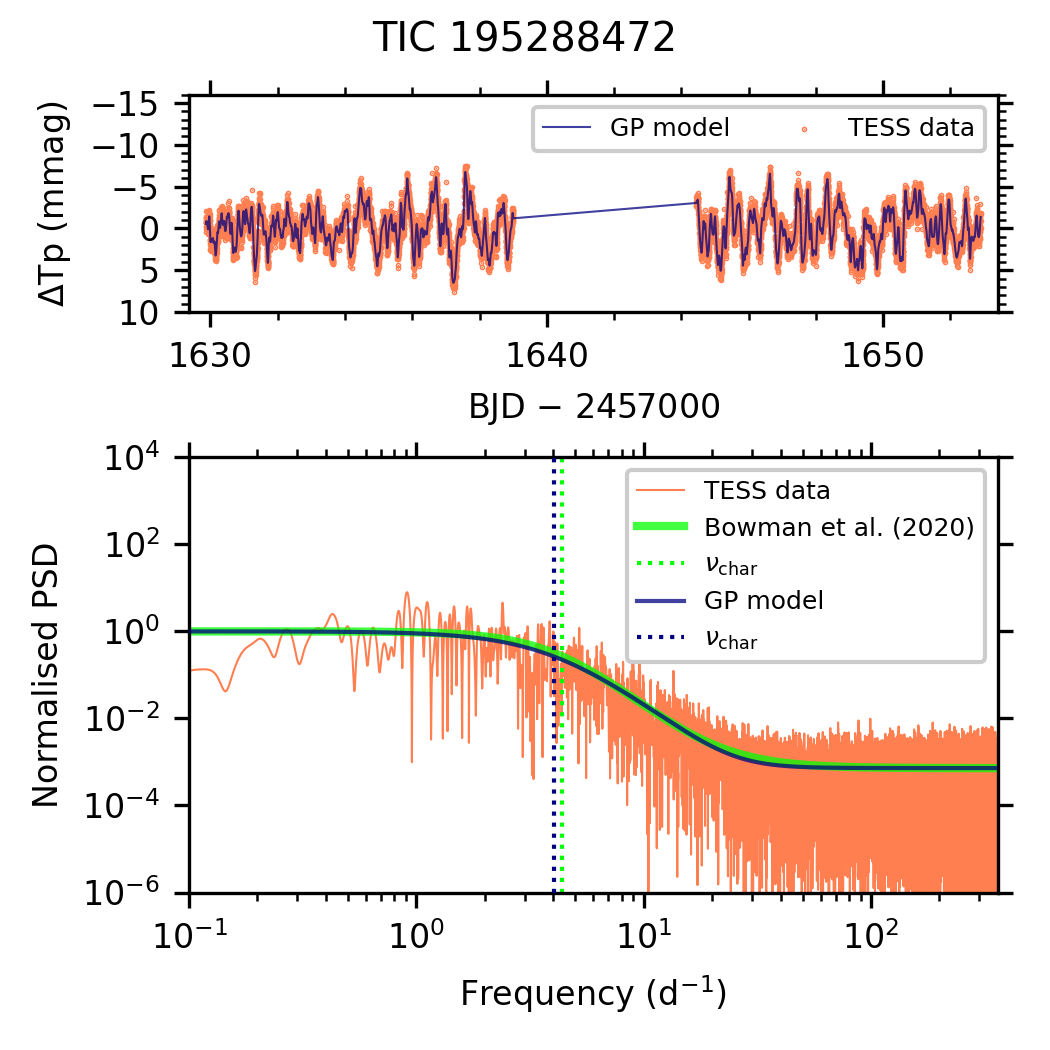}
\includegraphics[width=0.33\textwidth]{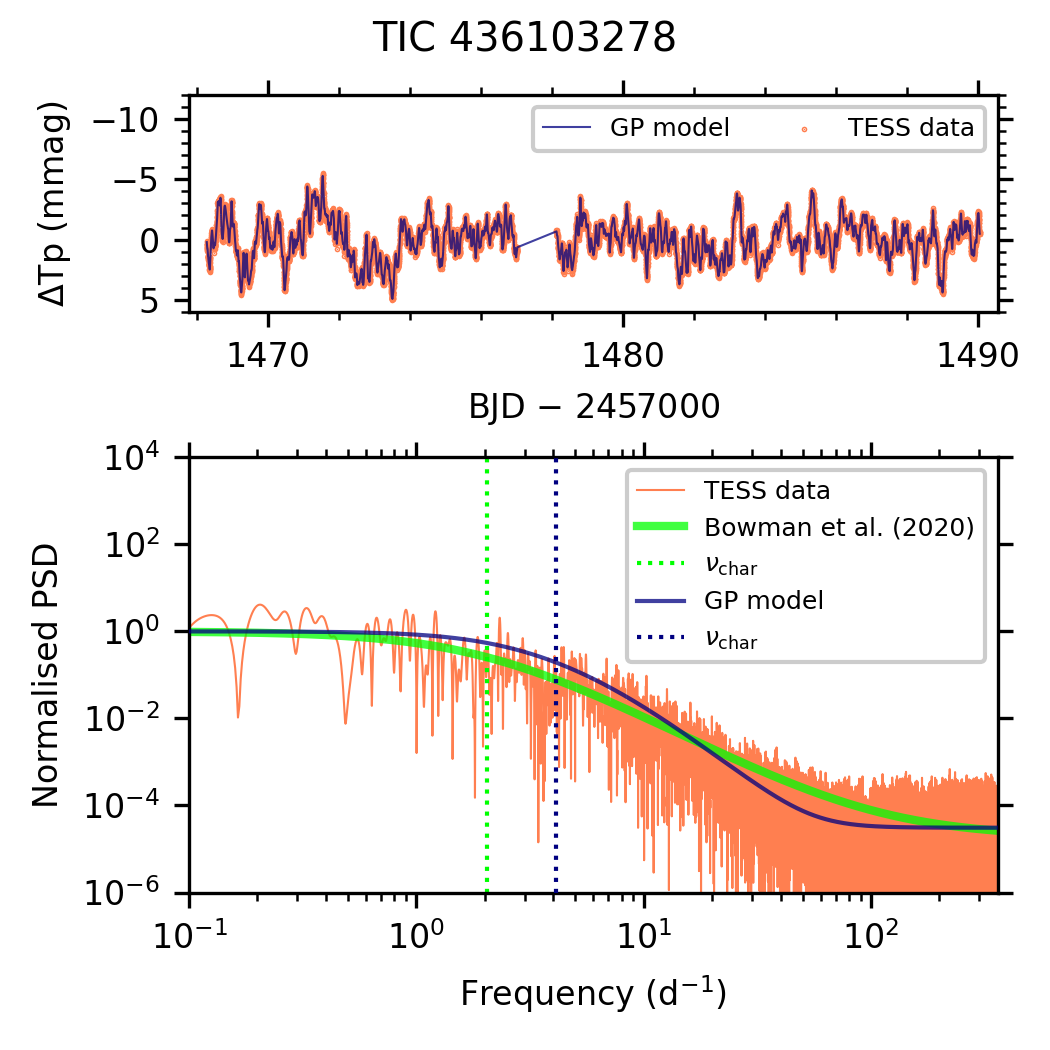}
\includegraphics[width=0.33\textwidth]{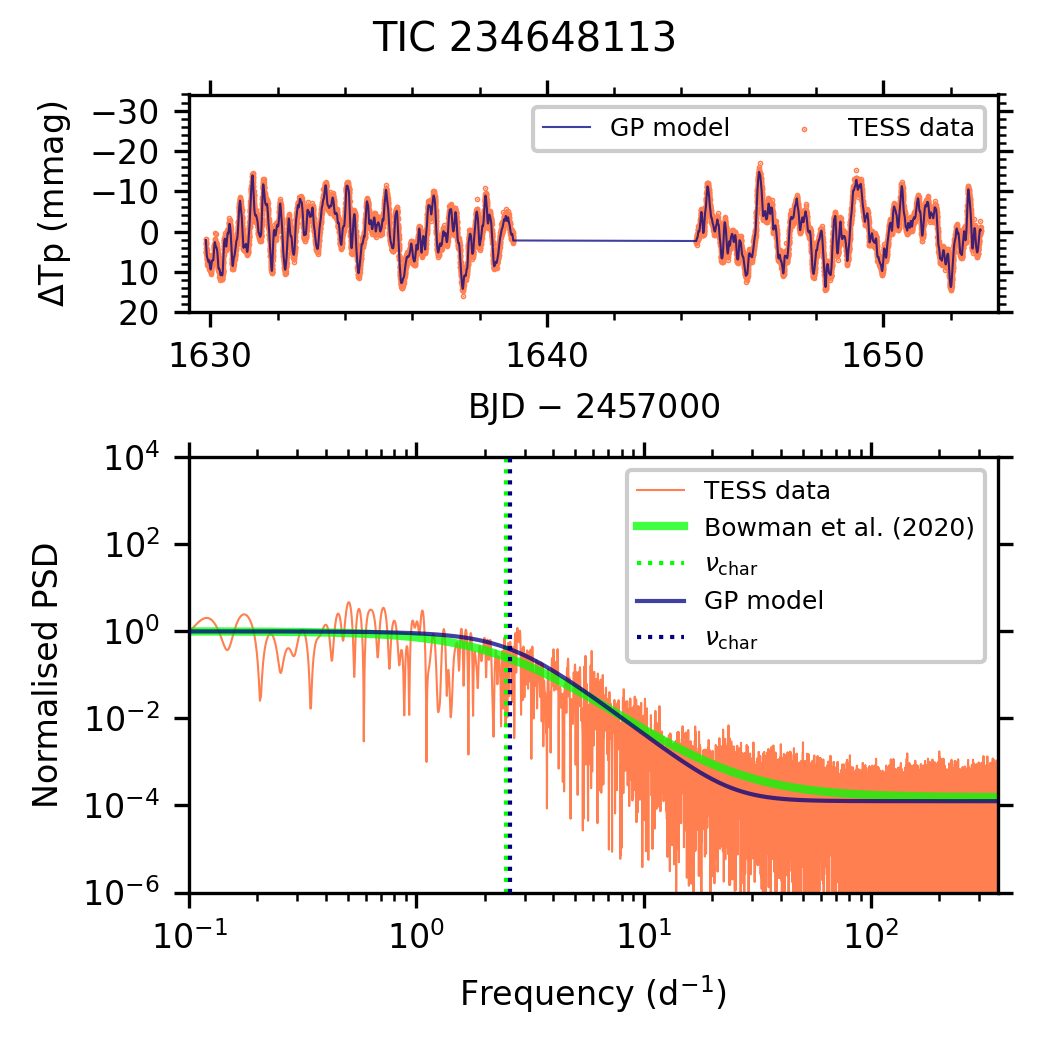}
\includegraphics[width=0.33\textwidth]{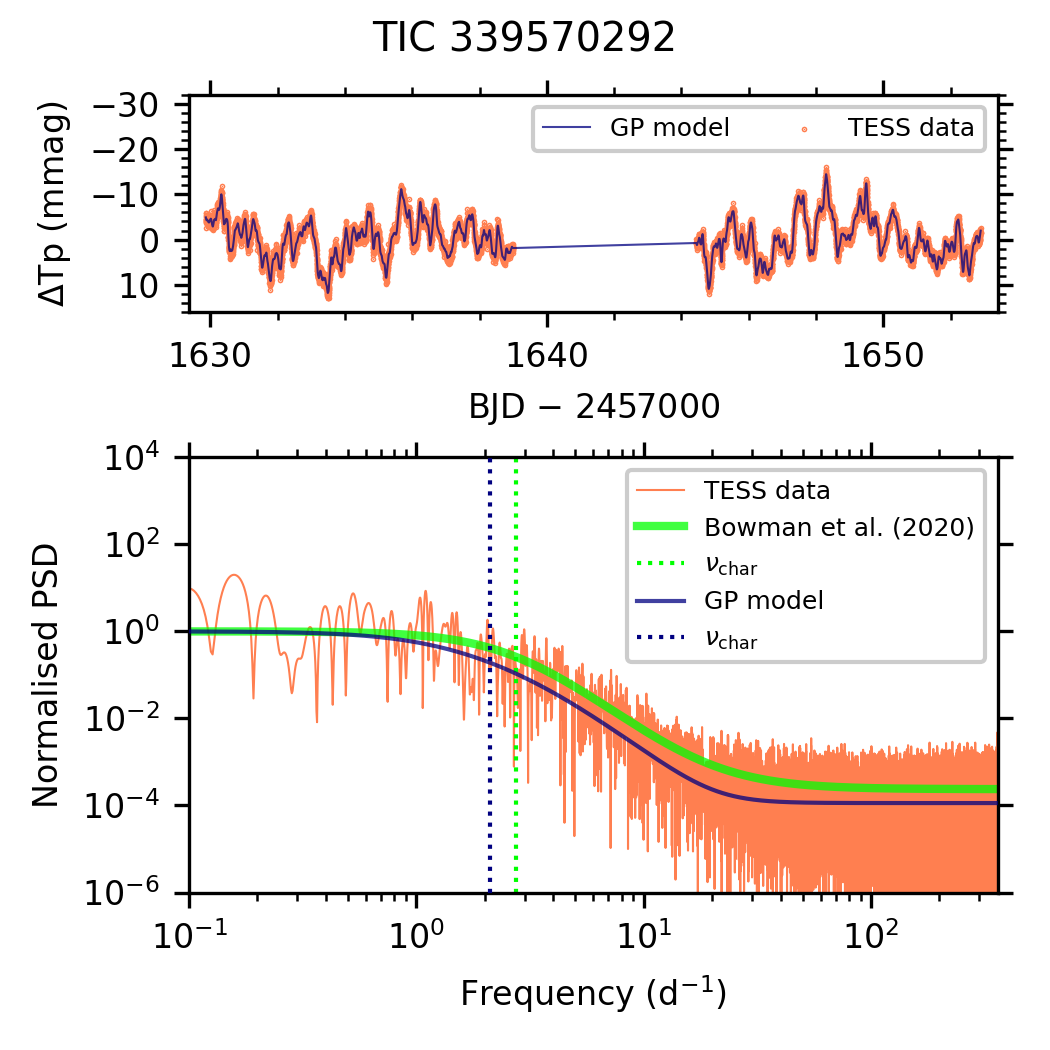}
\includegraphics[width=0.33\textwidth]{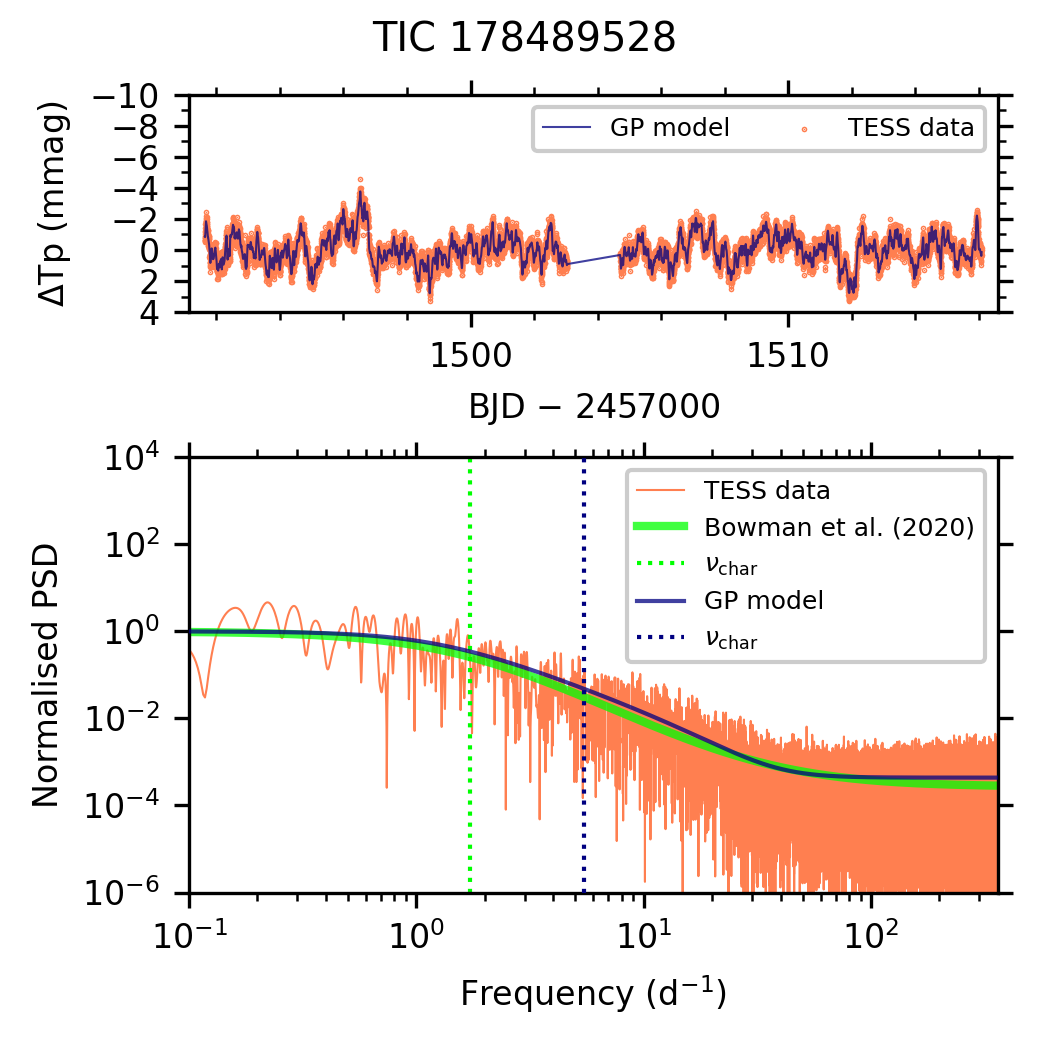}
\includegraphics[width=0.33\textwidth]{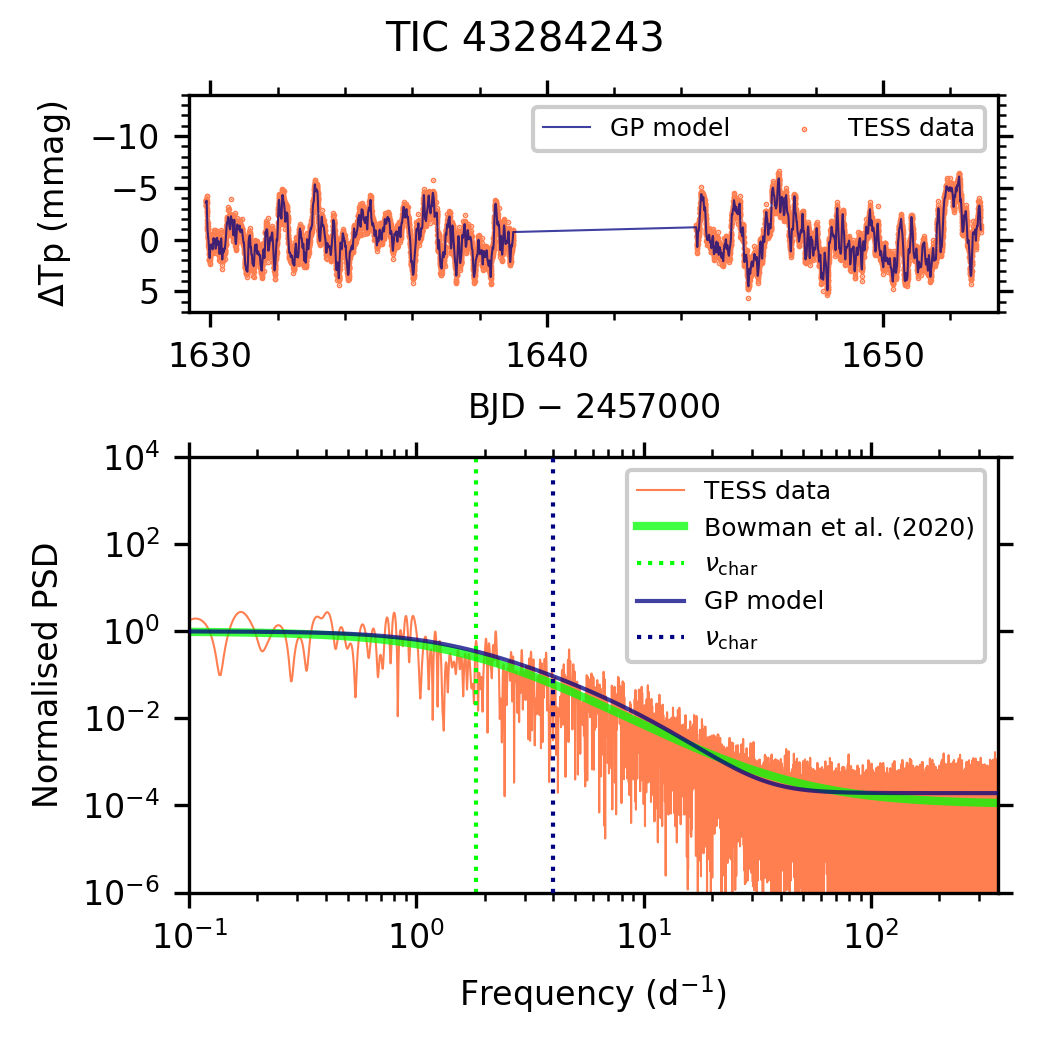}
\includegraphics[width=0.33\textwidth]{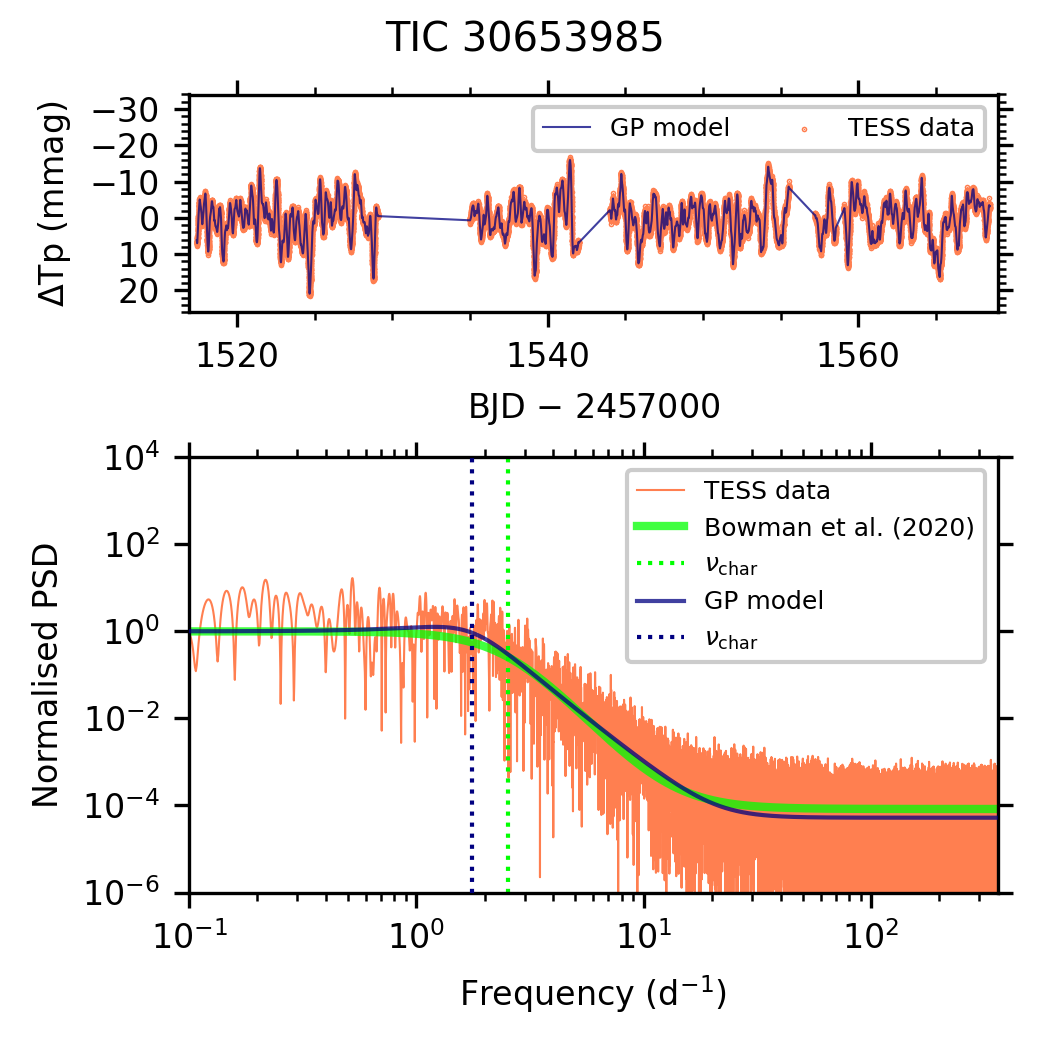}
\caption{Summary figures for massive stars analysed in this work.}
\label{figure: A2}
\end{figure*}

\begin{figure*}  
\centering
\includegraphics[width=0.33\textwidth]{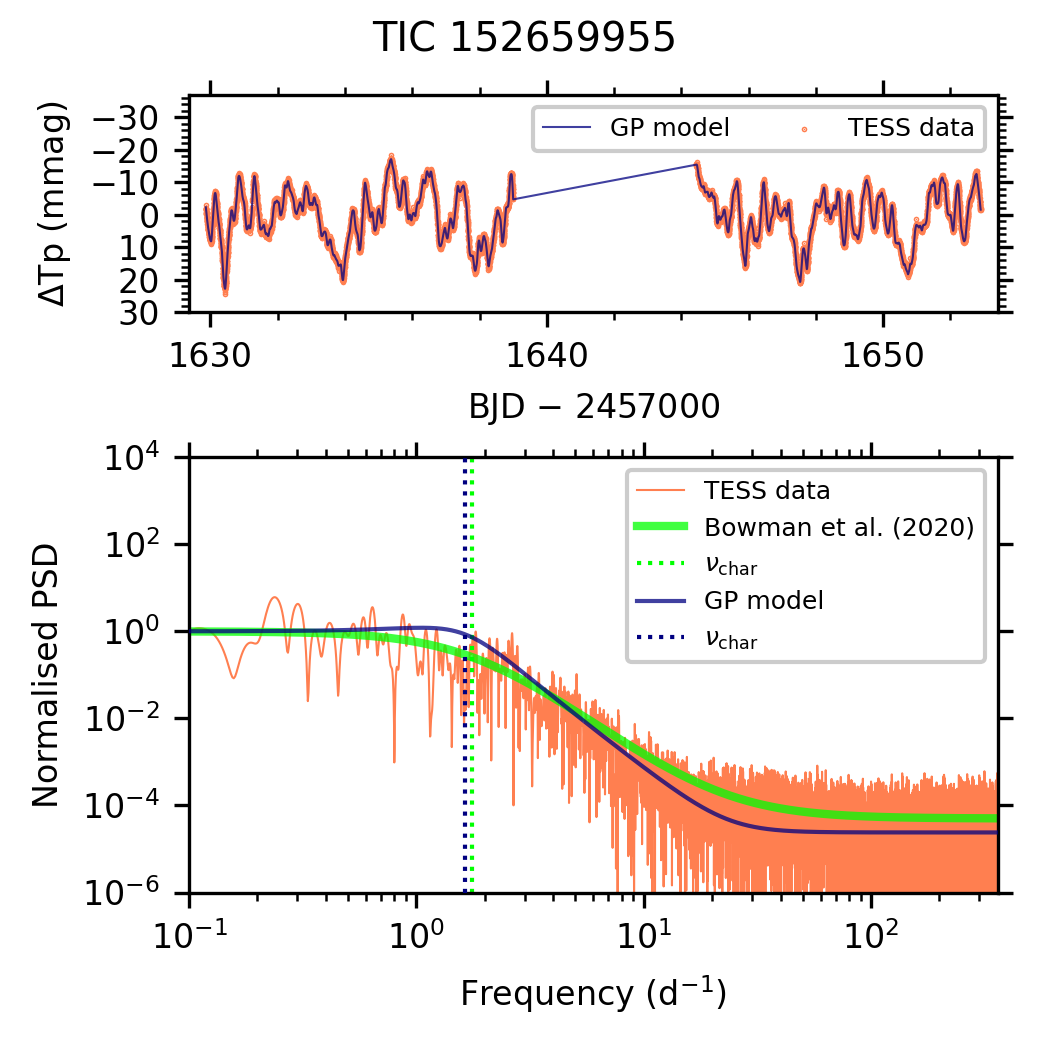}
\includegraphics[width=0.33\textwidth]{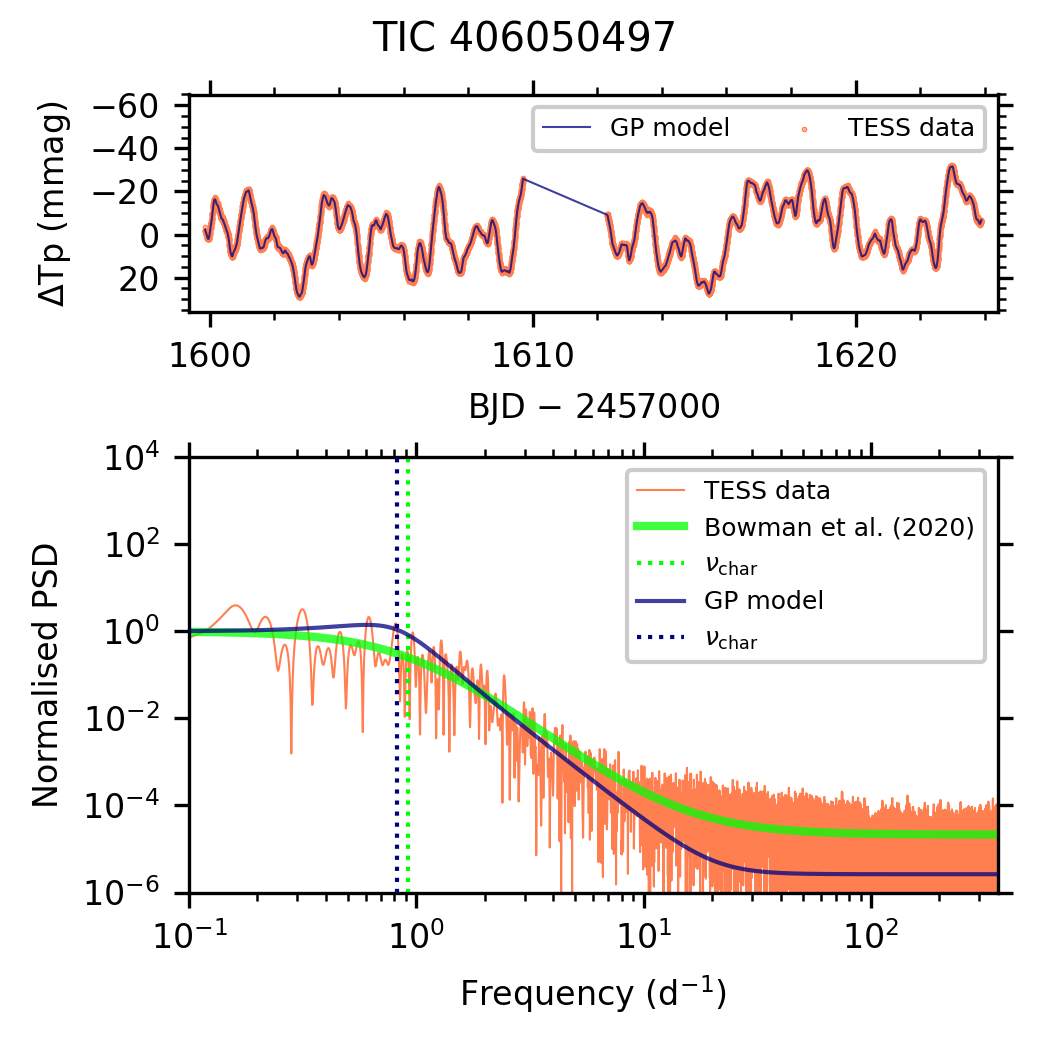}
\includegraphics[width=0.33\textwidth]{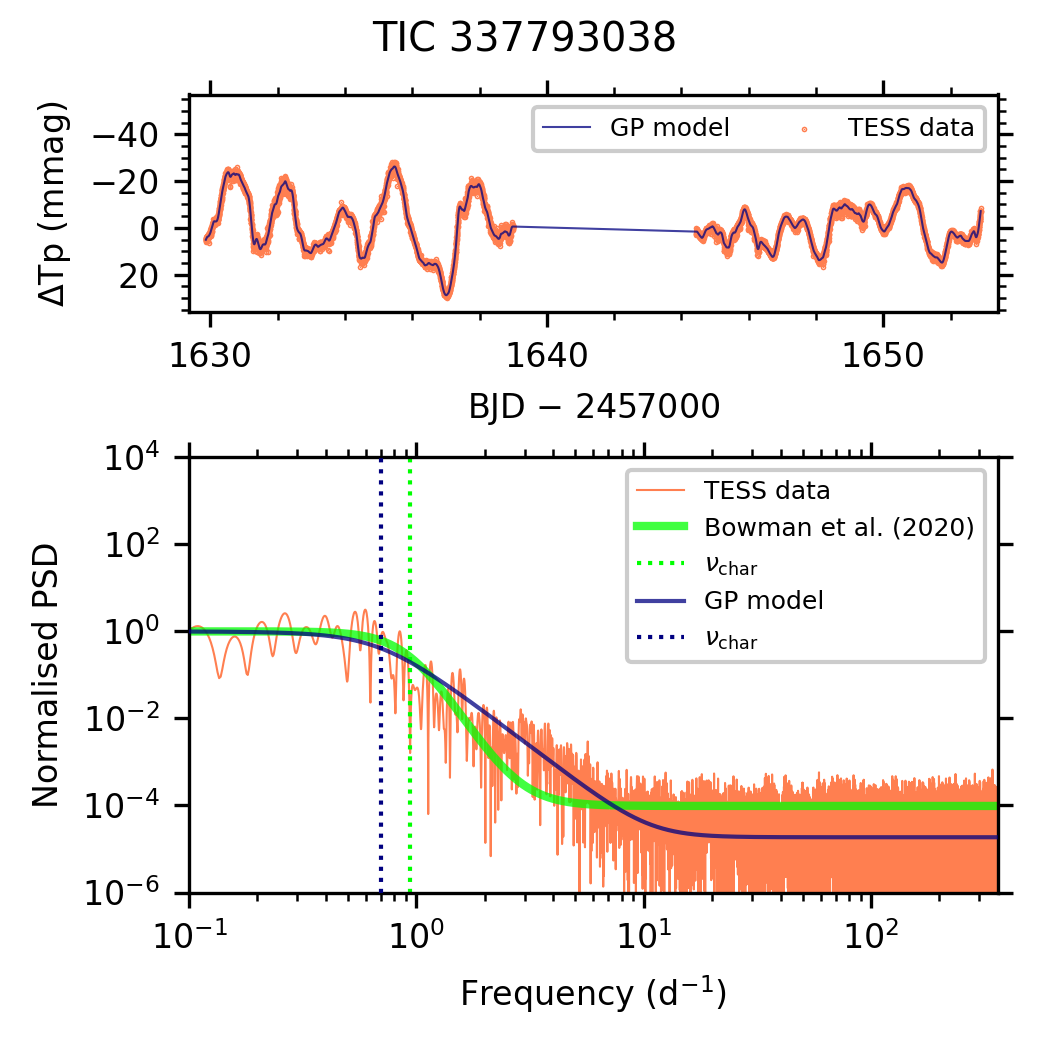}
\includegraphics[width=0.33\textwidth]{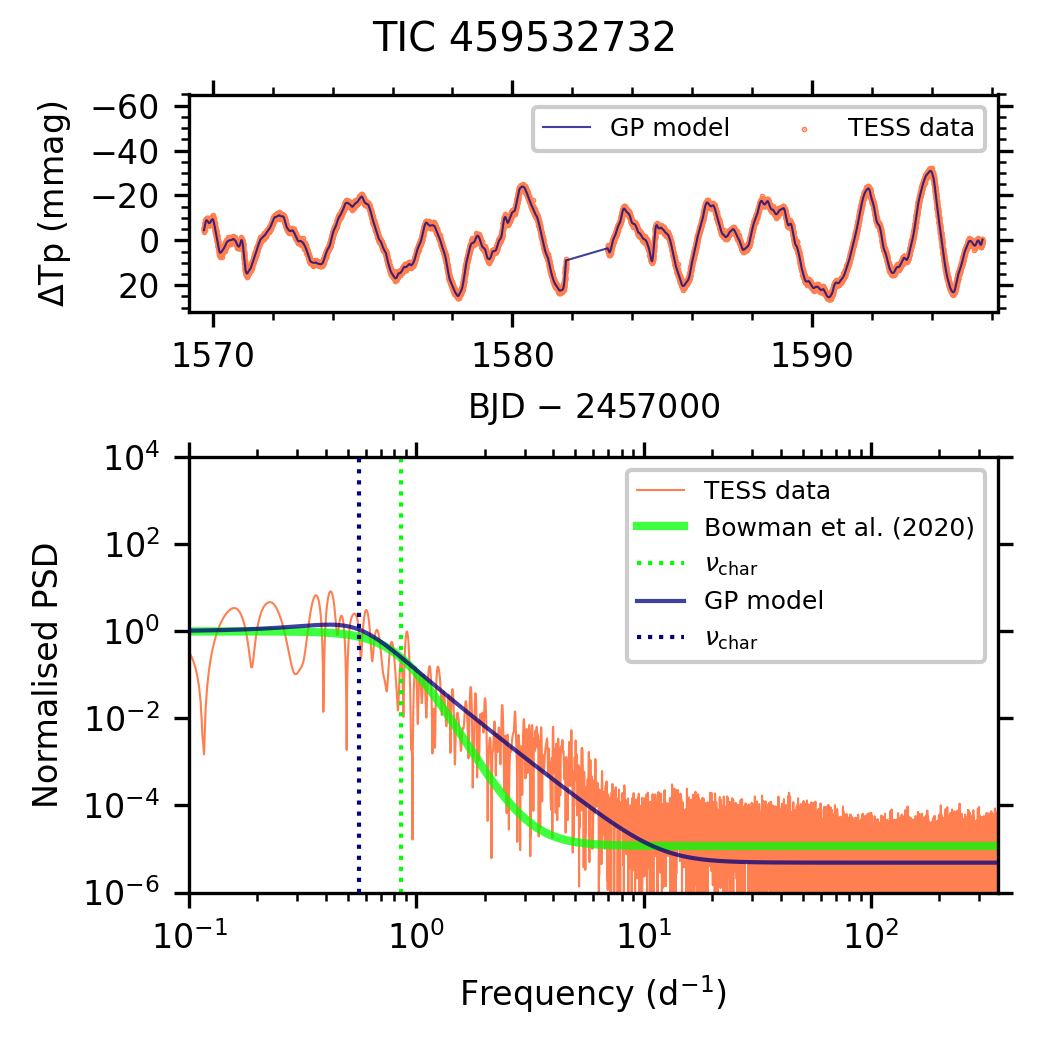}
\includegraphics[width=0.33\textwidth]{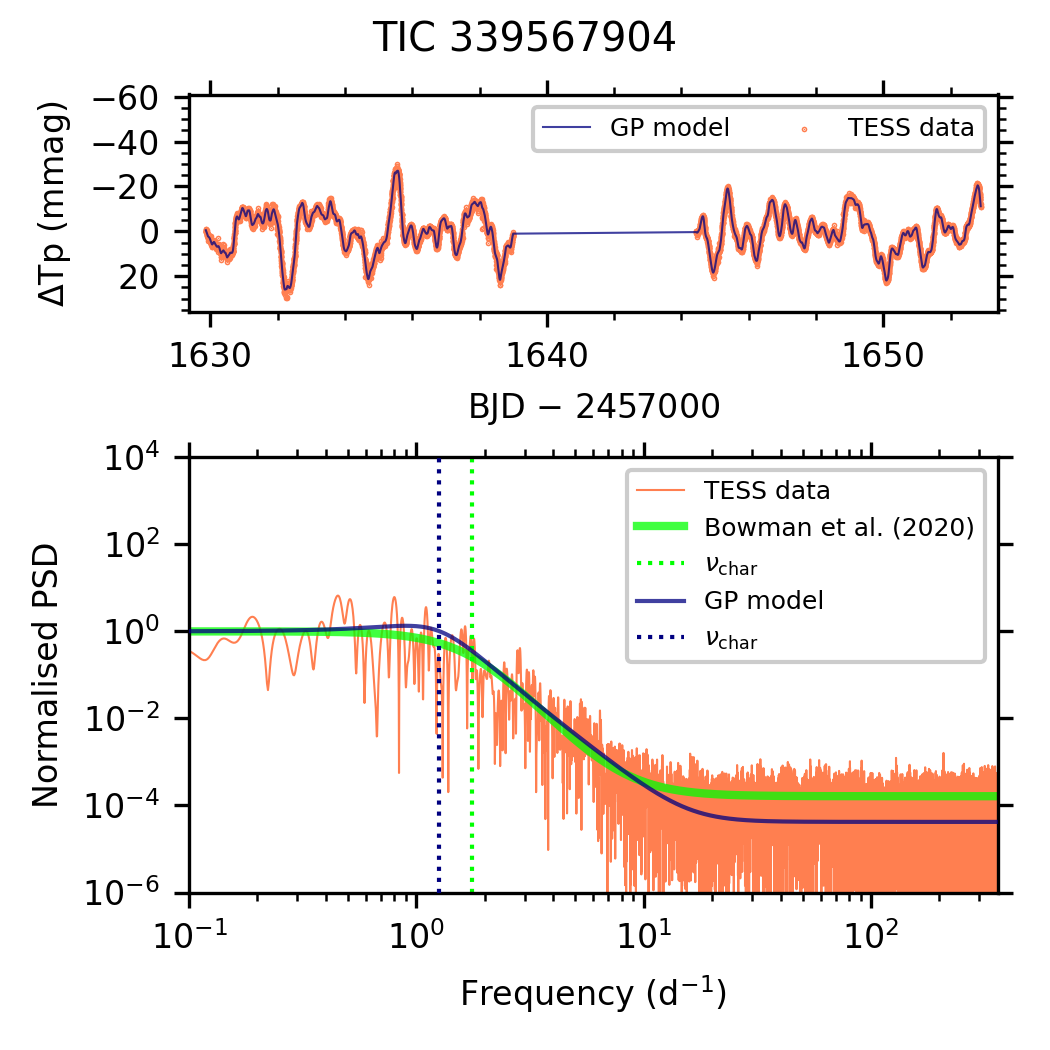}
\includegraphics[width=0.33\textwidth]{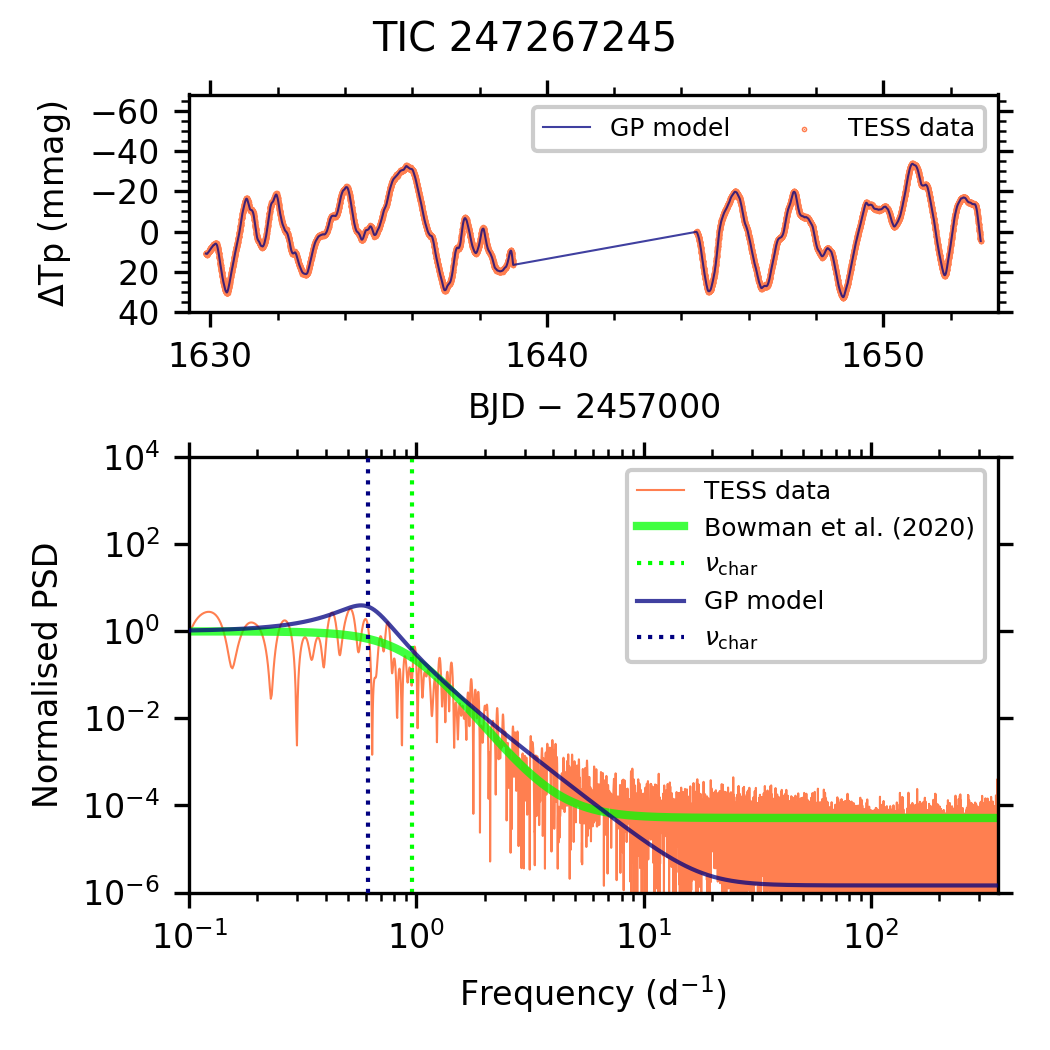}
\includegraphics[width=0.33\textwidth]{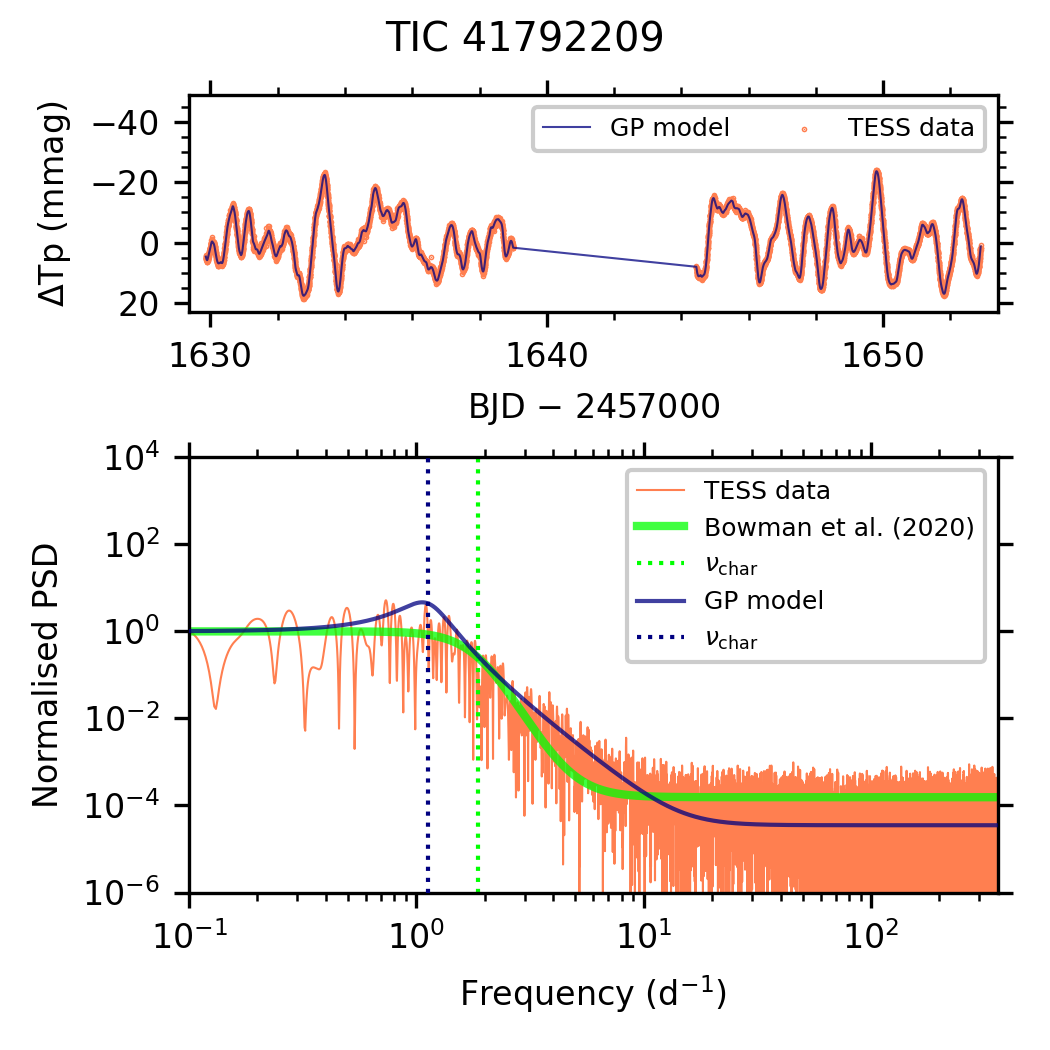}
\includegraphics[width=0.33\textwidth]{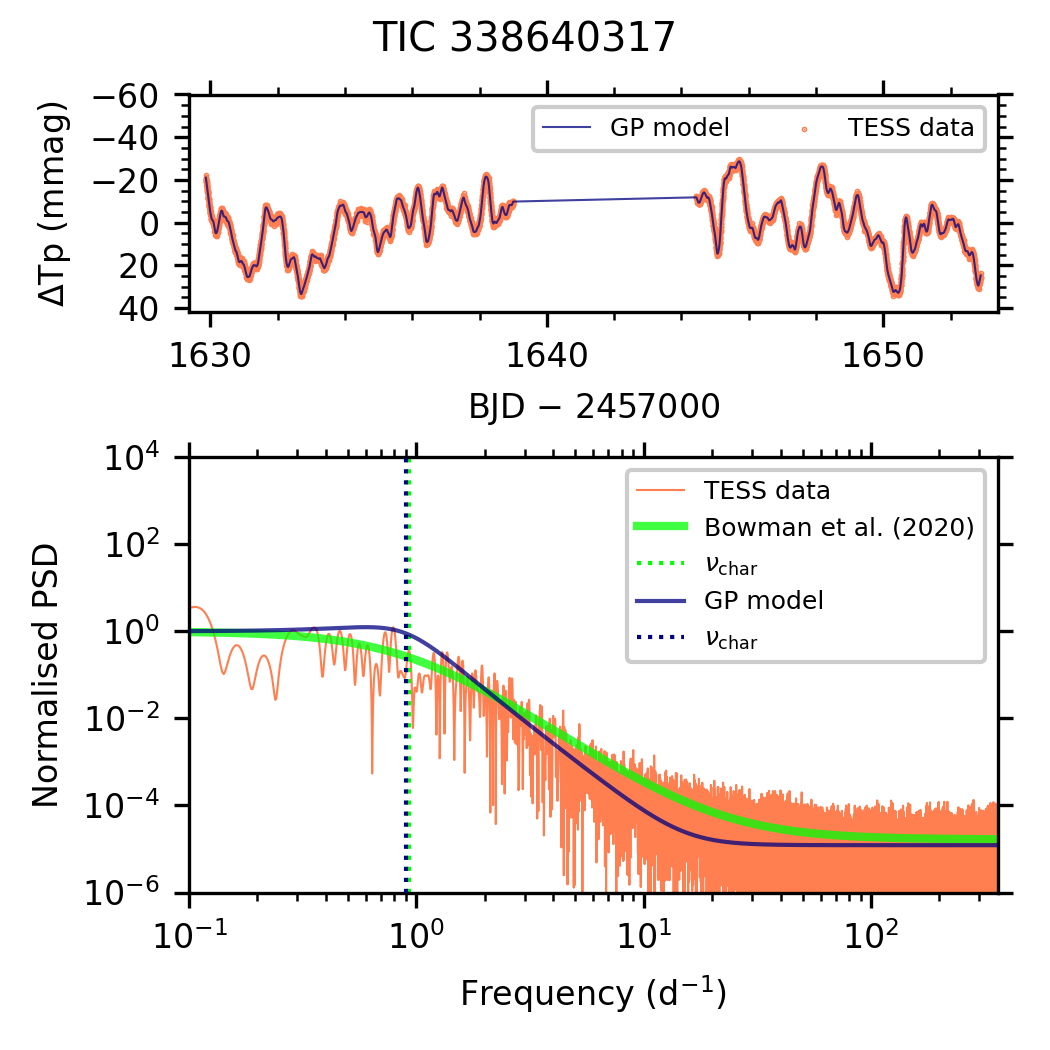}
\includegraphics[width=0.33\textwidth]{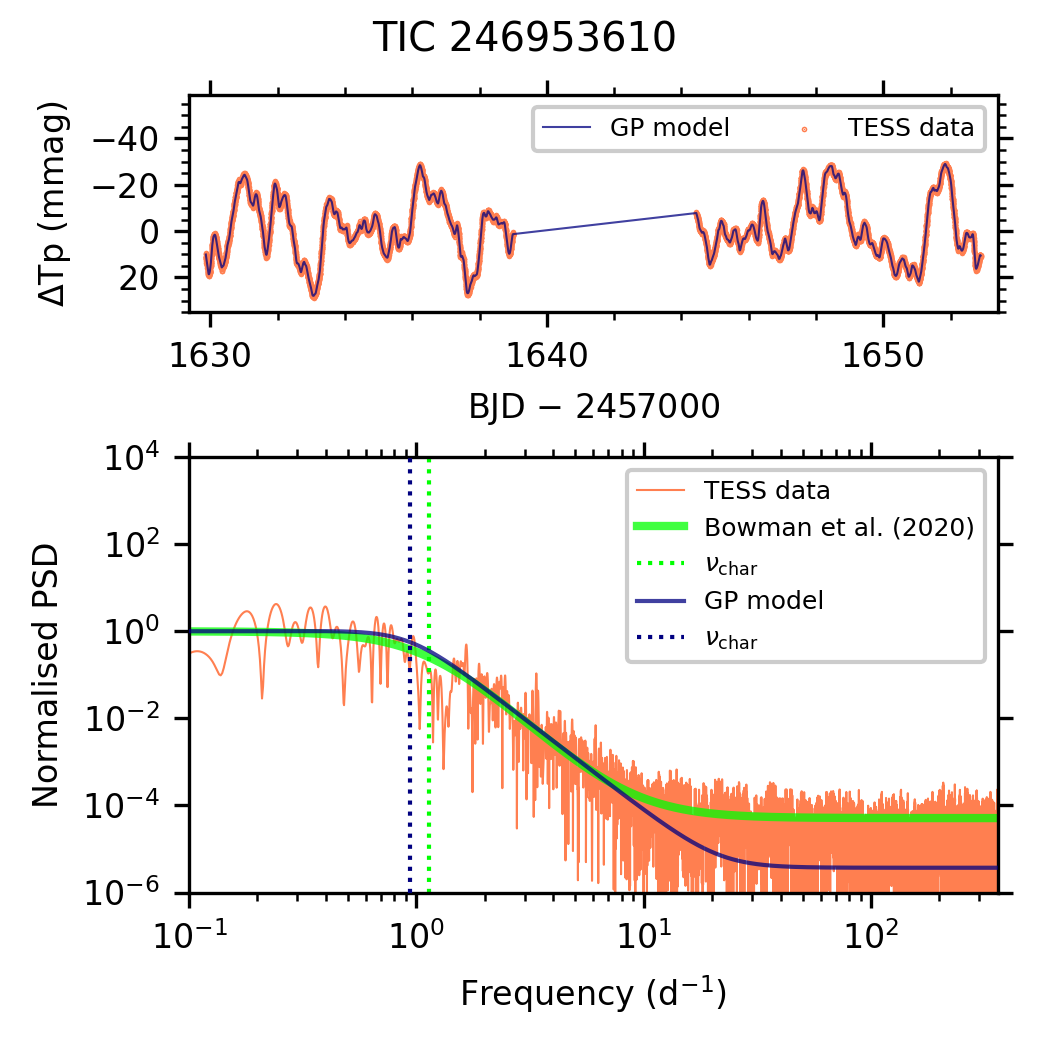}
\caption{Summary figures for massive stars analysed in this work.}
\label{figure: A3}
\end{figure*}

Summary figures for the 30 massive stars studied in this work containing the TESS light curve fitted using GP regression, the normalised power spectral density of the GP regression model, TESS data and the corresponding best-fit model determined by \citet{Bowman2020b} are given in Figs.~\ref{figure: A1}, \ref{figure: A2}, and \ref{figure: A3}. We note that for all summary figures that the minimal informative frequency (i.e. resolution) of the data set is set by the inverse of the time span, which is well below the chosen 0.1~d$^{-1}$ lower limit of the x-axes.

\section{Extended data tables}

The full parameter list of GP regression parameters including uncertainties determined by a Hamiltonian no U-turn MC sampler are given in Table~\ref{table: params} for the 30 massive stars studied in this work. In Table~\ref{table: params fixed Q}, we provide the GP regression parameters for when a fixed input value of $Q = 1 / \sqrt2$ is used.

\longtab{
\begin{landscape}

\begin{longtable}{l r c | c c c c c c c c}
\caption{\label{table: params} GP regression parameters for sample of massive stars studied in this work. } \\
\hline\hline
\multicolumn{2}{c}{} & \multicolumn{1}{c}{\citet{Bowman2020b}} && \multicolumn{7}{c}{GP regression (this work)} \\
\multicolumn{1}{c}{Name} & \multicolumn{1}{c}{TIC} & \multicolumn{1}{c}{$\nu_{\rm char}$} && \multicolumn{1}{c}{$\nu_{\rm char}$} & \multicolumn{1}{c}{$\rho_{\rm char}$} & \multicolumn{1}{c}{$\tau_{\rm damp}$} & \multicolumn{1}{c}{$\nu_{\rm damp}$} & \multicolumn{1}{c}{$\sigma_{\rm A}$} & \multicolumn{1}{c}{$Q$} & \multicolumn{1}{c}{$\ln(C_{\rm jitter})$} \\
\multicolumn{1}{c}{} & \multicolumn{1}{c}{} & \multicolumn{1}{c}{(d$^{-1}$)} && \multicolumn{1}{c}{(d$^{-1}$)} & \multicolumn{1}{c}{(d)} & \multicolumn{1}{c}{(d)} & \multicolumn{1}{c}{(d$^{-1}$)} & \multicolumn{1}{c}{(mag)} & \multicolumn{1}{c}{} & \multicolumn{1}{c}{} \\
\hline
\endfirsthead
\caption{\it continued.}\\
\hline\hline
\multicolumn{2}{c}{} & \multicolumn{1}{c}{\citet{Bowman2020b}} && \multicolumn{7}{c}{GP regression (this work)} \\
\multicolumn{1}{c}{Name} & \multicolumn{1}{c}{TIC} & \multicolumn{1}{c}{$\nu_{\rm char}$} && \multicolumn{1}{c}{$\nu_{\rm char}$} & \multicolumn{1}{c}{$\rho_{\rm char}$} & \multicolumn{1}{c}{$\tau_{\rm damp}$} & \multicolumn{1}{c}{$\nu_{\rm damp}$} & \multicolumn{1}{c}{$\sigma_{\rm A}$} & \multicolumn{1}{c}{$Q$} & \multicolumn{1}{c}{$\ln(C_{\rm jitter})$} \\
\multicolumn{1}{c}{} & \multicolumn{1}{c}{} & \multicolumn{1}{c}{(d$^{-1}$)} && \multicolumn{1}{c}{(d$^{-1}$)} & \multicolumn{1}{c}{(d)} & \multicolumn{1}{c}{(d)} & \multicolumn{1}{c}{(d$^{-1}$)} & \multicolumn{1}{c}{(mag)} & \multicolumn{1}{c}{} & \multicolumn{1}{c}{} \\
\hline
\endhead
\hline
\endfoot
HD~96715	&	306491594	&	$5.74308	\pm	0.00368$	&&	$7.94_{-0.39}^{+0.38}$	&	$0.126_{-0.006}^{+0.006}$	&	$0.017_{-0.002}^{+0.002}$	&	$59.4_{-5.9}^{+6.1}$		&	$0.00124_{-0.00006}^{+0.00006}$	&	$0.42_{-0.04}^{+0.04}$	&	$-7.619_{-0.008}^{+0.008}$	\\  
HD~46223	&	234881667	&	$3.99087	\pm	0.00135$	&&	$5.67_{-0.44}^{+0.45}$	&	$0.176_{-0.013}^{+0.014}$	&	$0.024_{-0.003}^{+0.003}$	&	$41.2_{-5.4}^{+5.6}$		&	$0.00174_{-0.00014}^{+0.00015}$	&	$0.43_{-0.06}^{+0.06}$	&	$-7.705_{-0.011}^{+0.012}$	\\   
HD~155913	&	216662610	&	$6.15743	\pm	0.00115$	&&	$5.26_{-0.46}^{+0.46}$	&	$0.190_{-0.015}^{+0.018}$	&	$0.027_{-0.004}^{+0.003}$	&	$37.3_{-4.9}^{+4.7}$		&	$0.00237_{-0.00020}^{+0.00022}$	&	$0.44_{-0.07}^{+0.06}$	&	$-7.671_{-0.012}^{+0.014}$	\\  
HD~46150	&	234840662	&	$3.95351	\pm	0.00181$	&&	$6.19_{-0.50}^{+0.48}$	&	$0.161_{-0.012}^{+0.013}$	&	$0.022_{-0.003}^{+0.003}$	&	$46.4_{-6.4}^{+6.3}$		&	$0.00120_{-0.00009}^{+0.00010}$	&	$0.42_{-0.06}^{+0.06}$	&	$-7.962_{-0.012}^{+0.012}$	\\  
HD~90273	&	464295672	&	$2.96658	\pm	0.00147$	&&	$5.92_{-0.37}^{+0.40}$	&	$0.169_{-0.011}^{+0.011}$	&	$0.018_{-0.002}^{+0.002}$	&	$56.2_{-6.2}^{+6.6}$		&	$0.00171_{-0.00010}^{+0.00010}$	&	$0.33_{-0.04}^{+0.03}$	&	$-7.249_{-0.008}^{+0.008}$	\\  
HD~53975	&	148506724	&	$2.77860	\pm	0.00194$	&&	$4.88_{-0.66}^{+0.65}$	&	$0.205_{-0.027}^{+0.029}$	&	$0.012_{-0.002}^{+0.002}$	&	$85.9_{-13.1}^{+13.8}$	&	$0.00126_{-0.00015}^{+0.00017}$	&	$0.18_{-0.03}^{+0.03}$	&	$-8.117_{-0.012}^{+0.012}$	\\  
HD~41997	&	294114621	&	$2.61022	\pm	0.00018$	&&	$2.44_{-0.28}^{+0.26}$	&	$0.409_{-0.044}^{+0.048}$	&	$0.060_{-0.008}^{+0.009}$	&	$16.7_{-2.3}^{+2.5}$		&	$0.00446_{-0.00052}^{+0.00058}$	&	$0.46_{-0.07}^{+0.09}$	&	$-7.613_{-0.011}^{+0.011}$	\\  
HD~46573	&	234947719	&	$2.00862	\pm	0.00051$	&&	$4.48_{-0.48}^{+0.47}$	&	$0.223_{-0.023}^{+0.024}$	&	$0.021_{-0.003}^{+0.003}$	&	$47.2_{-6.2}^{+6.3}$		&	$0.00201_{-0.00021}^{+0.00023}$	&	$0.30_{-0.05}^{+0.05}$	&	$-7.832_{-0.011}^{+0.013}$	\\  
HD~48279	&	234009943	&	$2.77586	\pm	0.00089$	&&	$4.09_{-0.43}^{+0.43}$	&	$0.244_{-0.023}^{+0.029}$	&	$0.032_{-0.006}^{+0.007}$	&	$31.6_{-5.9}^{+6.1}$		&	$0.00155_{-0.00015}^{+0.00017}$	&	$0.41_{-0.08}^{+0.07}$	&	$-6.891_{-0.011}^{+0.012}$	\\  
HD~74920	&	430625455	&	$3.66392	\pm	0.00053$	&&	$3.47_{-0.20}^{+0.20}$	&	$0.288_{-0.017}^{+0.016}$	&	$0.054_{-0.005}^{+0.005}$	&	$18.4_{-1.7}^{+1.7}$		&	$0.00327_{-0.00022}^{+0.00021}$	&	$0.59_{-0.06}^{+0.06}$	&	$-7.902_{-0.008}^{+0.009}$	\\  
HD~135591	&	455675248	&	$1.97700	\pm	0.00066$	&&	$4.24_{-0.36}^{+0.31}$	&	$0.236_{-0.018}^{+0.020}$	&	$0.029_{-0.003}^{+0.003}$	&	$34.8_{-3.2}^{+3.4}$		&	$0.00189_{-0.00016}^{+0.00016}$	&	$0.38_{-0.05}^{+0.05}$	&	$-8.588_{-0.012}^{+0.010}$	\\  
HD~326331	&	339568114	&	$2.20987	\pm	0.00017$	&&	$2.57_{-0.26}^{+0.27}$	&	$0.389_{-0.039}^{+0.042}$	&	$0.070_{-0.011}^{+0.011}$	&	$14.2_{-2.1}^{+2.4}$		&	$0.00541_{-0.00059}^{+0.00069}$	&	$0.57_{-0.10}^{+0.12}$	&	$-7.599_{-0.013}^{+0.013}$	\\  
HD~123056	&	330281456	&	$4.01672	\pm	0.00065$	&&	$3.09_{-0.40}^{+0.40}$	&	$0.324_{-0.037}^{+0.047}$	&	$0.029_{-0.004}^{+0.005}$	&	$34.2_{-5.2}^{+5.2}$		&	$0.00270_{-0.00032}^{+0.00040}$	&	$0.28_{-0.05}^{+0.05}$	&	$-7.567_{-0.012}^{+0.011}$	\\  
HD~97253	&	467065657	&	$2.20426	\pm	0.00019$	&&	$2.08_{-0.17}^{+0.16}$	&	$0.481_{-0.037}^{+0.040}$	&	$0.120_{-0.019}^{+0.019}$	&	$8.3_{-1.3}^{+1.4}$		&	$0.00451_{-0.00043}^{+0.00051}$	&	$0.78_{-0.13}^{+0.15}$	&	$-8.122_{-0.011}^{+0.010}$	\\  
HD~156738	&	195288472	&	$4.38608	\pm	0.00054$	&&	$4.01_{-0.37}^{+0.36}$	&	$0.249_{-0.022}^{+0.024}$	&	$0.040_{-0.006}^{+0.006}$	&	$24.8_{-3.8}^{+3.9}$		&	$0.00233_{-0.00023}^{+0.00024}$	&	$0.51_{-0.09}^{+0.09}$	&	$-7.476_{-0.012}^{+0.013}$	\\  
HD~36861	&	436103278	&	$2.05434	\pm	0.00077$	&&	$4.07_{-0.35}^{+0.36}$	&	$0.246_{-0.020}^{+0.023}$	&	$0.034_{-0.003}^{+0.003}$	&	$29.3_{-2.9}^{+2.9}$		&	$0.00160_{-0.00014}^{+0.00016}$	&	$0.44_{-0.06}^{+0.06}$	&	$-9.249_{-0.012}^{+0.013}$	\\  
HD~150574	&	234648113	&	$2.47023	\pm	0.00017$	&&	$2.56_{-0.27}^{+0.25}$	&	$0.391_{-0.040}^{+0.040}$	&	$0.077_{-0.012}^{+0.014}$	&	$13.1_{-2.2}^{+2.2}$		&	$0.00568_{-0.00060}^{+0.00074}$	&	$0.61_{-0.12}^{+0.12}$	&	$-7.340_{-0.012}^{+0.013}$	\\  
HD~152247	&	339570292	&	$2.75008	\pm	0.00027$	&&	$2.07_{-0.28}^{+0.29}$	&	$0.483_{-0.060}^{+0.075}$	&	$0.066_{-0.011}^{+0.011}$	&	$15.2_{-2.5}^{+2.6}$		&	$0.00442_{-0.00066}^{+0.00071}$	&	$0.43_{-0.09}^{+0.10}$	&	$-7.368_{-0.013}^{+0.012}$	\\  
HD~55879	&	178489528	&	$1.71966	\pm	0.00108$	&&	$5.41_{-0.62}^{+0.64}$	&	$0.185_{-0.020}^{+0.023}$	&	$0.013_{-0.002}^{+0.002}$	&	$79.7_{-12.8}^{+12.0}$	&	$0.00091_{-0.00009}^{+0.00011}$	&	$0.21_{-0.03}^{+0.03}$	&	$-8.286_{-0.012}^{+0.011}$	\\  
HD~154643	&	43284243		&	$1.83226	\pm	0.00041$	&&	$3.96_{-0.47}^{+0.48}$	&	$0.253_{-0.029}^{+0.033}$	&	$0.024_{-0.003}^{+0.003}$	&	$41.5_{-5.4}^{+6.2}$		&	$0.00208_{-0.00023}^{+0.00029}$	&	$0.30_{-0.05}^{+0.05}$	&	$-7.992_{-0.013}^{+0.013}$	\\  
CPD-47~2963	&	30653985		&	$2.53356	\pm	0.00019$	&&	$1.75_{-0.11}^{+0.12}$	&	$0.571_{-0.036}^{+0.038}$	&	$0.174_{-0.023}^{+0.026}$	&	$5.7_{-0.8}^{+0.8}$		&	$0.00527_{-0.00043}^{+0.00049}$	&	$0.96_{-0.14}^{+0.15}$	&	$-7.841_{-0.008}^{+0.008}$	\\  
HD~156154	&	152659955	&	$1.76469	\pm	0.00008$	&&	$1.62_{-0.18}^{+0.17}$	&	$0.618_{-0.059}^{+0.075}$	&	$0.181_{-0.035}^{+0.043}$	&	$5.5_{-1.1}^{+1.3}$		&	$0.00790_{-0.00104}^{+0.00120}$	&	$0.92_{-0.21}^{+0.25}$	&	$-7.811_{-0.012}^{+0.013}$	\\ 
HD~112244	&	406050497	&	$0.91870	\pm	0.00003$	&&	$0.81_{-0.10}^{+0.12}$	&	$1.236_{-0.156}^{+0.176}$	&	$0.408_{-0.097}^{+0.142}$	&	$2.4_{-0.6}^{+0.7}$		&	$0.01374_{-0.00249}^{+0.00285}$	&	$1.03_{-0.28}^{+0.39}$	&	$-8.012_{-0.010}^{+0.012}$	\\  
HD~151804	&	337793038	&	$0.93978	\pm	0.00001$	&&	$0.67_{-0.15}^{+0.13}$	&	$1.499_{-0.275}^{+0.390}$	&	$0.296_{-0.084}^{+0.103}$	&	$3.4_{-1.0}^{+1.1}$		&	$0.01138_{-0.00261}^{+0.00323}$	&	$0.62_{-0.22}^{+0.25}$	&	$-7.006_{-0.012}^{+0.013}$	\\  
HD~303492	&	459532732	&	$0.85247	\pm	0.00001$	&&	$0.54_{-0.09}^{+0.08}$	&	$1.836_{-0.261}^{+0.327}$	&	$0.613_{-0.174}^{+0.233}$	&	$1.6_{-0.5}^{+0.5}$		&	$0.01316_{-0.00266}^{+0.00329}$	&	$1.05_{-0.35}^{+0.41}$	&	$-7.540_{-0.010}^{+0.011}$	\\  
HD~152249	&	339567904	&	$1.76168	\pm	0.00005$	&&	$1.24_{-0.15}^{+0.15}$	&	$0.804_{-0.090}^{+0.101}$	&	$0.256_{-0.060}^{+0.073}$	&	$3.9_{-1.0}^{+0.9}$		&	$0.00931_{-0.00142}^{+0.00170}$	&	$1.00_{-0.27}^{+0.30}$	&	$-7.287_{-0.013}^{+0.012}$	\\  
HD~152424	&	247267245	&	$0.95329	\pm	0.00001$	&&	$0.61_{-0.08}^{+0.07}$	&	$1.652_{-0.188}^{+0.229}$	&	$1.058_{-0.426}^{+0.662}$	&	$0.9_{-0.5}^{+0.4}$		&	$0.01715_{-0.00383}^{+0.00550}$	&	$2.00_{-0.84}^{+1.31}$	&	$-8.352_{-0.012}^{+0.012}$	\\  
HD~154368	&	41792209		&	$1.85794	\pm	0.00004$	&&	$1.12_{-0.11}^{+0.08}$	&	$0.894_{-0.067}^{+0.086}$	&	$0.612_{-0.199}^{+0.275}$	&	$1.6_{-0.6}^{+0.6}$		&	$0.00857_{-0.00145}^{+0.00188}$	&	$2.15_{-0.71}^{+1.02}$	&	$-7.775_{-0.012}^{+0.012}$	\\  
HD~152003	&	338640317	&	$0.92626	\pm	0.00003$	&&	$0.88_{-0.13}^{+0.13}$	&	$1.139_{-0.160}^{+0.195}$	&	$0.341_{-0.087}^{+0.117}$	&	$2.9_{-0.9}^{+0.9}$		&	$0.01354_{-0.00256}^{+0.00297}$	&	$0.94_{-0.29}^{+0.38}$	&	$-7.339_{-0.011}^{+0.013}$	\\  
HD~152147	&	246953610	&	$1.13960	\pm	0.00002$	&&	$0.91_{-0.16}^{+0.13}$	&	$1.095_{-0.155}^{+0.200}$	&	$0.259_{-0.067}^{+0.070}$	&	$3.9_{-1.0}^{+1.0}$		&	$0.01296_{-0.00233}^{+0.00294}$	&	$0.74_{-0.21}^{+0.25}$	&	$-7.884_{-0.014}^{+0.012}$	\\  

\end{longtable}
\end{landscape}
}

\longtab{
\begin{landscape}

\begin{longtable}{l r c | c c c c c}
\caption{\label{table: params fixed Q} GP regression parameters for sample of massive stars studied in this work when fixed $Q = 1 / \sqrt2$. } \\
\hline\hline
\multicolumn{2}{c}{} & \multicolumn{1}{c}{\citet{Bowman2020b}} && \multicolumn{4}{c}{GP regression (this work)} \\
\multicolumn{1}{c}{Name} & \multicolumn{1}{c}{TIC} & \multicolumn{1}{c}{$\nu_{\rm char}$} && \multicolumn{1}{c}{$\nu_{\rm char}$} & \multicolumn{1}{c}{$\rho_{\rm char}$} & \multicolumn{1}{c}{$\sigma_{\rm A}$} & \multicolumn{1}{c}{$\ln(C_{\rm jitter})$} \\
\multicolumn{1}{c}{} & \multicolumn{1}{c}{} & \multicolumn{1}{c}{(d$^{-1}$)} && \multicolumn{1}{c}{(d$^{-1}$)} & \multicolumn{1}{c}{(d)} & \multicolumn{1}{c}{(mag)} & \multicolumn{1}{c}{} \\
\hline
\endfirsthead
\caption{\it continued.}\\
\hline\hline
\multicolumn{2}{c}{} & \multicolumn{1}{c}{\citet{Bowman2020b}} && \multicolumn{4}{c}{GP regression (this work)} \\
\multicolumn{1}{c}{Name} & \multicolumn{1}{c}{TIC} & \multicolumn{1}{c}{$\nu_{\rm char}$} && \multicolumn{1}{c}{$\nu_{\rm char}$} & \multicolumn{1}{c}{$\rho_{\rm char}$} & \multicolumn{1}{c}{$\sigma_{\rm A}$} & \multicolumn{1}{c}{$\ln(C_{\rm jitter})$} \\
\multicolumn{1}{c}{} & \multicolumn{1}{c}{} & \multicolumn{1}{c}{(d$^{-1}$)} && \multicolumn{1}{c}{(d$^{-1}$)} & \multicolumn{1}{c}{(d)} & \multicolumn{1}{c}{((mag)} & \multicolumn{1}{c}{} \\
\hline
\endhead
\hline
\endfoot
HD~96715	&	306491594	&	$5.74308	\pm	0.00368$	&&	$7.97_{-0.28}^{+0.30}$	&	$0.125_{-0.005}^{+0.005}$	&	$0.00124_{-0.00005}^{+0.00005}$	&	$-7.613_{-0.008}^{+0.008}$	\\
HD~46223	&	234881667	&	$3.99087	\pm	0.00135$	&&	$5.98_{-0.34}^{+0.35}$	&	$0.167_{-0.009}^{+0.010}$	&	$0.00173_{-0.00012}^{+0.00012}$	&	$-7.702_{-0.011}^{+0.012}$	\\
HD~155913	&	216662610	&	$6.15743	\pm	0.00115$	&&	$5.62_{-0.35}^{+0.35}$	&	$0.178_{-0.011}^{+0.011}$	&	$0.00235_{-0.00017}^{+0.00020}$	&	$-7.667_{-0.013}^{+0.012}$	\\
HD~46150	&	234840662	&	$3.95351	\pm	0.00181$	&&	$6.44_{-0.37}^{+0.35}$	&	$0.155_{-0.008}^{+0.009}$	&	$0.00120_{-0.00008}^{+0.00009}$	&	$-7.958_{-0.012}^{+0.012}$	\\
HD~90273	&	464295672	&	$2.96658	\pm	0.00147$	&&	$5.91_{-0.24}^{+0.25}$	&	$0.169_{-0.007}^{+0.007}$	&	$0.00171_{-0.00008}^{+0.00008}$	&	$-7.242_{-0.008}^{+0.007}$	\\
HD~53975	&	148506724	&	$2.77860	\pm	0.00194$	&&	$5.25_{-0.30}^{+0.31}$	&	$0.191_{-0.012}^{+0.011}$	&	$0.00125_{-0.00008}^{+0.00009}$	&	$-8.103_{-0.011}^{+0.012}$	\\
HD~41997	&	294114621	&	$2.61022	\pm	0.00018$	&&	$2.70_{-0.21}^{+0.20}$	&	$0.371_{-0.026}^{+0.031}$	&	$0.00443_{-0.00041}^{+0.00047}$	&	$-7.611_{-0.011}^{+0.012}$	\\
HD~46573	&	234947719	&	$2.00862	\pm	0.00051$	&&	$4.97_{-0.30}^{+0.30}$	&	$0.201_{-0.012}^{+0.013}$	&	$0.00199_{-0.00014}^{+0.00016}$	&	$-7.825_{-0.012}^{+0.012}$	\\
HD~48279	&	234009943	&	$2.77586	\pm	0.00089$	&&	$4.15_{-0.34}^{+0.33}$	&	$0.241_{-0.020}^{+0.018}$	&	$0.00154_{-0.00013}^{+0.00013}$	&	$-6.889_{-0.011}^{+0.012}$	\\
HD~74920	&	430625455	&	$3.66392	\pm	0.00053$	&&	$3.61_{-0.17}^{+0.17}$	&	$0.277_{-0.013}^{+0.013}$	&	$0.00326_{-0.00019}^{+0.00022}$	&	$-7.901_{-0.009}^{+0.008}$	\\
HD~135591	&	455675248	&	$1.97700	\pm	0.00066$	&&	$4.81_{-0.24}^{+0.25}$	&	$0.208_{-0.010}^{+0.011}$	&	$0.00189_{-0.00012}^{+0.00014}$	&	$-8.584_{-0.011}^{+0.011}$	\\
HD~326331	&	339568114	&	$2.20987	\pm	0.00017$	&&	$2.71_{-0.22}^{+0.24}$	&	$0.369_{-0.030}^{+0.033}$	&	$0.00540_{-0.00063}^{+0.00060}$	&	$-7.598_{-0.013}^{+0.012}$	\\
HD~123056	&	330281456	&	$4.01672	\pm	0.00065$	&&	$3.50_{-0.23}^{+0.26}$	&	$0.286_{-0.020}^{+0.020}$	&	$0.00267_{-0.00023}^{+0.00025}$	&	$-7.561_{-0.011}^{+0.011}$	\\
HD~97253	&	467065657	&	$2.20426	\pm	0.00019$	&&	$2.03_{-0.15}^{+0.16}$	&	$0.493_{-0.037}^{+0.039}$	&	$0.00449_{-0.00045}^{+0.00056}$	&	$-8.122_{-0.011}^{+0.011}$	\\
HD~156738	&	195288472	&	$4.38608	\pm	0.00054$	&&	$4.19_{-0.30}^{+0.29}$	&	$0.239_{-0.016}^{+0.017}$	&	$0.00232_{-0.00020}^{+0.00022}$	&	$-7.474_{-0.012}^{+0.012}$	\\
HD~36861	&	436103278	&	$2.05434	\pm	0.00077$	&&	$4.59_{-0.26}^{+0.26}$	&	$0.218_{-0.012}^{+0.013}$	&	$0.00159_{-0.00013}^{+0.00013}$	&	$-9.247_{-0.012}^{+0.013}$	\\
HD~150574	&	234648113	&	$2.47023	\pm	0.00017$	&&	$2.64_{-0.21}^{+0.23}$	&	$0.379_{-0.031}^{+0.032}$	&	$0.00565_{-0.00054}^{+0.00075}$	&	$-7.339_{-0.012}^{+0.013}$	\\
HD~152247	&	339570292	&	$2.75008	\pm	0.00027$	&&	$2.31_{-0.20}^{+0.20}$	&	$0.434_{-0.037}^{+0.038}$	&	$0.00437_{-0.00048}^{+0.00055}$	&	$-7.366_{-0.012}^{+0.013}$	\\
HD~55879	&	178489528	&	$1.71966	\pm	0.00108$	&&	$5.70_{-0.35}^{+0.32}$	&	$0.176_{-0.010}^{+0.010}$	&	$0.00091_{-0.00006}^{+0.00006}$	&	$-8.274_{-0.011}^{+0.012}$	\\
HD~154643	&	43284243		&	$1.83226	\pm	0.00041$	&&	$4.52_{-0.30}^{+0.30}$	&	$0.221_{-0.015}^{+0.015}$	&	$0.00206_{-0.00016}^{+0.00019}$	&	$-7.986_{-0.013}^{+0.013}$	\\
CPD-47~2963	&	30653985		&	$2.53356	\pm	0.00019$	&&	$1.62_{-0.11}^{+0.11}$	&	$0.616_{-0.041}^{+0.042}$	&	$0.00526_{-0.00048}^{+0.00053}$	&	$-7.842_{-0.008}^{+0.008}$	\\
HD~156154	&	152659955	&	$1.76469	\pm	0.00008$	&&	$1.51_{-0.17}^{+0.16}$	&	$0.664_{-0.067}^{+0.078}$	&	$0.00788_{-0.00108}^{+0.00138}$	&	$-7.811_{-0.013}^{+0.012}$	\\
HD~112244	&	406050497	&	$0.91870	\pm	0.00003$	&&	$0.72_{-0.10}^{+0.11}$	&	$1.388_{-0.180}^{+0.221}$	&	$0.01379_{-0.00260}^{+0.00310}$	&	$-8.012_{-0.012}^{+0.011}$	\\
HD~151804	&	337793038	&	$0.93978	\pm	0.00001$	&&	$0.70_{-0.12}^{+0.11}$	&	$1.431_{-0.219}^{+0.253}$	&	$0.01105_{-0.00219}^{+0.00286}$	&	$-7.005_{-0.012}^{+0.013}$	\\
HD~303492	&	459532732	&	$0.85247	\pm	0.00001$	&&	$0.49_{-0.08}^{+0.08}$	&	$2.031_{-0.305}^{+0.356}$	&	$0.01288_{-0.00283}^{+0.00330}$	&	$-7.540_{-0.010}^{+0.011}$	\\
HD~152249	&	339567904	&	$1.76168	\pm	0.00005$	&&	$1.14_{-0.13}^{+0.15}$	&	$0.879_{-0.105}^{+0.112}$	&	$0.00923_{-0.00149}^{+0.00178}$	&	$-7.287_{-0.012}^{+0.013}$	\\
HD~152424	&	247267245	&	$0.95329	\pm	0.00001$	&&	$0.43_{-0.09}^{+0.09}$	&	$2.308_{-0.430}^{+0.581}$	&	$0.01751_{-0.00459}^{+0.00672}$	&	$-8.352_{-0.013}^{+0.011}$	\\
HD~154368	&	41792209		&	$1.85794	\pm	0.00004$	&&	$0.83_{-0.12}^{+0.11}$	&	$1.201_{-0.165}^{+0.178}$	&	$0.00855_{-0.00162}^{+0.00190}$	&	$-7.776_{-0.013}^{+0.013}$	\\
HD~152003	&	338640317	&	$0.92626	\pm	0.00003$	&&	$0.81_{-0.12}^{+0.11}$	&	$1.241_{-0.169}^{+0.200}$	&	$0.01349_{-0.00253}^{+0.00319}$	&	$-7.339_{-0.012}^{+0.012}$	\\
HD~152147	&	246953610	&	$1.13960	\pm	0.00002$	&&	$0.90_{-0.13}^{+0.12}$	&	$1.106_{-0.151}^{+0.157}$	&	$0.01283_{-0.00258}^{+0.00269}$	&	$-7.885_{-0.012}^{+0.012}$	\\

\end{longtable}
\end{landscape}
}

\end{appendix}


\end{document}